\documentclass[aps,preprint,superscriptaddress]{revtex4-1}

\usepackage{graphicx}
\usepackage{amsmath}
\usepackage{amssymb}
\usepackage{slashed}

\newcommand{\tr}{\rm tr \,}

\newcommand{\der}{\partial}
\newcommand{\CD}{\ooalign{\hfil$\cdot$\hfil\crcr$\der$}}

\usepackage{array} 
\newcolumntype{C}{>{$}c<{$}}

\begin{document}


\title{Coupled-channel dynamics with chiral long-range forces \\ in the open-charm sector of QCD}


\author{Matthias F.M. Lutz}
\affiliation{GSI Helmholtzzentrum f\"ur Schwerionenforschung GmbH, \\Planckstra\ss e 1, 64291 Darmstadt, Germany}
\author{Xiao-Yu Guo}
\affiliation{Beijing University of Technology,\\
  Beijing 100124, China}
\author{Yonggoo Heo}
\affiliation{Bogoliubov Laboratory for Theoretical Physics, Joint Institute for Nuclear Research, RU-141980 Dubna, Moscow region, Russia}
\affiliation{GSI Helmholtzzentrum f\"ur Schwerionenforschung GmbH, \\Planckstra\ss e 1, 64291 Darmstadt, Germany}
\author{C.L. Korpa}
\affiliation{MTA-PTE High-Field Terahertz Research Group,\\ Ifj\'us\'ag \'utja 6,
7624 P\'ecs, Hungary, \\and \\Institute of Physics, University of P\'ecs,\\ Ifj\'us\'ag \'utja 6,
7624 P\'ecs, Hungary}
\date{\today}

\begin{abstract}

We perform an analysis of Lattice QCD data in the open-charm sector based on the chiral SU(3) Lagrangian. The low-energy constants are adjusted to recover the open-charm meson masses on Lattice QCD ensembles from HPQCD, ETMC and HSC with pion and kaon masses smaller than 550 MeV. A significant set of low-energy parameters is obtainable only if the most recent information from  HSC on scattering observables is included in our global fit. For the first time our analysis considers the effect of left-hand cuts as developed in terms of a generalized potential approach (GPA) previously by one of the authors. Here we use coupled-channel interaction terms at the one-loop level. 
The elastic s-wave and p-wave $D\,\pi$, $D K $  and $D \bar K  $ scattering phase shifts on ensembles with nominal pion masses of about 239 MeV and 391 MeV are reproduced faithfully. Based on such low-energy parameters we predict s- and p-wave phase shifts and inelasticities at physical quark masses, where the statistical uncertainties in the phase shifts are smaller than 1 degree always. 
Most striking would be the exotic s-wave $D_s \pi$ channel, for which we predict a resonance state at about 2.287 GeV where the phase shift passes through 90 degrees. 
\end{abstract}


\maketitle
\tableofcontents

\newpage


\section{Introduction}
\label{sec:1}

A decades long challenge of modern physics is the request to describe and predict strongly interacting coupled-channel systems in QCD. With recent progress in Lattice QCD and effective field theory  approaches such systems can be tackled successfully in a combination of the two approaches. 
In this work we wish to illustrate the enormous potential of this new strategy in hadron physics at the hands of systems with one heavy and one light quark \cite{Casalbuoni:1996pg,Lutz:2015ejy,Chen:2016spr}. 

Such open-charm systems are particularly useful, since they are constrained by 
two distinct approximate symmetries, the heavy-quark spin symmetry of the charm quark and  the chiral symmetry of the up, down and strange quarks. Both guide the construction of the effective field theory approaches pioneered in a series of works \cite{Kolomeitsev:2003ac,Hofmann:2003je,Lutz:2007sk}. 
Here we wish to reconsider the size of the counter terms in their chiral Lagrangian. First rough studies \cite{Hofmann:2003je,Lutz:2007sk} suffered from limited  empirical constraints.
Additional information from first Lattice QCD simulation on a set of s-wave scattering lengths was used in a series of later works \cite{Liu:2012zya,Altenbuchinger:2013vwa,Cleven:2014oka,Du:2016tgp,Huang:2022cag}. 
Results were obtained that in part showed unnaturally large counter terms and/or illustrated some residual dependence on how to  set up the coupled-channel computation. In addition one may worry that the used 
MILC Asqtad ensembles suffer from strange quark masses which are at the border or even beyond the applicability domain of the chiral approach. 

In a more recent work by the GSI group \cite{Guo:2018kno} it was pointed out that the 
relevant counter terms used in the description of the s-wave scattering lengths \cite{Liu:2012zya} 
impact also the quark-mass dependence of the $D$ meson masses, for which an even more significant data basis from Lattice QCD groups exists \cite{Aoki:2008sm,Mohler:2011ke,Na:2012iu,Kalinowski:2015bwa,Cichy:2016bci,Cheung:2016bym,Moir:2016srx}. In turn global fits considering both sources lead in part to quantitatively distinct results \cite{Guo:2021kdo}. Moreover, in  \cite{Guo:2018kno} dedicated predictions for scattering phase shifts on an ensemble of HSC at pion masses of about 
239 MeV were made.  First results by HSC were reported in  \cite{Cheung:2020mql,Gayer:2021xzv}, which we confronted with our predictions in  \cite{Guo:2021kdo}. A remarkably consistent picture emerged. 
Most striking we deem our successful prediction of the significant pion-mass dependence of the  s-wave $D\,\pi$ scattering phase shifts. On the other hand we uncovered a sizeable tension in the  elastic s-wave $D \bar K $ scattering phase shifts, as constrained by lattice studies on HSC and MILC Asqtad ensembles.

Since by now HSC delivered a quite large set of s- and p-wave scattering phases on two ensembles with  nominal  pion masses of about 239 MeV and 391 MeV, we abandoned the use of the older data set on the MILC Asqtad ensembles in this work. Combined fits to the HSC results with those of  ETMC and HPQCD \cite{Kalinowski:2015bwa,Na:2012iu} on charmed meson masses are performed. Since we include p-wave 
channels the schematic coupled-channel setup as introduced in \cite{Kolomeitsev:2003ac} and used in our previous works \cite{Lutz:2007sk,Guo:2018kno,Guo:2021kdo}
is not applicable any longer. In particular, the presence of u-channel exchange processes lead to close by left-hand cut contributions which require careful treatment. While this may not be crucial for 
s-wave channels it turns more and more important as the angular momentum of the coupled-channel states gets larger. Here we apply the generalized potential approach (GPA) that was established in 
\cite{Gasparyan:2010xz,Danilkin:2010xd,Danilkin:2011fz,Gasparyan:2012km}. It systematically extends the 
applicability domain of the chiral Lagrangian into the resonance region by using an expansion of the 
generalized potential in terms of conformal variables. The key observation was that the expansion coefficients require information only that is well accessible within Chiral Perturbation Theory 
($\chi$PT). 

The work is organized as follows. In Section 2 the part of the chiral Lagrangian that is relevant here is recalled. Section 3 details in depth how to perform a systematic one-loop computation using on-shell masses in loop expressions. Explict results at chiral order 3 are supplemented by order 4 results in the Appendix. The renormalization-scale dependence of the set of Low Energy Constants (LEC) is discussed in Section 4. 
It follows a primer on how to set up the  generalized potential approach GPA. 
The application to lattice data sets is presented in Sections 6 and 7. Lattice data taken on ensembles of HPQCD, ETMC and HSC are considered. We present our predictions for phase shifts and inelasticities based on a parameter set obtained form the considered lattice data.  
With a summary and outlook the paper is closed.


\newpage

\section{The chiral Lagrangian with open-charm meson fields}
\label{sec:2}

We recall the chiral Lagrangian as presented in \cite{Guo:2018kno} for the two antitriplets of $D$ mesons with $J^P =0^-$ and  $J^P =1^-$ 
quantum numbers \cite{Yan:1992gz,Casalbuoni:1996pg,Kolomeitsev:2003ac,Hofmann:2003je,Lutz:2007sk}. 
The kinetic terms read
\begin{eqnarray}
&& \mathcal{L}_{\mathrm{kin}}=(\CD_\mu D)\,(\CD^\mu \bar D)-  M^2\, D \, \bar D
-(\CD_\mu D^{\mu\alpha})\,(\CD^\nu \bar D_{\nu\alpha})+\frac{1}{2}\, \tilde M^2\,D^{\mu\alpha} \,\bar D_{\mu\alpha}
\nonumber\\
&& \qquad -\, f^2\,{\tr } U_\mu\,U^\mu +\frac{1}{2}\,f^2\,{\tr } \chi_+ \,,
\label{def-kin}
\end{eqnarray}
where
\begin{eqnarray}
&& U_\mu = {\textstyle \frac{1}{2}}\,e^{-i\,\frac{\Phi}{2\,f}} \left(
    \partial_\mu \,e^{i\,\frac{\Phi}{f}} \right) e^{-i\,\frac{\Phi}{2\,f}} \,, \qquad \qquad 
    \Gamma_\mu ={\textstyle \frac{1}{2}}\,e^{-i\,\frac{\Phi}{2\,f}} \,\partial_\mu  \,e^{+i\,\frac{\Phi}{2\,f}}
+{\textstyle \frac{1}{2}}\, e^{+i\,\frac{\Phi}{2\,f}} \,\partial_\mu \,e^{-i\,\frac{\Phi}{2\,f}}\,,
\nonumber\\
&& \chi_\pm = {\textstyle \frac{1}{2}} \left(
e^{+i\,\frac{\Phi}{2\,f}} \,\chi_0 \,e^{+i\,\frac{\Phi}{2\,f}}
\pm e^{-i\,\frac{\Phi}{2\,f}} \,\chi_0 \,e^{-i\,\frac{\Phi}{2\,f}}
\right) \,, \qquad \chi_0 =2\,B_0\, {\rm diag} (m_u,m_d,m_s) \,,
\nonumber\\
&& \CD_\mu \bar D = \partial_\mu \, \bar D + \Gamma_\mu\,\bar D \,, \qquad \qquad \qquad \quad \;\; 
\CD_\mu D = \partial_\mu \,D  - D\,\Gamma_\mu \,,
\label{def-chi}
\end{eqnarray}
with  $\chi^\dagger_\pm = \pm \chi_\pm $, $U_\mu ^\dagger = - U_\mu$ and  $\Gamma_\mu ^\dagger = - \Gamma_\mu$.
The $1^-$ states are interpolated in terms of an antisymmetric tensor fields $D_{\mu \nu}$.  The covariant 
derivative $\CD_\mu$ involves the chiral connection $\Gamma_\mu$, the quark masses enter via the symmetry breaking fields $\chi_\pm$ and
the octet of the Goldstone boson fields is encoded into the $3\times3$ matrix $\Phi$. The parameter $f$ is the chiral limit value of the 
pion-decay constant. Finally the parameters $M$ and $\tilde M$ give the masses of the $D$ and $D^*$ mesons at $m_u=m_d =m_s =0$.

We continue with first order interaction terms
\begin{eqnarray}
{\mathcal L}^{(1)} &=& 2\,g_P\,\Big\{D_{\mu \nu}\,U^\mu\,(\CD^\nu \bar D)
 - (\CD^\nu D )\,U^\mu\,\bar D_{\mu \nu} \Big\}
\nonumber\\
&-& \frac{i}{2}\,\tilde g_P\,\epsilon^{\mu \nu \alpha \beta}\,\Big\{
D_{\mu \nu}\,U_\alpha \,
(\CD^\tau \bar D_{\tau \beta} )
+ (\CD^\tau D_{\tau \beta})\,U_\alpha\,\bar D_{\mu \nu}) \Big\} \,,
\label{def-gP}
\end{eqnarray}
which upon an expansion in powers of the Goldstone boson fields provide the 3-point coupling constants of the Goldstone bosons to
the $D$ mesons. While the decay of the charged $D^*$ meson  implies
$ |g_P| = 0.57 \pm 0.07 $ 
the parameter $\tilde g_P$ in (\ref{def-gP}) cannot be extracted from empirical data directly. The size of $\tilde g_P \simeq g_P$
can be estimated using the heavy-quark spin symmetry of QCD \cite{Yan:1992gz,Casalbuoni:1996pg}. We use the leading order result in our work with $\tilde g_P = g_P $.

Second order terms of the chiral Lagrangian were first studied in \cite{Hofmann:2003je,Lutz:2007sk,Guo:2008gp}. The complete collection of relevant terms is
\begin{eqnarray}
&&\mathcal{L}^{(2)}=-\big( 4\,c_0-2\,c_1\big)\, D \,\bar{D}  \,{\tr} \chi_+ -2\,c_1\,D \,\chi _+\,\bar{D}
\nonumber\\
&& \qquad  - \, \big(8\,c_2+4\,c_3\big)\,D\,\bar{D}\,{\tr }U_{\mu }\,U^{\mu } + 4\,c_3\, D \,U_{\mu }\,U^{\mu  }\,\bar{D} 
\nonumber\\
&& \qquad  -\, \big( 4\,c_4+2\,c_5\big)\, ({\CD_\mu } D)\,({\CD_\nu }\bar{D}) \,{\tr} [ U^{\mu }, \,U^{\nu  }]_+ /M^2
+2\,c_5\,({\CD_\mu } D)\,[ U^{\mu }, \,U^{\nu  }]_+({\CD_\nu }\bar{D})/M^2
\nonumber\\
&& \qquad  -\,i\, c_6\,\epsilon ^{\mu \nu \rho \sigma }\,\big(D\,[U_{\mu },\, U_{\nu }^{ }]_-\bar{D}_{\rho \sigma }-D_{\rho \sigma }\,[U_{\nu }^{ }, U_{\mu }]_-\bar{D}\big)
\nonumber\\
&& \qquad  +\,\big(2\,\tilde{c}_0-\tilde{c}_1\big)\,D^{\mu \nu }\,\bar{D}_{\mu \nu }\,{\tr}\chi _+
 +\tilde{c}_1\,D^{\mu \nu }\,\chi _+\,\bar{D}_{\mu \nu }
\nonumber\\
&&\qquad + \big( 4\,\tilde{c}_2+2\,\tilde{c}_3\big)\,D^{\alpha \beta }\,\bar{D}_{\alpha \beta }\,{\tr }U_{\mu }\,U^{\mu }
-2\,\tilde{c}_3\,D^{\alpha \beta }\,U_{\mu }\,U^{\mu  }\,\bar{D}_{\alpha \beta }
\nonumber\\
&& \qquad  +\,\big(2\,\tilde{c}_4+\tilde{c}_5\big)\, ({\CD_\mu }D^{\alpha \beta })\,({\CD_\nu }\bar{D}_{\alpha \beta } )\,{\tr} [U^{\mu }, \,U^{\nu  }]_+/\tilde M^2
\nonumber\\
&&\qquad - \, \tilde{c}_5 \,({\CD_\mu } D^{\alpha \beta })\,[U^{\mu },\, U^{\nu  }]_+ ({\CD_\nu }\bar{D}_{\alpha \beta })/\tilde M^2
+4\,\tilde{c}_6\,D^{\mu \alpha }\,[U_{\mu },\, U^{\nu }]_-\bar{D}_{\nu \alpha } \,,
\end{eqnarray}
where in  the limit of a very large charm-quark mass a common mass arises with $\tilde M/ M \rightarrow 1 $. All parameters $c_i$ and $\tilde c_i$ are expected to scale linearly in the parameter $M$. It holds $\tilde c_i = c_i$  in the heavy-quark mass limit \cite{Lutz:2007sk}. A first estimate of the LEC can be found in \cite{Lutz:2007sk} based on the leading order large-$N_c$ relations 
\begin{eqnarray}
&& c_0 \simeq \frac{c_1}{2} \,, \qquad \qquad c_2 \simeq -\frac{c_3}{2}\,, \qquad \qquad c_4 \simeq -\frac{c_5}{2}\,,
\nonumber\\
&& \tilde c_0 \simeq \frac{\tilde c_1}{2} \,, \qquad \qquad \tilde c_2 \simeq -\frac{\tilde c_3}{2} \,,\qquad \qquad 
\tilde c_4 \simeq -\frac{\tilde c_5}{2}\,.
\label{large-Nc}
\end{eqnarray}
In the combined heavy-quark and large-$N_c$ limit we are left with 4 free parameters only, $c_1, c_3, c_5, c_6$. A reliable estimate of the correction terms is important in order to arrive at a detailed 
picture of this exotic open-charm sector of QCD \cite{Hofmann:2003je,Lutz:2007sk}.  
Additional terms relevant at chiral order three were considered in \cite{Geng:2010vw,Yao:2015qia,Du:2017ttu,Guo:2018kno,Jiang:2019hgs}. A complete list of such terms is 
\allowdisplaybreaks[1]
\begin{eqnarray}
&& \mathcal{L}^{(3)}=
 \, 4\,g_1\,{D}\,[\chi_-,\,{U}_\nu]_- (\CD^\nu \bar{D})/M
  - 4\,g_2\,{D}\,[{U}_\mu,\,([\CD_\nu,\,{U}^\mu]_- + [\CD^\mu,\,{U}_\nu]_-)]_-( \CD^\nu\bar{D})/M
\nonumber\\ 
&&   \qquad  - \,4\,g_3\,{D}\,[{U}_\mu,\,[\CD_\nu,\,{U}_\rho]_-]_-\,[\CD^\mu,\,[\CD^\nu,\,\CD^\rho]_+]_+\bar{D}/M^3
\nonumber\\
&&\qquad   -\,2\,i\,g_4 \,\epsilon_{\mu \nu \rho \sigma} \,(\CD_\alpha D )\, [U^\mu,\,([\CD^\alpha,\,U^\nu]_- +[\CD^\nu,U^\alpha]_-)]_+\bar D^{\rho\sigma}/M 
\nonumber\\
&&\qquad   -\,2\,i\,g_5 \,\epsilon_{\mu \nu \rho \sigma} \,(\CD_\alpha D ) \, {\tr } [U^\mu,\,([\CD^\alpha,\,U^\nu]_-+[\CD^\nu,U^\alpha]_-)]_+\bar D^{\rho\sigma}/M   
\nonumber\\ 
&&  \qquad  - \,2\,\tilde g_1\,  D_{\mu\nu} \,[\chi_-,\,U_\rho ]_- (\CD^\rho\bar D^{\mu\nu}) /\tilde M    + 2\,\tilde g_2\, D_{\mu\nu} \,[U^\sigma, \,([\CD_\rho,\, U_\sigma ]_- + [\CD_\sigma,\,U_\rho ]_-) ]_- ( \CD^\rho \bar D^{\mu\nu} ) /\tilde M 
\nonumber\\
&&  \qquad
    +\,2\,\tilde g_3\, D_{\alpha\beta}\,[U_\mu,\,[\CD_\rho,\,U_\nu]_-  ]_- [\CD^\mu,\,[\CD^\rho,\,\CD^\nu]_+ ]_+\bar D^{\alpha\beta} /\tilde M^3
\nonumber\\
&&\qquad  -\,2\,\tilde g_4\,\Big( D^{\alpha\beta}\,[U_\alpha,\,([\CD_\beta,\,U_\mu]_-+[\CD_\mu,\,U_\beta]_-)]_+( \CD_\nu\bar D^{\mu\nu} )  
\nonumber\\
&& \quad \qquad \qquad -\,D^{\alpha\beta}\,[U_\nu,\,([\CD_\beta,\,U_\mu]_-+[\CD_\mu,\,U_\beta]_-)]_+(\CD_\alpha\bar D^{\mu\nu} )\Big) /  \tilde M  
\nonumber\\
&&\qquad  -\,2\,\tilde g_5\,\Big( D^{\alpha\beta} \, {\tr }[U_\alpha,\,([\CD_\beta,\,U_\mu]_-+[\CD_\mu,\,U_\beta]_-)]_+(\CD_\nu\bar D^{\mu\nu}  ) 
\nonumber\\
&& \quad \qquad \qquad -\,D^{\alpha\beta}\, {\tr }[U_\nu,\,([\CD_\beta,\,U_\mu]_-+[\CD_\mu,\,U_\beta]_-)]_+(\CD_\alpha\bar D^{\mu\nu} )\Big) /  \tilde M  +{\rm h.c.}
   \,,
\label{def-gtilde}
\end{eqnarray}
where we note that the LEC with $\tilde g_i \simeq g_i$ do not contribute to the meson masses at the one-loop level. Rather they are instrumental to achieve a more accurate description of the coupled-channel systems considered here.

We close this Section with a collection of terms relevant at chiral order 4. While the Goldstone boson sector  \cite{Gasser:1984gg}  is well established 
\begin{eqnarray}
&& \mathcal{L}^{(4)}  =  16\,L_1\,({\tr}U_\mu\,U^\mu)^2 + 16\,L_2\,{\tr} U_\mu\,U_\nu\,{\tr}U^\mu\,U^\nu 
+ 16\,L_3\,{\tr}U_\mu\,U^\mu\,U_\nu\,U^\nu 
\nonumber\\
&& \qquad \, - \,8\,L_4\,{\tr}U_\mu\,U^\mu\,{\tr}\chi_+- 8\,L_5\,{\tr}U_\mu\,U^\mu\,\chi_+ 
+ 4\,L_6\,({\tr}\chi_+)^2
\nonumber\\
&& \qquad \,+\, 4\,L_7\,({\tr}\chi_-)^2
+2\,L_8\,{\tr}(\chi_+ \chi_+ + \chi_- \chi_- )\,,
\label{def-L42}
\end{eqnarray}
such terms are less well explored in the open-charm sector. The terms in (\ref{def-L42}) play an instrumental role in the translation of the quark-mass parameters to the masses of the pseudo-Goldstone bosons as measured on various QCD Lattice ensembles (see e.g. \cite{Lutz:2018cqo,Guo:2018kno,Bavontaweepanya:2018yds,Guo:2018zvl}). In the open-charm sector we write 
\allowdisplaybreaks[1]
\begin{eqnarray}
&& \mathcal{L}^{(4)}=- D \, \Big( {d}_1\,\chi _+^2+ {d}_2\, \chi _+\, {\tr } \chi _+ + {d}_3\, {\tr } \chi _+^2 +{d}_4\,( {\tr } \chi _+)^2 
 + {d}_5\,\chi _-^2 + {d}_6 \,\chi _-\, {\tr } \chi _-
 \nonumber\\
&& \qquad \qquad + \,
{d}_7\, {\tr } \chi_-^2  + {d}_8\,({\tr } \chi _-)^2  \Big) \,\bar D
\nonumber\\
&& \qquad - \,4\,D\,\Big(  d_9\, U_{\mu }\,\chi_+ \,U^{\mu  } 
+ \,d_{10}\, [ U_{\mu } \,U^{\mu  },\, \chi_+ ]_+ \
+ \,d_{11}\, U_{\mu } \,U^{\mu  } \,{ \tr } \chi_+
+ \,d_{12}\, \chi_+ \,{\tr} U_{\mu } \,U^{\mu  }
\nonumber\\
&& \qquad \qquad + \,
d_{13} \,{\tr} U_{\mu }\,\chi_+  \,U^{\mu  } 
+ d_{14}\,{\tr} U_{\mu } \,U^{\mu  } \, {\tr} \chi_+ \Big) \,\bar D
\nonumber\\
&& \qquad -\,2\,({\CD_\mu } D) \, \Big( d_{15}\,( U^{\mu }\,\chi_+ \,U^{\nu  }+  U^{\nu }\,\chi_+ \,U^{\mu  } )+  d_{16}\,[ [U^{\mu },\, U^{\nu  }]_+,\, \chi_+ ]_+ + d_{17}\,[ U^{\mu },\, U^{\nu  }]_+ {\tr }\chi_+ 
\nonumber\\
&& \qquad \qquad + \,
d_{18}\,\chi_+ \,{\tr }[ U^{\mu },\, U^{\nu  }]_+ + d_{19}\, {\tr} ( U^{\mu }\,\chi_+ \,U^{\nu  }+  U^{\nu }\,\chi_+ \,U^{\mu  } )
\nonumber\\
&& \qquad \qquad + \,
d_{20}\,{\tr} \chi_+ \,{\tr }[ U^{\mu },\, U^{\nu  }]_+
\Big)\, ({\CD_\nu }\bar{D})/M^2
  \nonumber\\ 
&& \qquad  + \, 4\,D\,\Big( d_{21}\, ( \CD^\sigma U_{\mu })\,(\CD_\sigma U^{\mu  }) - ( d_{21} +2\,d_{22} )\,{\tr }
( \CD^\sigma U_{\mu })\,(\CD_\sigma U^{\mu  } ) \Big)\,\bar{D}   
 \nonumber\\ 
&& \qquad  + \, 2\,(\CD_\mu D)\,\Big( d_{23}\, [( \CD^\sigma U^{\mu }),\,(\CD_\sigma U^{\nu  })]_+ -2\,( d_{23}+ 2\,d_{24})\,{\tr }
( \CD^\sigma U^{\mu })\,(\CD_\sigma U^{\nu  } ) \Big)\,( \CD_\nu \bar{D} )/M^2  
 \nonumber\\ 
&& \qquad  + \, 2\,(\CD_\mu \CD_\sigma D)\,\Big( d_{25}\, [( \CD^\sigma U^{\mu }),\,(\CD^\tau U^{\nu  })]_+ -2\,(d_{25}+2\,d_{26})\,{\tr }
( \CD^\sigma U^{\mu })\,(\CD^\tau U^{\nu  } ) \Big)\,( \CD_\nu \CD_\tau \bar{D} )/M^4 
\nonumber\\
&& \qquad  + \,\frac{1}{2}\,D^{\alpha \beta }\,\Big( \tilde{d}_1\,\chi _+^2+\tilde{d}_2\, \chi _+\, {\tr } \chi _+ + \tilde{d}_3\, {\tr } \chi _+^2 +\tilde{d}_4\,({\tr } \chi _+)^2 
 + \tilde{d}_5\,\chi_-^2 + \tilde{d}_6 \,\chi _-\, {\tr } \chi _-
 \nonumber\\
&& \qquad \qquad + \,
\tilde{d}_7\, {\tr } \chi_-^2  + \tilde{d}_8\,({\tr } \chi _-)^2  \Big) \,D_{\alpha \beta}
 \nonumber\\
&& \qquad + \,2\, D^{\alpha \beta }\,\Big( {\tilde d}_9\, U_{\mu }\,\chi_+ \,U^{\mu  }
+{\tilde d}_{10}\,[ U_{\mu } \,U^{\mu  },\, \chi_+ ]_+
+{\tilde d}_{11}\, U_{\mu } \,U^{\mu  } \, \tr \chi_+
\nonumber\\
&& \qquad \qquad + \,{\tilde d}_{12}\,\chi_+  \,{\tr} U_{\mu } \,U^{\mu  }
+ {\tilde d}_{13} \,{\tr} U_{\mu }\,\chi_+  \,U^{\mu  } 
+ {\tilde d}_{14}\, {\tr} U_{\mu } \,U^{\mu  } \, {\tr} \chi_+ \Big) \,\bar D_{\alpha \beta }
\nonumber\\
&& \qquad +\,({\CD_\mu } D^{\alpha \beta }) \, \Big( {\tilde d}_{15}\,( U^{\mu }\,\chi_+ \,U^{\nu  }+  U^{\nu }\,\chi_+ \,U^{\mu  } )+  {\tilde d}_{16}\,[ [ U^{\mu },\, U^{\nu  }]_+,\, \chi_+ ]_+ + \tilde d_{17}\,[ U^{\mu },\, U^{\nu  }]_+ {\tr }\chi_+ 
\nonumber\\
&& \qquad \qquad + \,
\tilde d_{18}\,\chi_+ \,{\tr }[ U^{\mu },\, U^{\nu  }]_+ + \tilde d_{19}\, {\tr} ( U^{\mu }\,\chi_+ \,U^{\nu  }+  U^{\nu }\,\chi_+ \,U^{\mu  } )
\nonumber\\
&& \qquad \qquad + \,
\tilde d_{20}\,{\tr} \chi_+ \,{\tr }[ U^{\mu },\, U^{\nu  }]_+
\Big)\, ({\CD_\nu }\bar{D}_{\alpha \beta })/\tilde M^2
\nonumber\\ 
&& \qquad  - \, 2\,D^{\alpha \beta }\,\Big( {\tilde d}_{21}\, ( \CD^\sigma U_{\mu })\,(\CD_\sigma U^{\mu  }) -( {\tilde d}_{21}+2\,{\tilde d}_{22})\,{\tr }
( \CD^\sigma U_{\mu })\,(\CD_\sigma U^{\mu  } ) \Big)\,\bar{D}_{\alpha \beta }   
 \nonumber\\ 
&& \qquad  - \, (\CD_\mu \CD_\sigma D^{\alpha \beta })\,\Big( {\tilde d}_{25}\, [( \CD^\sigma U^{\mu }),\,(\CD^\tau U^{\nu  })]_+ -2\,( {\tilde d}_{25} +2\,{\tilde d}_{26} )\,{\tr }
( \CD^\sigma U^{\mu })\,(\CD^\tau U^{\nu  } ) \Big)\,( \CD_\nu \CD_\tau \bar{D}_{\alpha \beta } )/\tilde M^4
 \nonumber\\ 
&& \qquad  - \, (\CD_\mu D^{\alpha \beta })\,\Big(\tilde d_{23}\, [( \CD^\sigma U^{\mu }),\,(\CD_\sigma U^{\nu  })]_+ -2\,({\tilde d}_{23} +2\, \tilde d_{24})\,{\tr }
( \CD^\sigma U^{\mu })\,(\CD_\sigma U^{\nu  } ) \Big)\,( \CD_\nu \bar{D}_{\alpha \beta } )/\tilde M^2  
\,,
\label{def-L4}
\end{eqnarray}
where we follow our notation request again, in which the heavy-quark mass limit 
implies $ \tilde d_i = d_i $. There are $16 =2\times 8$ symmetry breaking counter terms, $d_{1-8}$ and $\tilde d_{1-8}$ proportional to the product of two light quark masses. The first half of them are relevant in the chiral extrapolation of the $D$ meson masses at chiral order $Q^4$ but at that order also for the scattering of Goldstone bosons off the $D$ mesons. All remaining LEC, $d_{5-26}$ and $\tilde d_{5-26}$ contribute to the two-body scattering processes only. Their impact on the $D$ meson masses starts at chiral order $Q^6$ via tadpole type contributions. 
Additional symmetry breaking terms proportional to $d_{9-20}$ and $\tilde d_{9-20}$ are linear in the light quark masses. There remain the twelve symmetry conserving $d_{21-26}$ and $\tilde d_{21-26}$. 

The rather large set of unknown LEC at this order will be reduced systematically by the neglect of specific structures, terms involving single, double and triple flavor traces, that are suppressed  in the large-$N_c$ limit of QCD.

\begin{figure}[t]
\center{
\includegraphics[keepaspectratio,width=0.8\textwidth]{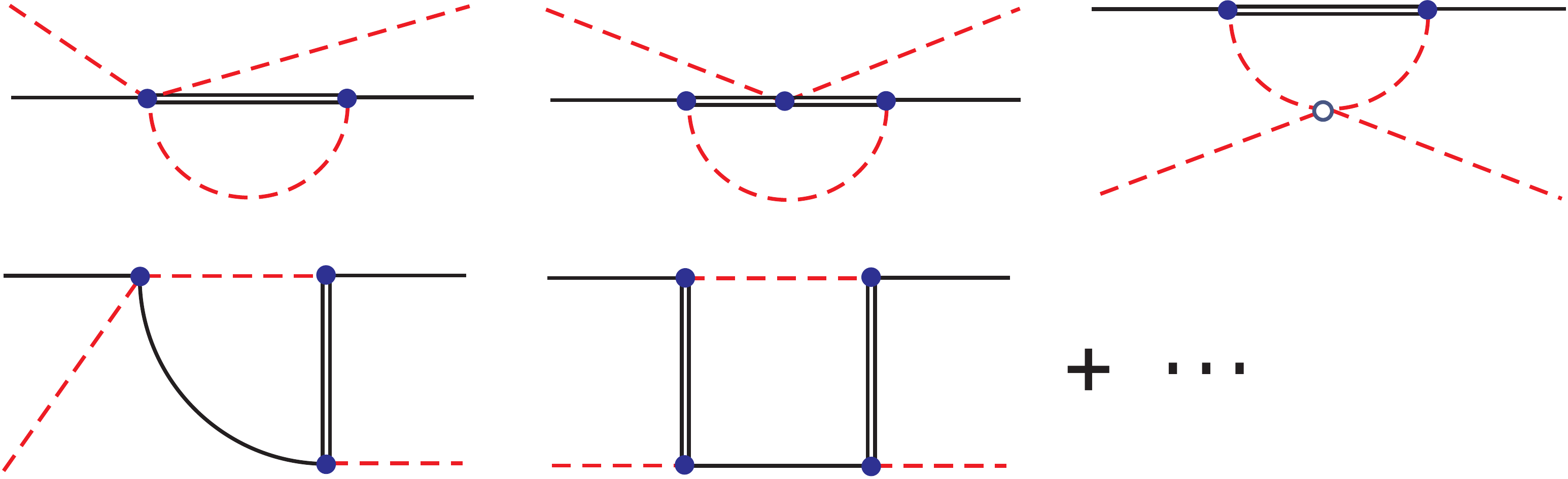} }
\vskip-0.2cm
\caption{\label{fig:4} One-loop triangle and box diagrams at chiral order three and four. Solid vertices stand for order one and open vertices for order 2 structures. We use solid and double solid lines for the open-charm  mesons, single dashed for the Goldsone bosons. Further diagrams at that order which are related to a renormalization of $g_P$ or $\tilde g_P$ are not shown.  }
\end{figure}

While our computation of the $D$ meson masses is complete at chiral order 4, this is not yet fully the case for the two-body forces used in our coupled-channel approach. Here we are complete at order 4 in the limit $g_P =\tilde g_P \to 0$ at least. The LEC $g_i$ and $ \tilde g_i$  enter at that order also via one-loop triangle and box  contributions, which are beyond the scope of the current work. In our work we consider one-loop contributions to the scattering kernel in the sectors involving $D$ and $D^*$ mesons in the initial and final states for the first time. 
Some loop contributions were considered previously in the $D$ meson sector only \cite{Yao:2015qia,Du:2017ttu}. Here we can only partially reproduce that previous study \cite{Yao:2015qia}. While the various Clebsch we can relate to in our computation, the specific form of the loop functions we cannot reproduce. In particular we find a scale dependence of the LEC $g_1$, which was overlooked previously. 
The pertinent contributions are documented in the next Section, where we focus on the contributions that are fully controlled at this order 3. Contributions at chiral order 4 can be found in Appendix A.  
As a consequence of our scheme we arrive at a rather compact final form of such contributions in both needed sectors that are particularly convenient for a code implementation. 
We do not consider a further set of one loop contributions from (\ref{def-gP}) that arise at $g_P =\tilde g_P \neq 0$.  We would argue that in our particular coupled-channel scheme their contributions can be absorbed into a  renormalization of $g_i, \tilde g_i$  and $d_i,\tilde d_i$, at least on a phenomenological level. Such a procedure is substantiated by the finding of \cite{Yao:2015qia}, where the effect of such diagrams was claimed to be marginal, at least in the s-wave scattering channels.

\clearpage

\section{Scattering at the one-loop level}

\begin{table}
\tabcolsep=6.7mm
\begin{tabular}{|cccccccccccc|}
\hline
\multicolumn{3}{|c|}{$(\frac{1}{2},+2)$} &
\multicolumn{3}{c|}{$(0,+1)$} &
\multicolumn{3}{c|}{$(1,+1)$} &
\multicolumn{3}{c|}{$(\frac12,0)$}
\\ \hline
\multicolumn{3}{|c|}{$(D_s\,K)$} &
\multicolumn{3}{c|}{$\left(\begin{array}{c} ({\textstyle{1\over
\sqrt{2}}}\,D^t \, i\,\sigma_2\, K)\\ (D_s\,\eta )
\end{array}\right)$}&
\multicolumn{3}{c|}{$\left(\begin{array}{c} (D_s \, \pi)\\ (
{\textstyle{1\over
\sqrt{2}}}\,D^t \, i\,\sigma_2\,\sigma\, K)
\end{array}\right)$} &
\multicolumn{3}{c|}{$\left(\begin{array}{c} ({\textstyle{1\over
\sqrt{3}}}\,\pi \cdot
\sigma\, D)\\ ( \eta \,D) \\ (D_s\,i\,\sigma_2\,\bar K^t)
\end{array}\right)$}
\\\hline\hline
\multicolumn{4}{|c|}{\phantom{xxxxxx}$(\frac32,0)$\phantom{xxxxxx}} &
\multicolumn{4}{c|}{\phantom{xxxxxx}$(0,-1)$\phantom{xxxxxx}} &
\multicolumn{4}{c|}{$(1,-1)$}
\\\hline
\multicolumn{4}{|c|}{$( \pi \cdot T\,D)$} &
\multicolumn{4}{c|}{$( {\textstyle{1\over \sqrt{2}}}\,\bar K
\,D)$} &
\multicolumn{4}{c|}{$({\textstyle{1\over \sqrt{2}}}\,\bar K
\,\sigma\,D)$}
\\\hline
\end{tabular}
\caption{Coupled-channel states with $(I,S)$ as introduced in \cite{Kolomeitsev:2003ac}. Here $K$ and $\bar K$ are isospin doublet fields. The matrices $\vec \sigma$ and $\vec T$ act in isospin space. }
\label{tab:states}
\end{table}

\begin{figure}[t]
\center{
\includegraphics[keepaspectratio,width=1.0\textwidth]{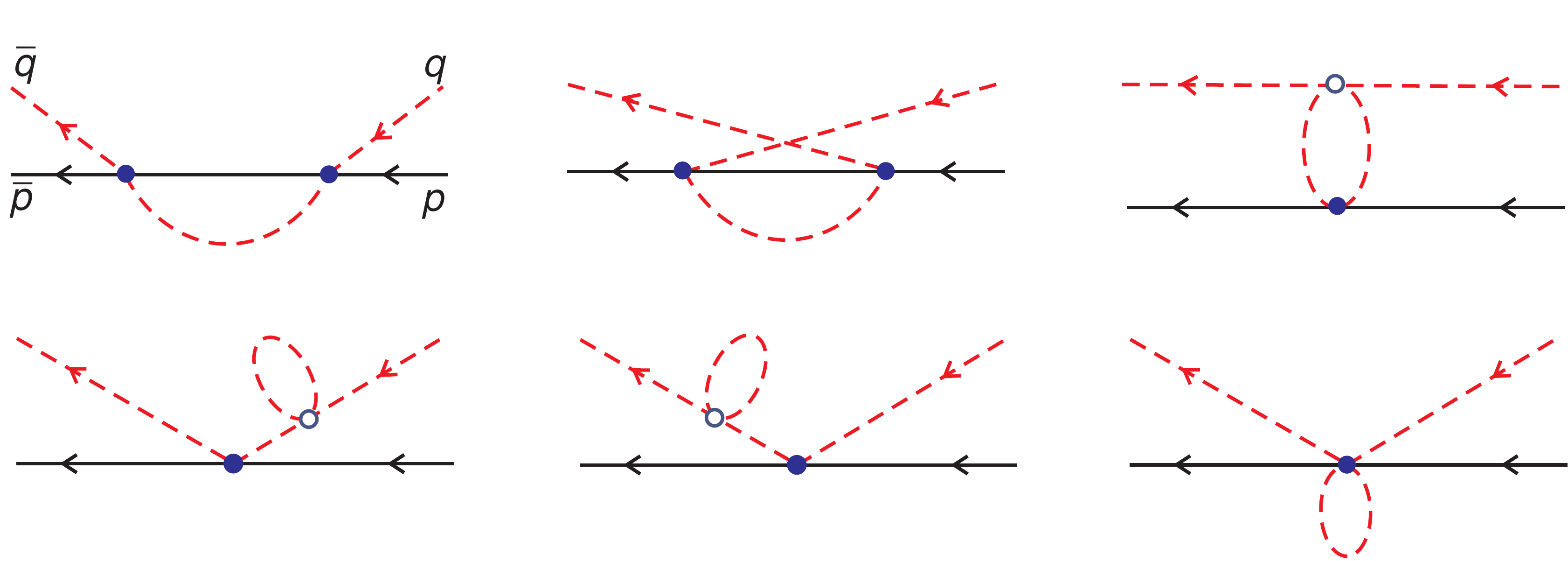} }
\vskip-1.2cm
\caption{\label{fig:1} One-loop diagrams at chiral order three. Solid vertices stand for order one and open  vertices for order 2 structures. We use solid lines for the open-charm  mesons, dashed for the Goldsone bosons.  }
\end{figure}

We consider one-loop contributions to the two-body scattering amplitudes in the isospin-strangeness basis $(I,S)$. At chiral order 3 we focus on the Tomozawa-Weinberg interaction terms in (\ref{def-kin}). There are 5 types of diagrams to be evaluated, as is illustrated in Fig. \ref{fig:1}, with  a generic tadpole diagram from a 6-point vertex and bubble-type contributions in the s-, t- and u-channel. In addition there are contributions from the wave-function renormalization factors of the Goldstone bosons.
In application of the 
Passarino-Veltman scheme \cite{Passarino:1978jh} the results can be most economically expressed in terms of two scalar loop functions
\begin{eqnarray}
&& I_a \;=\int \frac{d^d l}{(2\pi)^d}\frac{ i\,\mu^{d-4} }{l^2- m_a^2+i\,\epsilon}\,,
\nonumber\\
&& I_{ab} =  \int \frac{d^d l}{(2\pi)^d}\frac{ -i\,\mu^{d-4} }{l^2- m_a^2+i\,\epsilon}
 \frac{1}{(l+p)^2-m_b^2 +i\,\epsilon} = I_{ab}(p^2) \,,
 \label{def-tadpole}
\end{eqnarray}
properly introduced in dimensional regularization. There are a few well known technical issues to consider. A straightforward evaluation of the set of diagrams leads to results that suffer from terms that are at odds with their expected chiral power. There are terms of too low but also too high orders, both of which need to be eliminated as to arrive at consistent results. 
In our work we apply a recent scheme proved particulary efficient for both tasks \cite{Lutz:2018cqo,Bavontaweepanya:2018yds,Lutz:2020dfi,Sauerwein:2021jxb}.

In a first step we insist on a slight change in the choice of the basis loops. Renormalized scalar bubble-loop contributions, $\bar I_{ab}$, are introduced that are independent on the renormalization scale. Here it is instrumental to carefully discriminate the light from the heavy particles. Following our previous works  
\cite{Lutz:2018cqo,Bavontaweepanya:2018yds,Lutz:2020dfi,Sauerwein:2021jxb} we introduce
\begin{eqnarray}
&& I_{QH}(s) = \bar I_{QH}(s) + \frac{1}{16\,\pi^2} - \frac{I_H}{M_H^2}\,,
 \nonumber\\
&& \bar I_{QH}(s) = \frac{1}{16\,\pi^2}
\left\{ -\frac{1}{2} \left( 1+ \frac{m_Q^2-M_H^2}{s} \right) \,\log \left( \frac{m_Q^2}{M_H^2}\right)
\right.
\nonumber\\
&& \;\quad \;\,+\left.
\frac{p_{Q H}}{\sqrt{s}}\,
\left( \log \left(1-\frac{s-2\,p_{Q H}\,\sqrt{s}}{m_Q^2+M_H^2} \right)
-\log \left(1-\frac{s+2\,p_{Q H}\,\sqrt{s}}{m_Q^2+M_H^2} \right)\right)
\right\} \, ,
\nonumber\\
&& p_{Q H}^2 =
\frac{s}{4}-\frac{m_Q^2+M_H^2}{2}+\frac{(m_Q^2-M_H^2)^2}{4\,s}  \,,
\label{def-bubble}
\end{eqnarray} 
and
\begin{eqnarray}
&& I_{PQ}(t) = \bar I_{PQ}(t) - \frac{I_Q}{2\,m_Q^2}-  \frac{I_P}{2\,m_P^2}\,, 
 \nonumber\\
&& \bar I_{PQ}(t) = \frac{1}{16\,\pi^2}
\left\{ 1- \frac{m_P^2-m_Q^2}{2\,t}  \,\log \left( \frac{m_P^2}{m_Q^2}\right)
\right.
\nonumber\\
&& \;\quad \;\,+\left.
\frac{p_{P Q}}{\sqrt{t}}\,
\left( \log \left(1-\frac{t-2\,p_{P Q}\,\sqrt{t}}{m_P^2+m_Q^2} \right)
-\log \left(1-\frac{t+2\,p_{P Q}\,\sqrt{t}}{m_P^2+m_Q^2} \right)\right)
\right\} \, ,
\nonumber\\
&& p_{P Q}^2 =
\frac{t}{4}-\frac{m_P^2+m_Q^2}{2}+\frac{(m_P^2-m_Q^2)^2}{4\,t}  \,,
\label{def-bubble-PQ}
\end{eqnarray} 
where we use $P, Q$ as placeholders for the light fields (Goldstone bosons) but $H$ as a placeholder for the heavy fields (charmed mesons). For clarity of the presentation we use the Mandelstam variables $s= (p+q)^2$, $u= (p-\bar q)^2$ and $t= (\bar q -q)^2$ in (\ref{def-bubble}, \ref{def-bubble-PQ}). Our final expressions will be given in terms of the renormalized bubbles $\bar I_{QH}(s)$ and $\bar I_{QH}(u)$ and $\bar I_{PQ}(t)$.

We emphasize that a subtraction scheme for the loop functions 
if performed at the level of the Passarino-Veltman functions  is symmetry conserving \cite{Lutz:1999yr,Semke:2005sn,Lutz:2014oxa,Lutz:2020dfi}. As long as 
there is an unambiguous prescription how to represent all one-loop contributions  in terms of a well defined basis we do not expect any 
violation of chiral Ward identities. Given that observation we can eliminate the 
power-counting violating terms quite efficiently. It suffices to insist on the simple replacement rules
\begin{eqnarray}
I_{QH}\to \bar I_{QH} \,, \qquad \qquad \qquad I_H \to 0 \,,
\label{def-replacement}
\end{eqnarray}
after which the expectation of dimensional counting rules is implemented. In particular it holds $\bar I_{QH} \sim Q$ and $\bar I_{QP} \sim Q^0$, where we use $Q^n$ to indicate the chiral order $n$ of a given term. 
To avoid a proliferation of our notations $Q$ is used also as a placeholder index for a Goldstone boson field in (\ref{def-replacement}).
In turn it is left to renormalize the remaining tadpole contribution from the light fields, $I_Q$, for which a conventional $\overline{MS}$ subtraction scheme may be used, leading to the renormalized tadpole function
\begin{eqnarray}
 \bar I_Q = \frac{m_Q^2}{(4 \pi )^ 2} \, \log \frac{m_Q^2}{ \mu^2}\,,
 \label{def-IQ}
\end{eqnarray}
with the renormalization scale $\mu$ of dimensional regularization. 

Yet, after these steps there does remain an important issue. The renormalization-scale dependence in the one-loop contributions cannot be balanced by the available counter terms. This is so, since such a computation generates besides the order $Q^3$ terms further $Q^4$ terms that would ask for an additional  set of LEC. A simple remedy is available by a chiral expansion of the kinematical coefficients that accompany the scalar basis functions.  

\begin{table}
\tabcolsep=2.1mm
\begin{center}
\begin{tabular}{|C||C|C|C|C||C|C|C||C|C|C|}
\hline
(I,\,S)&\text{Channel}&
C_{\rm WT} &\; C_{2}\; & C_{3} & C^{(1)}_{\pi} & C^{(1)}_{K} & C^{(1)}_{\eta}  & C^{(all)}_\pi & C^{(all)}_K & C^{(all)}_{\eta}  \\
\hline
\hline
(\frac12,\,+2)& 11    &
 -1 & 2& \frac{1}{2} &0 & -4 & 0 & -\frac{3}{16} & -\frac{3}{8} & -\frac{3}{16} \\
\hline
\hline
(0,\,+1)      & 11    &
 2 & 2& 0 & 0 & 8 & 0 & \frac{3}{8} & \frac{3}{4} & \frac{3}{8} \\
\hline
              & 12    &
 \sqrt{3} & 0&-\frac{1}{2\sqrt{3}}& 0 & 2 \sqrt{3} & 2 \sqrt{3} & \frac{3 \sqrt{3}}{32} & \frac{9 \sqrt{3}}{16} & \frac{3 \sqrt{3}}{32} \\
\hline
              & 22    &
 0 & 2 &\frac{1}{3} &  0 & 0 & 0 & 0 & 0 & 0 \\
\hline
\hline
(1,\,+1)      & 11    &
 0 & 2&   1& 0 & 0 & 0 & 0 & 0 & 0 \\
\hline
              & 12    &
 1 & 0 &\frac{1}{2} & 2 & 2 & 0 & \frac{11}{32} & \frac{5}{16} & \frac{3}{32} \\
\hline
              & 22    &
 0 & 2  & 1& 0 & 0 & 0 & 0 & 0 & 0 \\
\hline
\hline
(\frac12,\,0) & 11    &
 2 & 2 & \frac{1}{2} &8 & 0 & 0 & 1 & \frac{1}{2} & 0 \\
\hline
              & 12    &
 0 &0 & \frac{1}{2} &0 & 0 & 0 & 0 & 0 & 0 \\
\hline
              & 13    &
 \sqrt{\frac{3}{2}} & 0 & -\sqrt{\frac{3}{8}} &\sqrt{6} & \sqrt{6} & 0 & \frac{11}{32}\sqrt{\frac{3}{2}} & \frac{5}{16}\sqrt{\frac{3}{2}} & \frac{3}{32}\sqrt{\frac{3}{2}} \\
\hline
              & 22    &
 0 & 2 & \frac{5}{6} &0 & 0 & 0 & 0 & 0 & 0 \\
\hline
              & 23    &
 -\sqrt{\frac{3}{2}} & 0 & -\frac{1}{2\sqrt{6}} &0 & -\sqrt{6} & -\sqrt{6} & -\frac{3}{32}\sqrt{\frac{3}{2}} & -\frac{9}{16}\sqrt{\frac{3}{2}} & -\frac{3}{32}\sqrt{\frac{3}{2}} \\
\hline
              & 33    &
 1 & 2 & \frac{1}{2} & 0 & 4 & 0 & \frac{3}{16} & \frac{3}{8} & \frac{3}{16} \\
\hline
\hline
(\frac32,\,0) & 11    &
 -1 & 2 & \frac{1}{2} &-4 & 0 & 0 & -\frac{1}{2} & -\frac{1}{4} & 0 \\
\hline
\hline
(0,\,-1)      & 11    &
 1 & 2 & \frac{3}{2} & 0 & 4 & 0 & \frac{3}{16} & \frac{3}{8} & \frac{3}{16} \\
\hline
\hline
(1,\,-1)      & 11    &
 -1 & 2 & \frac{1}{2} & 0 & -4 & 0 & -\frac{3}{16} & -\frac{3}{8} & -\frac{3}{16} \\
\hline
\end{tabular}
\caption{The coefficients $C^{(I,S)}$ that characterize the  interaction of Goldstone bosons with
heavy-meson fields $H$ as introduced in (\ref{res-T3}) for
given isospin (I) and strangeness (S). The channel ordering is specified in
Tab. \ref{tab:states}. }
\label{tab:coeffA}
\end{center}
\end{table}

In order to do so we first formulate our power counting in terms of on-shell masses, for which we introduce 
\begin{eqnarray}
&& M^2_H \sim (s + u)/2 \sim Q^0  \ ,  \qquad  \qquad \frac{s- u}{4\,M_H} \sim m_{\rm Goldstone} \sim Q\,,
\nonumber\\
&& m_{\rm Quark} \sim t \sim s+u-2\,M^2_H \sim Q^2\,.
 \label{def-power-counting}
\end{eqnarray}
We illustrate the impact of such counting rules. 
For instance, a term $t\, \bar I_Q \sim Q^4$ or $(s+u-2\,M_H^2)\,\bar I_Q \sim Q^4$ is considered at chiral order 4.  In contrast 
contributions $(s-u)\, \bar I_Q \sim Q^3$ matter at chiral order 3. Note that we need to count $M_H-M_R\sim Q$ or $M_H-M_R\sim Q^2$ depending on the nature of the fields $H$ and $R$,
where we use $R$ as a further placeholder for a heavy field \cite{Guo:2018kno}.

Having set the scene we can write down our final expression as follows. 
There are two distinct sectors we need to consider. First we detail our results for the processes involving two $J^P=0^-$ $D$ mesons. Our expressions for the corresponding sector involving two  $J^P=1^-$ $D$ mesons will easily  be implied by formal replacement rules that are a consequence of the heavy-quark symmetry. The details of how to do so are skechted at the end of this Section.   

We consider a scattering amplitude in a channel with well defined isospin and strangeness $I$ and $S$, which implies specific meson masses $(m_b, M_b)$ and $(m_a, M_a)$  in the initial and final states respectively. The contributions at order $Q^2$ have been documented ample times in the literature (first and correctly in \cite{Kolomeitsev:2003ac,Hofmann:2003je}). Like in our previous works we do consider the s- and u-channel $D$ meson exchange terms. 
For the readers convenience we display here $Q$, $Q^2$ and $Q^3$ contributions at $g_P = \tilde g_P= 0$ in  
a unified fashion
\begin{eqnarray}
&& f^2\,T^{(1)}_{ab}(s,t,u)= \frac{s-u}{4}\,C_{\rm WT} \,,
\nonumber\\
&& f^2\,T^{(2)}_{ab}(s,t,u)= 2\, c_0 \,B_0 \,\Big( 2\,m\,C^{(\pi)}_0 + (m+m_s)\,C^{(K)}_0 \Big) + 
2\, c_1 \,B_0\, \Big( 2\,m\,C^{(\pi)}_1 + (m+m_s)\,C^{(K)}_1 \Big) 
\nonumber\\
&& \qquad +\,4\,( \bar q\cdot q)\,\Big(c_2\,C_2 + c_3\,C_3 \Big)+
\frac{(s-u)^2}{4\,M^2}\,\Big( c_4\, C_{2} + c_5\,C_3 \Big) \,,
\nonumber\\
&& f^2\,T^{(3)}_{ab}(s,t,u) = 2\,t\,\frac{s-u}{f^2}\, \Big( \frac{f^2\,g_2}{M} -   \frac{1}{64}\,
\Big( \frac{\bar I_a}{m_a^2} + \frac{\bar I_b}{m_b^2}\Big) \Big) \,C_{\rm WT} 
\nonumber\\
&& \qquad -\,
 2\,\frac{s-u}{f^2}\,\Big( \big(  m_a^2+m_b^2 \big) \,\frac{f^2\,g_2}{M} -  \frac{1}{32}\,
\big( \bar I_a+ \bar I_b\big) \Big) \,C_{\rm WT} 
\nonumber\\
&& \qquad +\, \frac{s-u}{f^2}\,\sum_Q \Big( m_Q^2 \,\frac{f^2\,g_1}{M}\, - \frac{1}{32}\, \bar I_Q \Big)\,C^{(1)}_{Q}\,
 + (s-u)^3\, \frac{g_3}{2\,M^3}\,C_{\rm WT}
\nonumber\\
&& \qquad +\,  \sum_{Q}\,\frac{s-u}{f^2}\, \Big( \bar I_Q \,C^{(all)}_Q - m_Q^2\,(C_{Q}^{(4)}\, L_4 + C^{(5)}_Q \,L_5 ) \Big)
\nonumber\\
&& \qquad +\, \sum_{QH}\,\frac{(s-u)^2}{f^2}\,\Big( \bar I_{QH}(s) \,C_{QH}^{(s)}+  \bar I_{QH}(u) \,C_{QH}^{(u)} \Big)
\nonumber\\
&& \qquad +\, 
\sum_{PQ}\,\frac{s-u}{f^2}\,\Big( p^2_{PQ}(t)\, \bar I_{PQ}(t) - \frac{(m_P^2-m_Q^2)^2 }{4\,t} \, \bar I_{PQ}(0) - 
\frac{m_P^2- m_Q^2}{4 \,(4\,\pi)^2}\,\log \frac{m_P^2}{m_Q^2}
\nonumber\\
&& \qquad \qquad -\,
\frac{t}{8 \,(4\,\pi)^2}\,\Big( \log \frac{m_P^2}{m_a^2}+\log \frac{m_Q^2}{m_b^2} \Big)
- \frac{3\,(m_P^2+m_Q^2)- t }{ 6\,(4\,\pi)^2}\Big)\, C_{PQ}^{(t)}\,,
\label{res-T3}
\end{eqnarray}
with $P\in \{\pi,K, \eta \}$, $Q\in \{\pi,K, \eta \}$ and $H \in \{D, D_s \}$. 
The important merit of (\ref{res-T3}) lies in its explict renormalization-scale invariance. The $\mu$ dependence in the LEC $g_1, g_2$ and $L_4,L_5$ is absorbed by the 
tadpole integrals $\bar I_Q, \bar I_P $ and $ \bar I_a, \bar I_b$ fully. That cancellation mechanism holds line by line in (\ref{res-T3}) in accordance with
\begin{eqnarray}
 && \mu^2\, \frac{\text{d}}{\text{d}\mu^2}\, L_4 = - \frac{1}{256\, \pi^2} \,,  \qquad \quad
 \mu^2 \,\frac{\text{d}}{\text{d}\mu^2}\, L_5 = -\frac{3}{256\,\pi^2} \,,
 \qquad \quad
\nonumber\\ 
&&  \frac{f^2\, \mu^2}{M}\,\frac{\text{d}}{\text{d}\mu^2}\, g_1 = -\frac{1}{512\,\pi^2}  \,,\qquad 
\frac{f^2\, \mu^2}{M}\,\frac{\text{d}}{\text{d}\mu^2}\, g_2 = -\frac{1}{512\,\pi^2} \,,
\qquad \frac{\text{d}}{\text{d}\mu^2}\, g_{3,4,5} = 0\,.
\label{res-L45-running}
\end{eqnarray}
Our results are expressed in terms of various Clebsch coefficients as introduced in part already in \cite{Kolomeitsev:2003ac,Hofmann:2003je}. The terms relevant at tree-level, $C_{\rm WT}, C_2, C_3$ and $C^ {(1)}_Q$, are recalled in  Tab. \ref{tab:coeffA} for the readers convenience as to settle the 
well known phase-convention ambiguities in flavor SU(3) approaches. The most involved 
Clebsch, $C^{(all)}_Q$, that is also included in Tab. \ref{tab:coeffA}, picks up contributions from all loop diagrams, i.e. the scalar tadpole part of the s-, t- and u-channel bubble exchanges, the wave-function renormalization factors of the Goldstone bosons and, not to forget, that contributions from the 6-point vertex in the Tomozawa-Weinberg term, as discussed above. The latter contributions are indispensable as to arrive at renormalization-scale invariant results, and cannot be found in the literature so far. 

We checked that our results for $C^{(s)}_{QH}$, $C^{(u)}_{QH}$  and $C^{(t)}_{PQ}$ are consistent with the findings in \cite{Yao:2015qia}. The results for $C^{(s)}_{QH}$ and $C^{(u)}_{QH}$ can be derived from $C_{\rm WT}$ quite directly. 
For the readers convenience, we show the somewhat more tedious t-channel Clebsch, $C^{(t)}_{PQ}$, in Tab. \ref{tab:coeffB}, where in our convention the sum over $PQ$ in (\ref{res-T3}) runs over six entries only. The table includes our results for $C^{(4)}_Q$ in addition.  The remaining $C^{(5)}_Q$ in  (\ref{res-T3}) is determined by 
\begin{eqnarray}
C_Q^{(5)}=  \frac{16}{3}\,C_Q^{(all)} - \frac{1}{3}\, C_Q^{(4)} \,  .
\end{eqnarray}
Here we should point at a subtle issue. The contribution of $L_4$ and $L_5$ to the wave-function renormalization factors of the Goldstone bosons and also the contributions of $g_1$ are accompanied by either 
$B_0\,m$ or $B_0\,m_s$. 
In our result (\ref{res-T3}) we distribute such terms into  $m_\pi^2$, $m_K^2$ or $m_\eta^2$ terms in the particular manner, 
\begin{eqnarray}
&& f^2\,(Z_\pi - 1 )=    - 8\,L_4 \,(m_\pi^2 +2\,m_K^2  ) - 8\,L_5\,m_\pi^2 + \frac{2}{3}\,\bar I_\pi + \frac{1}{3}\,\bar I_K\,,
\nonumber\\
&& f^2\,(Z_K -1) =  - 12\,L_4 \,(m_\pi^2 +m_\eta^2 ) - 8\,L_5\,m_K^2 + \frac{1}{4}\,\big( \bar I_\pi + \bar I_\eta + 2\,\bar I_K \big)\,,
\nonumber\\
&& f^2\,(Z_\eta \,\, - 1) =   - 24\,L_4 \,(2\,m_K^2 - m_\eta^2 ) - 8\,L_5\,m^2_\eta +\bar I_K \,,
\label{ref-wave-function}
\end{eqnarray}
such that our claim of the renormalization-scale invariance of the 7th line in  (\ref{res-T3}) holds. This is in line with the scheme we introduced in our previous works \cite{Lutz:2018cqo,Guo:2018kno,Bavontaweepanya:2018yds}.   

It may be useful for the reader that we illustrate the impact of such a strategy more explicitly. We reconsider the pion and kaon decays which are the key observable quantities to determine $L_4$ and $L_5$. From \cite{Bavontaweepanya:2018yds} we recall
\begin{eqnarray}
&&f\, f_\pi = f^2 -\bar I_\pi   - \frac{1}{2}\,\bar I_K + 4\,m_\pi^2\, L_5 + 4\,(m_\pi^2 +2\,m_K^2)\,L_4\,,
\nonumber\\
&& f\,f_K = f^2- \frac{3}{8}\,\bar I_\pi   - \frac{3}{4}\,\bar I_K - \frac{3}{8}\,\bar I_\eta  +  4\,m_K^2\, L_5 +  6\,(m_\pi^2 +m_\eta^2 )\,L_4 \,,
\label{def-fP-chiPT}
\end{eqnarray}
where the quark-mass dependence is eliminated in favor of suitable combinations of the squared pseudo-Goldstone boson masses as to arrive at renormalization-scale invariant expressions. 
For a given value of $f$ we will adjust $L_4$ and $L_5$ with (\ref{def-fP-chiPT}) to recover the empirical values of the pion and kaon decay constants.

\begin{table}
\tabcolsep=2.3mm
\begin{center}
\begin{tabular}{|C||C||C|C|C|C|C|C||C|C|C|}
\hline
(I,\,S)&\text{Channel}&
C^{(t)}_{\pi \pi} & C^{(t)}_{\pi K} & C^{(t)}_{\pi \eta} & C^{(t)}_{K K} & C^{(t)}_{K \eta} & C^{(t)}_{\eta \eta} & C^{(4)}_\pi &  C^{(4)}_{K} & C^{(4)}_{\eta} \\
\hline
\hline
(\frac12,\,+2)& 11    &
 0 & 0 & 0 & -\frac{1}{4} & 0 & 0 & -3 & 0 & -3 \\
\hline
\hline
(0,\,+1)      & 11    &
 \frac{1}{4} & 0 & 0 & \frac{1}{4} & 0 & 0 & 6 & 0 & 6 \\
\hline
              & 12    &
 0 & \frac{\sqrt{3}}{8} & 0 & 0 & \frac{\sqrt{3}}{8} & 0 & \frac{3 \sqrt{3}}{2} & 6 \sqrt{3} & -\frac{3 \sqrt{3}}{2} \\
\hline
              & 22    &
 0 & 0 & 0 & 0 & 0 & 0 & 0 & 0 & 0 \\
\hline
\hline
(1,\,+1)      & 11    &
 0 & 0 & 0 & 0 & 0 & 0 & 0 & 0 & 0 \\
\hline
              & 12    &
 0 & \frac{1}{8} & 0 & 0 & \frac{1}{8} & 0 & \frac{5}{2} & 2 & \frac{3}{2} \\
\hline
              & 22    &
 -\frac{1}{12} & 0 & 0 & \frac{1}{12} & 0 & 0 & 0 & 0 & 0 \\
\hline
\hline
(\frac12,\,0) & 11    &
 \frac{1}{3} & 0 & 0 & \frac{1}{6} & 0 & 0 & 4 & 8 & 0 \\
\hline
              & 12    &
 0 & 0 & 0 & 0 & 0 & 0 & 0 & 0 & 0 \\
\hline
              & 13    &
 0 & \frac{1}{8}\sqrt{\frac{3}{2}} & 0 & 0 & \frac{1}{8}\sqrt{\frac{3}{2}} & 0 & \frac{5}{2}\sqrt{\frac{3}{2}} & \sqrt{6} & \frac{3}{2}\sqrt{\frac{3}{2}} \\
\hline
              & 22    &
 0 & 0 & 0 & 0 & 0 & 0 & 0 & 0 & 0 \\
\hline
              & 23    &
 0 & -\frac{1}{8}\sqrt{\frac{3}{2}} & 0 & 0 & -\frac{1}{8}\sqrt{\frac{3}{2}} & 0 & -\frac{3}{2}\sqrt{\frac{3}{2}} & -3 \sqrt{6} & \frac{3}{2}\sqrt{\frac{3}{2}} \\
\hline
              & 33    &
 0 & 0 & 0 & \frac{1}{4} & 0 & 0 & 3 & 0 & 3 \\
\hline
\hline
(\frac32,\,0) & 11    &
 -\frac{1}{6} & 0 & 0 & -\frac{1}{12} & 0 & 0 & -2 & -4 & 0 \\
\hline
\hline
(0,\,-1)      & 11    &
 \frac{1}{4} & 0 & 0 & 0 & 0 & 0 & 3 & 0 & 3 \\
\hline
\hline
(1,\,-1)      & 11    &
 -\frac{1}{12} & 0 & 0 & -\frac{1}{6} & 0 & 0 & -3 & 0 & -3 \\
\hline
\end{tabular}
\caption{The coefficients $C^{(I,S)}$ that characterize the  interaction of Goldstone bosons with
heavy-meson fields $H$ as introduced in (\ref{res-T3}) for
given isospin (I) and strangeness (S). The channel ordering is specified in
Tab. \ref{tab:states}.}
\label{tab:coeffB}
\end{center}
\end{table}

We close this Section by a short discussion on how to translate our result (\ref{res-T3}) into the sectors involving two $J^P=1^-$ $D$ mesons as implied by the hevay-quark symmetry. Though in the chiral Lagrangian we have chosen to interpolate the $1^-$ mesons in terms of anti-symmetric tensor fields, for the readers convenience,  we express the scattering amplitudes in terms of 
the more conventional wave functions $\epsilon^\mu( p , \lambda)$ with momentum $p$ and polarization $\lambda $. Then an application of the partial-wave decomposition scheme as detailed in \cite{Lutz:2011xc} is accessible more directly. 
We exemplify the scattering amplitude by the leading order Tomozawa-Weinberg interactions from (\ref{def-kin}) for which we recall
\begin{eqnarray}
&& f^2 \,T^{(1^- \to 1^-)}_{ab}(s,t,u) = - \epsilon^\dagger_\mu( \bar p , \bar \lambda ) \Bigg[ \Big(   \frac{M_a^2+ M_b^2}{2\,M_a\,M_b}\,\frac{s-u}{4} +  \frac{M_a^2-M_b^2}{2\,M_a\,M_b}\,\frac{m_a^2-m_b^2}{4} \Big) g^{\mu \nu} 
\nonumber\\
&& \;   +\, \Big(\frac{M_a}{4\,M_b}-\frac{M_b}{4\,M_a} \Big)  \,\big( q^\mu \,q^\nu  - \bar q^\nu \,\bar q^\mu\big) + \Big(\frac{M_a}{4\,M_b}+\frac{M_b}{4\,M_a} \Big)  
\,\big( q^\mu \,\bar q^\nu  -q^\nu \,\bar q^\mu\big)\,\Bigg]\,\epsilon_\nu( p , \lambda)\,C_{\rm WT}\,,
\label{res-WT-spin-1}
\end{eqnarray}
where we use $\bar p_\mu $ and $\bar \lambda $ for the momentum and polarization of the final $D$ meson. If we expand (\ref{res-WT-spin-1}) according to our power counting rules (\ref{def-power-counting}), the leading order $Q$-term in the square bracket is given by $(s-u) \,g^{\mu \nu}/4$. As expected this structure corresponds to the analogous term $(s-u)/4$ in the first line of (\ref{res-T3}). We checked that this correspondence holds also for the $Q^2$ and $Q^3$ terms in (\ref{res-T3}) with the obvious replacements  $
 c_{i} \to \tilde c_i $, $
 g_{i} \to \tilde g_i $ and $ M \to \tilde M$. 
 
Yet, it remains to detail the contributions from the LEC $g_{4,5}$ and $\tilde g_{4,5}$. We find 
\begin{eqnarray}
&& f^2 \,T^{(3)}_{1^- \to 1 ^-}(s,t,u) =  \frac{s-u}{M_a\,M_b/\tilde M} \,\Big(  (\tilde g_4 + \tilde g_5) \,C_2 - 2\,\tilde g_4\,C_3\Big)\,\epsilon^\dagger_\mu( \bar p , \bar \lambda ) \Big[ 2\, \big(
\,q^\mu \,\bar q^\nu  - q^\nu \,\bar q^\mu \big) \Big]\,\epsilon_\nu( p , \lambda) \,,
\nonumber\\
&& f^2 \,T^{(3)}_{0^- \to 1 ^-}(s,t,u) = -2\,\frac{s-u}{M\,M_a} \,
\Big(  ( g_4 + g_5) \,C_2 -  2\,g_4\,C_3\Big)\,\epsilon^{ \alpha \beta \mu \nu }\,\bar q_\alpha \,q_\beta  \,\epsilon^\dagger_\mu( \bar p , \bar \lambda )\,p_\nu\, ,
\label{res-WT-spin-2}
\end{eqnarray}
where we use $M_a = \sqrt{\bar p ^2 }, M_b = \sqrt{p^2} $ and  the Clebsch matrices $C_2$ and $C_3$ as introduced in the previous works 
\cite{Hofmann:2003je,Lutz:2007sk}. While the $\tilde g_4$ and $\tilde g_5$ contribute to the $ 0^-\,1^- \to 0^-\,1 ^-$ processes, the heavy-quark symmetry related $g_4$ and $g_5$ contribute to the production processes $ 0^-\,0^- \to 0^-\,1 ^-$. In the heavy-quark mass limit it holds $\tilde g_n = g_n$, in particular for $n=4,5$.

\section{Scale dependence of LEC and some sum rules}

A brief discussion on the role of the renormalization scale $\mu$ dependence of the LEC is given. 
While we do not encounter a scale dependence in the leading order $Q^0$ and $Q^2$ this is no longer the case for the $Q^3$ and $Q^4$ counter terms. 
Well established is such a running in the light meson sector, with for instance $L_4$ and $L_5$ in (\ref{res-L45-running}), the cases which are most relevant for our current study. The running of the counter terms $g_n$ was already presented in (\ref{res-L45-running}).

In our previous work \cite{Guo:2018kno} we established results for $d_{1-4}$ and $\tilde d_{1-4}$ of the form 
\begin{eqnarray}
&& \qquad \qquad \qquad \mu^2 \frac{d}{d\,\mu^2}\,d_n = -\frac{\Gamma_{d_n}}{(4\,\pi\,f)^2}\,,
\nonumber\\
&&  \Gamma_{d_1}= \frac{1}{24}\, \big(4\, c_1+12 \,c_3+3\, c_5\big)\,,\qquad \qquad 
\Gamma_{d_2}= \frac{1}{36}\, \big(44 \,c_1-52 \,c_3-13\, c_5\big),\nonumber\\
&& \Gamma_{d_3}= \frac{1}{72}\, \big(240\, c_0-84 \,c_1+240 \,c_2+68\, c_3+60 \,c_4+17 \,c_5\big)\,,\nonumber\\
&& \Gamma_{d_4}= \frac{1}{108} \,\big(264\, c_0-132\, c_1+264 \,c_2+140\, c_3+66 \,c_4+35\, c_5\big)\,,
\label{res-Gamma-dn}
\end{eqnarray}
where identical results hold for the $\tilde c_n$ and $\tilde d_n$ coupling constants.
Our results were derived by insisting on the renormalization scale invariance of the $D$ meson masses. Here the $Q^2$ counter terms $c_n$ and $\tilde c_n$ contribute via tadpole-type integrals that depend on $\mu$. Such a dependence is cancelled identically with (\ref{res-Gamma-dn}). As already emphasized the scale dependent $d_n$ and $\tilde d_n$ contribute also to the two-body scattering amplitudes at tree-level. Thus, there must be a set of one-loop 
contributions that balance their $\mu$ dependence also here. Indeed, with the set of diagrams collected in Fig. \ref{fig:2} this is accomplished.  
We verify this in the Appendix A, where explicit and concise results are collected for all one-loop contributions proportional to $c_n$.

\begin{figure}[t]
\center{
\includegraphics[keepaspectratio,width=1.0\textwidth]{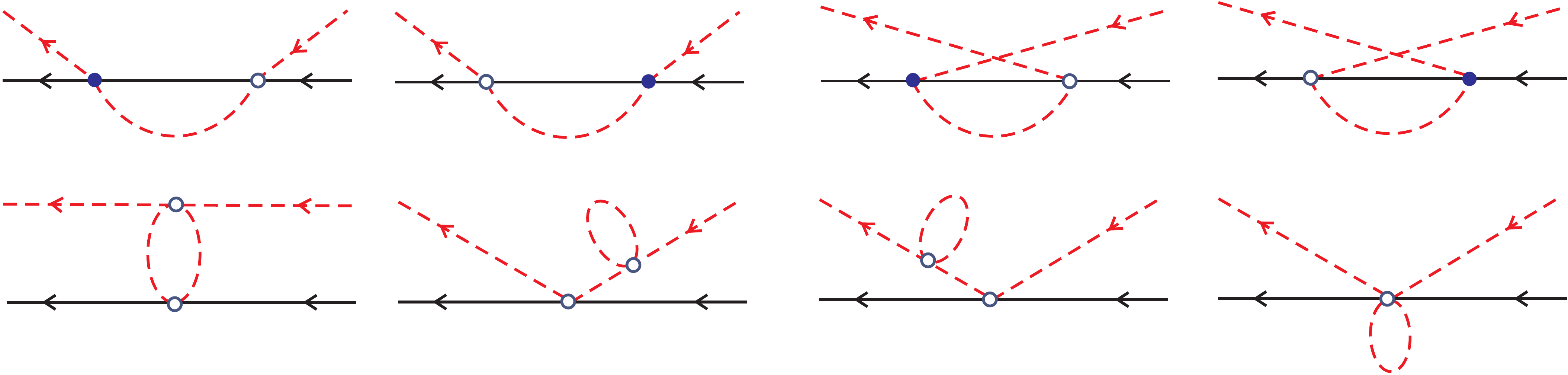} }
\vskip-1.0cm
\caption{\label{fig:2} One-loop diagrams at chiral order four. Solid vertices stand for order one and open  vertices for order 2 structures.  }
\end{figure}

How about the scale dependence of the remaining $d_{5-26}$ and $\tilde d_{5-26}$? In our current work we derived those for the first time.  
For the readers convenience we detail the contribution of the LEC as introduced in (\ref{def-L4})
to the scattering amplitudes at tree-level
\begin{eqnarray}
&& f^2\,T^{(4)}_{ab}(s,t,u)  =\frac{(s-u)^4}{64\,M^4}\,   \Big( d_{25}\,C_3 + d_{26}\,C_2 \Big)
+\,  (\bar q \cdot q)\, \frac{(s-u)^2}{4\,M^2}   \Big( d_{23}\,C_3 + d_{24}\,C_2 \Big)
\nonumber\\
&& \qquad+\,
4\, (\bar q \cdot q)^2\, \Big( d_{21}\,C_3 + d_{22}\,C_2 \Big)
 +\,
\sum_{n=1}^{6} \,C^{\chi}_n\,\Big(  d_{8+n}\,(\bar q\cdot q) + d_{14 +n}\,\frac{(s-u)^2}{16\,M^2}    \Big) 
+ \sum_{n=1}^8 \, d_n\,C^{\chi\chi }_n
\nonumber\\
&& \qquad +\,\Big( [ ( \bar q\cdot q) - m_a^2 ]\,[( \bar q\cdot q) - m_b^2] - \frac{1}{4}\,( M_a^2- M_b^2)^2  \Big) \Big(c_4\,C_2+ c_5\,C_3  \Big)/M^2
\nonumber\\
&& \qquad +\,g_1\,( M_a^2- M_b^2)  \,C^{\chi}_0/M \,,
\nonumber\\
&&  C^{ \chi}_n \,\,=2\,m \,B_0 \,C^{\pi}_n + (m+ m_s )\,B_0 \,C^{K }_n \,,
\nonumber\\
&& C^{\chi \chi}_n =(2\,m )^2\,B_0^2 \,C^{\pi \pi}_n + (m+ m_s )^2\,B_0^2 \,C^{KK }_n 
+2\, m\, (m+ m_s )\,B_0^2 \,C^{\pi K }_n\,,
\label{res-T4-counter}
\end{eqnarray}
where we apply the previously introduced Clebsch $C_2$ and $C_3$ properly supplemented by additional Clebsch $C^{\chi}_n$ and $C^{\chi \chi}_n$ that reflect the explict symmetry breaking impact of their associated LEC. Details on their specific form can be found in the  Appendix. Analogous results can be easily derived for the $ 0^-\,1^- \to 0^-\,1 ^-$ processes. We note that it is convenient to consider part of the third-line terms in (\ref{res-T4-counter}) as a renormalization of the $d_n \to d'_n$ in its first two lines. This goes with
\begin{eqnarray}
&& d'_n = d_n + \frac{c_4}{M^2}\,\gamma^{(4)}_{n} + \frac{c_5}{M^2}\,\gamma^{(5)}_{n} \qquad{\rm with} \qquad \gamma^{(4)}_{n}= \gamma^{(5)}_{n} = 0\qquad {\rm but}   
\nonumber\\
&& \gamma^{(4)}_3=  2\,\gamma^{(5)}_3= \frac{1}{2}\,,\qquad\; \;\gamma^{(5)}_5 = -\frac{1}{4}\,,
\qquad \gamma^{(5)}_6 = \frac{1}{6}\,,\qquad \;\;\gamma^{(5)}_8 = \frac{2}{3}\, \gamma^{(4)}_8 =-\frac{1}{9}\,,
\nonumber\\
&&\gamma^{(5)}_{11} = 2\, \gamma^{(5)}_{9}= -4\, \gamma^{(5)}_{10}= \frac{1}{6}\,,\qquad \quad \;\;\;\;
\gamma^{(5)}_{12} = -\frac{1}{4}\, \gamma^{(5)}_{13}= - \gamma^{(5)}_{14}=\frac{1}{3}\,\gamma^{(5)}_{21}= \frac{1}{12} \,,
\nonumber\\
&& \gamma^{(4)}_{22} =-\frac{1}{4}\, \gamma^{(4)}_{13}=\frac{1}{4}\,,
\label{res-dprime}
\end{eqnarray}
where we will omit the prime in $d'_n$ for notational clarity further below. The remaining terms proportional to $M_a^2-M_b^2 \sim (m -m_s)\,c_1$, if approximated to order $Q^2$ cannot be considered as an additional  renormalization of the $d_{1-8}$. It is important to realize that this is also not the case 
for the term in the 4th line of (\ref{res-T4-counter}). The scale-depdendence in $g_1$ is crucial to 
balance the scale dependence in our set of loop contributions.

Our results for the scale dependence of the $d_n$ and $\tilde d_n$  follow from the detailed analysis of the one-loop diagrams proportional to $c_n$ in the Appendix. For the symmetry conserving terms with $n> 20$ this leads to
\begin{eqnarray}
&& \mu^2\,\frac{d}{d\,\mu^2}\, d^{}_{21} = \Big(\frac{3}{2}\,c_3+  \frac{1}{4}\,c_5  \Big)\,\frac{1}{(4\,\pi\,f)^2}\,,
\nonumber\\
&& \mu^2\,\frac{d}{d\,\mu^2}\, d^{}_{22} =  \Big( 3\,c_2 + \frac{1}{2}\,c_3+  \frac{1}{2}\,c_4 + \frac{1}{12}\,c_5 \Big)\frac{1}{(4\,\pi\,f)^2} \,,\qquad  
\nonumber\\
&& 
\mu^2\,\frac{d}{d\,\mu^2}\, d^{}_n\, =0\qquad \qquad \qquad \;  
\,\qquad  {\rm for}\qquad 23\leq n \leq 26  \,.
\label{d-running-symmetric}
\end{eqnarray}
Further results for the scale dependence of the symmetry breaking terms $d_n$  with $n < 21$ are efficiently summarized with 
\begin{eqnarray}
&& \mu^2\,\frac{d}{d\,\mu^2}\, d^{\,\chi}_n\, =
\mu^2\,\frac{d}{d\,\mu^2}\, d^{\,t}_n = \frac{c_n}{(4\,\pi\,f)^2} \,\qquad  
{\rm for}\qquad n \in \{1,3,5\} \,,
\nonumber\\
&& \mu^2\,\frac{d}{d\,\mu^2}\, d^{\,\chi}_2 \,=\frac{c_2 -2\, c_0}{(4\,\pi\,f)^2}  \,,\qquad \;\, \qquad
\mu^2\,\frac{d}{d\,\mu^2}\, d^{\,t}_2 = \frac{c_2 - c_0/2}{(4\,\pi\,f)^2} \,,\qquad  
\nonumber\\
&& \mu^2\,\frac{d}{d\,\mu^2}\, d^{\,\chi}_4\, =\frac{c_4 +12\, c_0}{(4\,\pi\,f)^2}  \,,\qquad \qquad 
\mu^2\,\frac{d}{d\,\mu^2}\, d^{\,t}_4 = \frac{c_4 + 3\,c_0}{(4\,\pi\,f)^2} \,,\qquad  
\nonumber\\
&& 
\mu^2\,\frac{d}{d\,\mu^2}\, d^{su}_n = \frac{c_n}{(4\,\pi\,f)^2} \qquad \qquad \quad \;  \, \,
\,\qquad  {\rm for}\qquad n \in \{3,4,5\} \,, 
\nonumber\\
&& 
\mu^2\,\frac{d}{d\,\mu^2}\, d^{su}_6=
\mu^2\,\frac{d}{d\,\mu^2}\, d^{\,t}_6 = \frac{1}{(4\,\pi\,f)^2} \,,
\label{d-running}
\end{eqnarray}
in terms of the convenient linear combinations
\begin{eqnarray}
&& d^{\,\chi}_1 = -\frac{39}{23}\,d_1- \frac{27}{46}\,d_2\,,\qquad  \qquad  \quad \,\,
   d^{\,\chi}_2 = \frac{43}{115}\,d_1 +\frac{12}{23}\,d_2+\frac{3}{5}\,d_3 +\frac{111}{80}\,d_5+\frac{9}{16}\,d_7\,,
\nonumber\\
&& d^{\,\chi}_3= \frac{66}{23}\,d_1-\frac{9}{23}\,d_2- \frac{18}{5}\,d_5\,, \qquad \;\;
 d^{\,\chi}_4 = -\frac{258}{115}\,d_1 - \frac{72}{23}\,d_2 - \frac{18}{5}\,d_3
 -\frac{111}{20}\,d_5-\frac{9}{4}\,d_7\,,
\nonumber\\
&& d^{\,\chi}_5 = -\frac{396}{23}\,d_1+\frac{54}{23}\,d_{2} +\frac{72}{5}\,d_5\,,
\nonumber\\ 
\label{def-chitsu}\\
&& d^{\,t}_1 = \frac{43}{51}\,d_{9}+\frac{74}{17}\,d_{10} - \frac{295}{153}\,d_{11}+ \frac{116}{17}\,d_{12}\,, \qquad \; \,\,\,
 d^{\,t}_3 =-\frac{3}{17}\,d_{9}- \frac{6}{17}\,d_{10}+ \frac{29}{17}\,d_{11}-\frac{48}{17}\,d_{12}\,, 
\nonumber\\
&& d^{\,t}_2 = \frac{1}{102}\,d_9 -  \frac{11}{17}\,d_{10} - \frac{4}{9}\,d_{11}+ \frac{11}{68}\,d_{12}-\frac{9}{68}\,d_{13}-  \frac{7}{68}\,d_{14}\,,
 \nonumber\\
&& d^{\,t}_4 = -2\,d_9+ \frac{217}{51}\,d_{11} - \frac{159}{34}\,d_{12} +\frac{9}{34}\,d_{13}+\frac{75}{34}\,d_{14}\,,
\nonumber\\
&& d^{\,t}_5 = \frac{108}{17}\,d_{9}+ \frac{216}{17}\,d_{10}- \frac{228}{17}\,d_{11} + \frac{504}{17}\,d_{12} \,, \qquad 
 d^{\,t}_6 = \frac{332}{51}\,d_{9}+ \frac{40}{17}\,d_{10}- \frac{1196}{153}\,d_{11} + \frac{320}{17}\,d_{12} \,,
\nonumber\\ 
\nonumber\\
&& d^{su}_3 =  -\frac{12}{25}\,d_{15}- \frac{84}{25}\,d_{16}-\frac{26}{25}\,d_{19}+\frac{78}{25}\,d_{20} \,, \qquad 
 d^{su}_4 = \frac{19}{75}\,d_{15} -\frac{142}{75}\,d_{16} -\frac{21}{25}\,d_{19}+\frac{38}{25}\,d_{20}\,, 
 \nonumber\\
&& d^{su}_5 = -\frac{16}{25}\,d_{15} +\frac{88}{25}\,d_{16}+ \frac{32}{25}\,d_{19}-\frac{96}{25}\,d_{20}\,, \qquad \,
 d^{su}_6 = -\frac{136}{75}\,d_{15} +\frac{448}{75}\,d_{16}+ \frac{24}{25}\,d_{19}-\frac{72}{25}\,d_{20}\,,
\nonumber
\end{eqnarray}
where we note that our set of linearly combined LEC does not fully match the original set of LEC in (\ref{def-L4}). This is a consequence 
of the existence of five particular linear combinations that are renormalization-scale invariant in our current setup:
\begin{eqnarray}
&& d^{\chi }_6\, =  -\frac{11}{45} \,d_1 + \frac{1}{3} \,d_2 - \frac{11}{15}\,d_3 + d_4  \,, \qquad \quad \quad \; \;
d^{\chi}_7 =  \frac{1}{3}\,d_5 + d_6\,, \qquad \quad \quad d^{\chi}_8 =  \frac{1}{3}\,d_7 + d_8\,, 
\nonumber\\
&& d^{su}_1 = -\frac{3}{5}\,d_{15} -\frac{6}{5} \,d_{16}+ d_{17} -\frac{4}{5}\,d_{19} +\frac{12}{5}\,d_{20}\,,\qquad \qquad \qquad  \;  
\nonumber\\
&& d^{su}_2 = \frac{4}{15}\,d_{15} + \frac{8}{15}\,d_{16} + d_{18}-\frac{1}{5}\,d_{19}+\frac{3}{5}\,d_{20}\,.
\label{def-d-inv}
\end{eqnarray}
It is emphasized that corresponding expressions for the renormalization-scale dependence of $\tilde d_n$ in the $ 0^-\,1^- \to 0^-\,1 ^-$ processes are generated by the systematic replacements $d_n \to \tilde d_n$ and $c_n \to \tilde c_n$ in (\ref{d-running-symmetric}-\ref{def-d-inv}).

The implications of large-$N_c$ QCD for the renormalization-scale dependent $d_n$ and $\tilde d_n$  are more intricate as compared to those  for the renormalization-scale invariant LEC like $c_{0-5}$. Sum rules for the former can be established only for a natural choice of the renormalization scale $\mu$. Our further studies rest on the set of  approximations
\begin{eqnarray}
&& d_6 = d_7=d_8 = 0 \,,\qquad d_{22}=- d_{21}/2\,\qquad 
d_{24}=- d_{23}/2 \,,\qquad d_{26}=-d_{25}/2\,,
\nonumber\\
&& d_{11}=d_{12}=d_{13} = d_{14} = 0\,,\qquad \qquad  d_{17}=d_{18}=d_{19} = d_{20} = 0\,,\qquad 
\label{large-Nc-ds-A}
\end{eqnarray}
used at $\mu \sim m_\rho$. Clearly, at $\mu \neq m_\rho$, the LEC may then start to depart from (\ref{large-Nc-ds-A}). Such relations are easily derived by looking at the number of flavor traces  in the corresponding interaction terms in (\ref{def-L4}). The larger the number of flavor traces, the less important a given term turns in the large-$N_c$ hierarchy of QCD. 
It is useful to translate the relations (\ref{large-Nc-ds-A}) into the more convenient set of LEC introduced in (\ref{def-chitsu}) and (\ref{def-d-inv}). This leads to 
\begin{eqnarray}
&& d^\chi_5=  -\frac{576}{37}\,d^\chi_2 -6\, d^\chi_3 -\frac{96}{37}\, d^ \chi_4 \,, \qquad  \qquad \,
 d^\chi_7= \frac{80}{111}\,d^\chi_2 +\frac{40}{333}\, d^\chi_4\,, \qquad \qquad d^\chi_8 =0 \,,\qquad
\nonumber\\
&&  d^{\,t}_1= -\frac{37}{3} \,d^{\,t}_3 +\frac{2}{3} \,d^{\,t}_4 \,,\qquad \,
d^{\,t}_2= \frac{11}{6} \,d^{\,t}_3-\frac{1}{6}\,d^{\,t}_4 \,, \qquad \,
d^{\,t}_5=  -36 \,d^{\,t}_3 \,, \qquad \,
 d^{\,t}_6=-\frac{20}{3} \,d^{\,t}_3   -\frac{8}{3} \,d^{\,t}_4  \,,
\nonumber\\
&& d^{su}_1 = \frac{9}{11}\,d^{su}_3-\frac{9}{11}\,d^{su}_4\,,\qquad \quad \,\, d^{su}_2 = -\frac{4}{11}\,d^{su}_3 + \frac{4}{11}\,d^{su}_4 \,,
\qquad  \quad \,\,d^{su}_5 = \frac{2}{11}\,d^{su}_3 -\frac{24}{11}\,d^{su}_4 \,,
\nonumber\\
&&d^{su}_6 = \frac{12}{11}\,d^{su}_3 -\frac{56}{11}\,d^{su}_4\,,
\label{large-Nc-ds-B}
\end{eqnarray}
again at $\mu \sim m_\rho$, in terms of which it is justified to consider $d^\chi_{1-4,6}, d^{\,t}_{3-4}$ and $d^{su }_{3-4}$ together with $d_{21}, d_{23}, d_{25}$ as a set of independent $Q^4$ LEC to be determined in our work. For the convenience of the reader we provide the set of inverse relations with
\begin{eqnarray}
&& d_1 = -\frac{1}{6}\,d^{\chi}_1 + \frac{72}{37}\, d^{\chi}_2 + \frac{1}{4}\,d^{\chi}_3 + \frac{12}{37}\,d^\chi_4 \,,
\qquad \qquad \;\;\,
d_2 = -\frac{11}{9}\,d^{\chi}_1 - \frac{208}{37}\, d^{\chi}_2 - \frac{13}{18}\,d^{\chi}_3 - \frac{104}{111}\,d^\chi_4 \,,
\nonumber\\
&& d_3 = \frac{7}{6}\,d^{\chi}_1 + \frac{38}{111}\, d^{\chi}_2 + \frac{17}{36}\,d^{\chi}_3 - \frac{49}{222}\,d^\chi_4 \,,
\qquad \qquad 
d_4 = \frac{11}{9}\,d^{\chi}_1 + \frac{866}{333}\, d^{\chi}_2 + \frac{35}{54}\,d^{\chi}_3 + \frac{17}{74}\,d^\chi_4 + d^\chi_6 \,,
\nonumber\\
&& d_5 = \frac{80}{37}\,d^\chi_2+ \frac{40}{111}\, d^\chi_4\,,
\qquad \qquad \;\;\,
d_9 = - \frac{1}{2} \,d^{\,t}_4\,,\qquad \qquad d_{10} = -\frac{17}{6} \,d^{\,t}_3 +\frac{1}{4} \,d^{\,t}_4 \,, 
\nonumber\\ 
&& d_{15} = -\frac{71}{66}\,d^{su}_3 + \frac{21}{11}\,d^{su}_4 \,,\qquad \qquad 
d_{16} = -\frac{19}{132}\,d^{su}_3 - \frac{3}{11}\,d^{su}_4 \,,
\label{large-Nc-ds-B}
\end{eqnarray}
to be used in combination of (\ref{large-Nc-ds-A}).

\clearpage

\section{Coupled-channel scattering amplitudes}

Partial-wave amplitudes are introduced by suitable averages of the the two-body scattering amplitude $T_{ab}$ 
over the center-of-mass scattering angle $\theta$. 
For simplicity we set the scene with the simplest reaction, where a Goldstone boson hits a pseudoscalar $D$ meson. Since both carry spin zero, the math required is particularly simple. For $(-1)^{J} = P $ with total angular momentum, $J$, and parity, $P$, it holds
\begin{eqnarray}
T_{ab}^J(s) = \int_{-1}^{+1} \frac{d \cos \theta}{2} \,
\left(\frac{s}{p_a\,p_b} \right)^J\,T_{ab}(s,t,u)\,
P_J(\cos \theta) \,, \label{def-TJ}
\end{eqnarray}
where the relative
momenta of the initial and final states are denoted by $p_b = p$ and $p_a =\bar p$. The total angular momentum is $J$ and $P_J(\cos \theta )$ the Legendre polynominal. The scattering amplitude $T_{ab}$ 
depends on the Mandelstam variables $s,t$ and $u$, whose sum 
$ s+ t+ u = m_a^2 + M_a^2 +  m_b^2 + M_b^2 $ is fixed by the specifics of the given channel characterized by the indices $a$ and $b$. While in our convention the $m_a$ selects a pseudo-Goldstone boson, the $M_a$ follows from a $D$ meson with $J^P = 0^-$ (and further below also for $J^P =1^-$).

The coupled-channel partial-wave amplitudes $T^J_{ab}(s)$ are characterized by so-called left-hand and right-hand cuts, where the right-hand cuts are implied by the coupled-channel unitarity condition. 
It is instrumental to introduce a generalized potential $U^J_{ab}(s)$, which is determined by the left-hand cut contributions only. The separation may be introduced by the nonlinear integral equation
\begin{eqnarray}
T^J_{ab}(s)=U^J_{ab}(s)
+\sum_{c,d}\int^{\infty}_{\mu_{thr}^2}\frac{d\bar{s}}{\pi}\frac{s-\mu_M^2}{\bar{s}-\mu_M^2}\,
\frac{T^J_{ac}(\bar s)\,\rho^J_{cd}(\bar s)\,T^{J*}_{db}(\bar s)}{\bar{s}-s-i\epsilon}\,,
\label{def-non-linear}
\end{eqnarray}
with the phase-space matrix
\begin{eqnarray}\label{rho}
\rho^J_{ab} (s)=\frac{1}{8 \pi } \left(\frac{p_a}{\sqrt{s}}\right)^{2\,J+1}\,\delta_{ab}\,.
\label{def-phase-space}
\end{eqnarray}
Given a suitably approximated generalized potential $U^J_{ab}(s)$ the coupled-channel amplitudes are determined as solutions of the nonlinear set of integral equations (\ref{def-non-linear}). By construction any such solution satisfies the coupled-channel unitarity condition.  
The matching scale 
\begin{eqnarray}
&& m_1^2 + M_1^2 < \mu_M^2 <  (m_1+ M_1)^2  \,, \qquad \qquad
\mu_M^2  \equiv \frac{1}{2}\,\Big(  m_1^2 + M_1^2 +  (m_1+ M_1)^2 \Big) \,,
\label{def-matching}
\end{eqnarray}
in (\ref{def-non-linear}) specifies 
where we expect a strict $\chi$PT approach to coincide with the coupled-channel approach followed here. It should be slightly below the smallest two-body threshold at $m_1+ M_1$ accessible in a sector with given isospin and strangeness. Given our approximation scheme it cannot be moved much further left, as the unitarity effects from the crossed u-channel will turn more and more important. A useful condition
follows from a simple kinematical consideration. Consider elastic scattering in forward direction at $t=0$. Then s-channel and u-channel unitarized scattering amplitudes should coincide at 
$s=u = m_1^2+ M_1^2$. In our work we insist on $\mu_M$ as specified in (\ref{def-matching}).

Before going into the details of how to obtain solutions to (\ref{def-non-linear}) we need to detour on a technical issue implied by spin effects. Matters turn more complicated once a Goldstone boson interacts with a $D$ meson with $J^P = 1^-$. In this case a coupled-channel state comes in two helicity variants \cite{Jacob:1959at,Lutz:2003fm,Lutz:2011xc}. Such channels turn relevant as we consider p-wave phase shifts. For instance a p-wave $D\,\pi$ channel may couple to two $D^*\,\pi$ channels with different helicities. This is not possible for 
an s-wave $D\,\pi$ channel. It is long known that in this case helicity partial-wave amplitudes  of distinct angular momenta $J$ are correlated at pseudothresholds \cite{Jacob:1959at,Jackson:1968rfn,Hara:1964zza}. If such correlations are ignored 
kinematical singularities would arise that essentially prohibit the use of partial-wave dispersion relations, exactly those our GPA rests on. Therefore it is important to use 'covariant' partial-wave amplitudes as introduced first in \cite{Lutz:2003fm,Lutz:2011xc}. The merit of those is 
that they are uncorrelated at pseudothresholds and therefore are suitable to be used in our GPA. It may be of interest to recall that the 'covariant' partial-pwave amplitudes arise naturally if the Bethe-Salpeter coupled-channel equation is solved for spin systems in the presence of short range forces \cite{Lutz:2001yb,Stoica:2011cy,Lutz:2011xc,Heo:2014cja}. An unavoidable consequence of this request is that the associated phase-space matrix receives nondiagonal matrix elements. Non-trivial spin effects are seen in a sector with parity $P$ and total angular momentum $J > 0$  satisfying the condition 
$(-1)^{J +1} = P $.  In this case the two-component spin structure of the phase-space matrix is
\begin{eqnarray}
&& \rho^J_{ab,11} (s) = \frac{1}{p_a^2}\,\Big[1+\frac{J}{J+1}\,\frac{s\,\kappa_-^2}{4\,M_a^2}\Big] \rho^J_{ab}(s) \,,
\nonumber\\
&& \rho_{ab,12}^J(s) = \rho_{ab,21}^J(s)= \sqrt{\frac{J}{J+1}}\,\frac{\kappa_-}{2\,M_a^2} \,\rho^J_{ab}(s) \,,
\nonumber\\
&& \rho_{ab,22}^J(s) =  \frac{p_a^2}{M_a^2\,s}\, \rho^J_{ab}(s) 
 \qquad \quad {\rm with} \qquad \quad \kappa_ \pm = 1 \pm \frac{m_a^2 - M_a^2}{s}\,.
\end{eqnarray}
For the other case with $(-1)^{J} = P $ the phase-space in (\ref{def-phase-space}) applies. 
The associated generalizations of (\ref{def-TJ}) are detailed in \cite{Lutz:2011xc}. 

Given a generalized potential $U^J_{ab}(s)$ the nonlinear
integral equation implies partial-wave scattering amplitudes
$T^J_{ab}(s)$ that comply with the coupled-channel unitarity condition. 
From the form of (\ref{def-non-linear}) it follows that
the existence of a solution requires the generalized potential to
be bounded asymptotically, modulo some possibly logarithmic terms.
Therefore, a direct evaluation of $U^J_{ab}(s)$ in $\chi$PT is not possible. Any
finite order truncation leads to an unbounded potential,
characterized by an asymptotic growth in some power of $s$. In turn the equation 
(\ref{def-non-linear}) had to be tamed by some cutoff and physical results would almost unavoidably suffer from significant cutoff effects, which should not be accepted on our way towards 
an effective field theory approach for coupled-channel dynamics. In  principle, such cutoff effects have to be absorbed into a renormalization of the generalized potential. We choose to set up an approximation scheme for the renormalized potential directly. 

Fortunately, there is an elegant and efficient solution to this problem implied by the use of conformal variables. This is so since a numerical solution of (\ref{def-non-linear}) requires
the knowledge of the generalized potential, $U^J_{ab}(s)$, for energies larger than the maximum of the initial and final thresholds only. In this domain the generalized potential does not suffer from neither left- nor right-hand branch points being liberated from the s-channel unitarity cuts by construction. As was pointed out in \cite{Gasparyan:2010xz,Danilkin:2010xd}, the required generalized potential, $U_{ab}^J(s)$, can be reconstructed unambiguously in this domain in terms of its
derivatives at a chosen point $s= \mu_{ab,E}^2$, where the results of a conventional $\chi$PT approach are reliable. We note that matters can turn more complicated in the presence of anomalous thresholds  \cite{Lutz:2015lca,Lutz:2018kaz}. In our current work we avoid such anomalous systems.

Following \cite{Gasparyan:2010xz} we identify $\mu_{ab,E}$ with the mean of initial, $m_b + M_b$, and final, $m_a + M_a$, thresholds. A Taylor expansion of $U(s)$ around $ s= \mu_E^2$ has a rather small convergence radius, that is determined by the distance to the closest
left-hand cut branch point. In order to extend the convergence up to some cutoff scale $\Lambda_s$, we
apply the conformal map, that was constructed in \cite{Gasparyan:2010xz}. It is recalled with 
\begin{eqnarray}
\xi(s)=\frac{\alpha\,(\Lambda^2_s-s)^2-1}{(\alpha-2\,\beta)(\Lambda^2_s-s)^2+1}\,,\qquad \alpha = \frac{1}{(\Lambda^2_s-\mu_E^2)^2},
 \qquad \beta = \frac{1}{(\Lambda^2_s-\Lambda^2_0)^2}\,,
 \label{def-conformal}
\end{eqnarray}
where the parameter $\Lambda_0$ is identified such that the
mapping domain of the conformal map touches that left-hand branch point. Within this domain, i.e. $\Lambda^2_0 < s < \Lambda^2_s $, the generalized potential can be reconstructed in terms of its derivatives at the expansion point $\mu_E^2$. It holds
\begin{eqnarray}
U(s )=\sum_{k=0}^{\infty} \,c_k \,\xi^k(s) \qquad {\rm for} \quad \Lambda_0^2 < s < \Lambda_s^2 \,,
\label{def-U-xipanded}
\end{eqnarray}
where the coefficient $c_k$ is determined by the first $k$ derivatives of $U(s)$ at $\mu_E^2$.
In our analysis the values of $c_k$  are all derived  with on-shell meson masses and the LEC of the chiral Lagrangian. Our strategy is to integrate out the physics at $s > \Lambda_s^2$. In order not to  induce large effects close to $s \simeq \Lambda_s^2$ we do so by insisting the potential 
in (\ref{def-U-xipanded}) to be continuous and reach a constant value at $s > \Lambda_s^2$. 
Our choice for $\Lambda_s$ is well constrained. On the one hand $\Lambda_s$ cannot be smaller than the 
maxium of the coupled-channel thresholds, 
but it cannot be much larger either, since typically there are further channels that are not considered explicitly. Therefore it is natural to insist on
\begin{eqnarray}
\Lambda_s = \Lambda_{\rm max} + \delta \Lambda_s  \qquad {\rm with} \qquad \Lambda_{\rm max} = {\rm Max} \{ m_a+ M_a \} \,, 
\label{def-Lambda-max}
\end{eqnarray}
with  a universal channel and quark-mass independent value of $\delta \Lambda_s$. 
With a particular choice of $\delta \Lambda_s$ we may alter the high-energy behavior of our 
coupled-channel recation amplitudes in a highly correlated manner. We expect the optimal choice for $\delta \Lambda_s$ to increase as the accuracy of a truncation in the  expansion in (\ref{def-U-xipanded}) delivers the generalized potential reliable up to larger and larger energies.

We wish to emphasize two important issues. First, given the enormous efficiency of the chiral Lagrangian to determine close-by left-hand cuts, it is advantageous to keep the latter explicitly and expand only the far-distant left-hand cut contributions in terms of a conformal expansion. This leads to the general form
\begin{eqnarray}
&& U(s )= U_{\rm close-by}(s) + U_{\rm far-distant}(s)   \,,
\nonumber\\
&& {\rm with } \qquad U_{\rm far-distant}(s) = \sum_{k}^{} \,c_k \,\xi^k(s)  \,.
\label{def-U-xipanded}
\end{eqnarray}
For a given channel we characterize this division by t-channel and u-channel cutoff parameters $ \Lambda_t$ and $\Lambda_u$. In turn, the parameter $\Lambda_0$ in our conformal map (\ref{def-conformal}) is determined  by the condition that for chosen $\Lambda_s$, $\Lambda_t$ and $ \Lambda_u$, the residual left-hand cut branch point coincides with $\Lambda_0$. Note that it is not always trivial to derive its proper value. Here the general results established in \cite{Lutz:2015lca} are instrumental. 
In our current study we identify $\Lambda_t$ with the smallest t-channel two-body unitarity 
branch point active in the given channel. The analogous identification is assumed for the u-channel. 

Given our construction loop contributions to the generalized potential enter via their derivatives at the expansion point $\mu_E^2$ exclusively. All left-hand cut contributions implied by t- and u-channel unitarity branch points are integrated out systematically.
Via a truncation of (\ref{def-U-xipanded}) we obtain an approximate generalized potential for energies $\Lambda_0 < \sqrt{s} < \Lambda_s$. For energies larger than the cutoff scale $\Lambda_s$ the generalized potentials $U_{\rm far-distant}(s) $ are set to a constant \cite{Gasparyan:2010xz}. By virtue of the specific form of the conformal map this is a smooth procedure. It remains to specify the expansion order at which the sum in (\ref{def-U-xipanded}) is truncated. This is naturally implied by the accuracy level of the used chiral Lagrangian. 

How to find numerical solutions of (\ref{def-non-linear}).
This is readily achieved in application of the $N/D$ technique
\cite{Chew:1960iv}. The partial-wave scattering amplitude is
decomposed
\begin{eqnarray}
T_{ab}(s)=\sum_{c}\,D^{-1}_{ac}(s)\,N_{cb}(s)\,,
\label{def-NoverD}
\end{eqnarray}
in terms of a matrix-valued function $D_{ab}(s)$ with only right-hand cuts and  a matrix-valued function $N_{ab}(s)$ with only left-hand cuts. With the ansatz
\begin{eqnarray}
&&D_{ab}(s)=\delta_{ab}  + R_{ab} (s)
 -\sum_{c}\int_{\mu_{thr}^2}^{\infty}\frac{d\bar{s}}{\pi}\frac{s-\mu_M^2}{\bar{s}-\mu_M^2}
\frac{N_{ac}(\bar{s})\rho_{cb}(\bar{s})}{\bar{s}-s}\,,
\label{def-D}
\end{eqnarray}
the coupled-channel unitarity is ensured for any real-valued rational functions $R_{ab}(s)$  ~\cite{Castillejo:1955ed,Gasparyan:2010xz}. As it is, the $N/D$ technique does not provide a solution to (\ref{def-non-linear}). Only after we specify an ansatz for the rational functions $R_{ab}(s)$ this may or may not be possible. Note that for an unfortunately chosen potential it may well be that the nonlinar system 
does not allow any solution. 

For the particular choice $R_{ab}(s) =0$,  the ansatz (\ref{def-NoverD}) to represent a solution of the nonlinear integral equation (\ref{def-non-linear}), the matrix function $N_{ab}(s)$ has to satisfy the linear integral equation
\begin{eqnarray}
&& N_{ab}(s)=U^{}_{ab}(s)+\,\sum_{c,d}\int_{\mu_{\rm thr}^2}^\infty \frac{d\bar s}{\pi}\,\frac{s-\mu^2_M}{\bar s-\mu^2_M}\,
\frac{N_{ac}(\bar s)\,\rho_{cd}(\bar s)\,[U^{}_{db}(\bar s)-U^{}_{db}(s)]}{\bar s-s}\,.
\label{def-N-linear}
\end{eqnarray}
Note that while the linear equation (\ref{def-N-linear}) can always be solved numerically, there is no guarantee that its solution respects the nonlinear equation (\ref{def-non-linear}) also. If this is 
not the case a more suitable form of $R_{ab}(s)$ needs to be found. It is important to realize that the GPA is defined necessarily by the nonlinear system (\ref{def-non-linear}) and not by the linear system 
(\ref{def-N-linear}). Only then it is possible to derive a coupled-channel scattering amplitude from a given chiral Lagrangian.  

In our case we use the ansatz $R_{ab}(s) = 0$ in all sectors with the exception of the $J^P = 1^-$ sector in which there is a s-channel vector $D^*$ meson exchange possible. Here we discriminate two cases. Consider first a $D^*$ meson that is stable against strong decays. Here the potential $U^{}_{ab}(s)$ shows a pole on the real axis right at the on-shell mass of the $D^*$ meson. A priori there is no reason to depart from  $R_{ab}(s) = 0$. However, as was emphasized and illustrated in ~\cite{Gasparyan:2010xz} the set of equations (\ref{def-D},\ref{def-non-linear}) can be cast into an alternative form in terms of an effective potential $ U^{\rm eff}_{ab}(s)$ with
\begin{eqnarray}
 && U^{\rm eff}_{ab}(s) = U_{ab}(s) + \frac{g_a\,M_{1^-}^2\,g_b}{s-M^2_{1^-}}\,\frac{s-\mu^2_M}{M_{1^-}^2-\mu^2_M}\,,
 \label{def-Ueff}
\end{eqnarray}
that is regular at the pole mass, $M_{1^-}$, by construction. Suitable values of the coupling constants $g_a$ in (\ref{def-Ueff}) guarantee that regularity. To compensate for the modification of the potential nontrivial rational functions are required
\begin{eqnarray}
R_{ab}(s) =  -\frac{s-\mu^2_M}{s-M^2_{CDD}}\,R^{(D)}_{ab} \,,
\end{eqnarray}
where the CDD pole mass, $M_{CDD}$, and its residua, $R^{(D)}_{ab}$, parameters are to be determined by the request that the results of the original system are recovered. The specific form of this condition were derived in ~\cite{Gasparyan:2010xz} with
\begin{eqnarray}
&& N_{ab}(s)=U^{\rm eff}_{ab}(s)
-\frac{s-\mu^2_M}{s-M^2_{CDD}}\,\Big[R^{(B)}_{ab}+\sum_c R^{(D)}_{ac}\,U^{\rm eff}_{cb} (s)\Big]
\nonumber\\
&&\qquad  \;\,+\,\sum_{c,d}\int_{\mu_{\rm thr}^2}^\infty \frac{d\bar s}{\pi}\,\frac{s-\mu^2_M}{\bar s-\mu^2_M}\,
\frac{N_{ac}(\bar s)\,\rho_{cd}(\bar s)\,[U^{\rm eff}_{db}(\bar s)-U^{\rm eff}_{db}(s)]}{\bar s-s}\,,
\label{CDD-ansatz}
\end{eqnarray}
and
\begin{eqnarray}
&& R^{(D)}_{ab}=\frac{M_{1^-}^2-M_{CDD}^2}{M_{1^-}^2-\mu^2_M}\left(\delta_{ab}-\sum_c\,\int_{\mu_{\rm thr}^2}^\infty \frac{d\bar s}{\pi}\,\frac{M_{1^-}^2-\mu^2_M}{\bar s-\mu^2_M}\,\frac{N_{ac}(\bar s)\,\rho_{cb}(\bar s)}{\bar s-M_{1^-}^2}\right)\,,
\nonumber\\
&& R^{(B)}_{ab}=-\frac{\mu^2_M-M_{CDD}^2}{(\mu^2_M-M_{1^-}^2)^2}\,g_a\,M_{1^-}^2\,g_b
\nonumber\\
&&\qquad \,-\,\sum_{c,d}\,\int_{\mu_{thr}^2}^\infty \frac{d\bar s}{\pi}\,\frac{\bar s-M_{CDD}}{\bar s-\mu^2_M}\,
\frac{N_{ac}(\bar s)\,\rho_{cd}(\bar s)}{(\bar s-M_{1^-}^2)^2}\,g_d\,M_{1^-}^2\,g_b\,
\nonumber\\
&&\qquad \,+\,(M_{1^-}^2-M_{CDD}^2)\,\sum_{c,d}\,\int_{\mu_{ \rm thr}^2}^\infty \frac{d\bar s}{\pi}\,
\frac{N_{ac}(\bar s)\,\rho_{cd}(\bar s)\,U^{\rm eff}_{db}(\bar s)}{(\bar s-\mu^2_M)\,(\bar s-M_{1^-}^2)}\,.
\label{identify-R}
\end{eqnarray}
The merit of the results (\ref{CDD-ansatz}, \ref{identify-R}) lies
in its specification of the residua parameters, $R^{(D)}$ in terms of the parameters, $g_a, M_{1^-}$, characterizing a possible pole term in the generalized potential. By construction, the scattering amplitude, which results from (\ref{def-NoverD}-\ref{identify-R}), does not depend
on the choice of CDD-pole mass $M_{CDD}$.  

With this we can proceed to discuss the second case in which the $D^*$ meson is not stable against strong decays. In this case the state manifests itself in terms of a pole in $T_{ab}^J(s)$ on the 2nd Riemann sheet. By construction the generalized potential does not show this pole. Being void of right-hand cuts the generalized potential does not have that second Riemann sheet. 
Therefore the rewrite (\ref{CDD-ansatz}, \ref{identify-R}) is instrumental, since here 
the effective potential $U^{\rm eff}_{ab}(s)$ does not exhibit the $J^P =1 ^-$ pole either. So here  we can simply apply the results  (\ref{CDD-ansatz}, \ref{identify-R}). 
There is  subtle point, however, the parameter $M_{1^-}$ must still be real, being 
a quasi-particle approximation to the complex pole mass in this case. 

The system (\ref{CDD-ansatz}, \ref{identify-R}) can be solved numerically by matrix inversion techniques. Once we obtain a solution of $N_{ab}(s)$, we can compute $D_{ab}(s)$ via (\ref{def-D}), and  a well defined result for the partial-wave scattering amplitude is obtained with (\ref{def-NoverD}).

\clearpage

\section{From Lattice QCD data to low-energy constants}

Open-charm meson masses are available on various Lattice QCDs with specific unphysical quark masses
\cite{Aubin:2005ar,Follana:2007uv,Bazavov:2011aa,Mohler:2013rwa,Moir:2013ub,Lang:2014yfa,Bazavov:2014wgs,Kalinowski:2015bwa,Aoki:2016frl}. We recall the various available data sets here. Mohler and Woloshyn \cite{Mohler:2011ke} generated a data set based on the PACS-CS ensembles \cite{Aoki:2008sm}. The group of Marc Wagner analyzed a large set of ensembles from the European Twisted Mass Collaboration (ETMC) \cite{Kalinowski:2015bwa,Cichy:2016bci}. Further results are available from the HPQCD 
collaboration \cite{Na:2012iu} and from \cite{Liu:2012zya,Na:2012iu} based on \cite{WalkerLoud:2008bp} and  MILC Asqtad ensembles \cite{Orginos:1999cr,Orginos:1998ue,MILC2001,MILC2004}. The latest results are from the Hadron Spectrum Collaboration (HSC) \cite{Moir:2016srx,Cheung:2020mql,Gayer:2021xzv}.

Following \cite{Guo:2018kno,Guo:2021kdo} we consider lattice ensembles only where the pion and kaon masses are smaller than a critical value that reflects the expected convergence domain of the chiral extrapolation approach. In contrast to our previous 
study it is now possible to perform global fits that use a more stringent selection criterium for the definition
of our chisquare function, in which ensembles with pion and kaon masses smaller than 550 MeV are used only. That leaves the data set from ETMC, HPQCD and HSC only. The $D$ meson masses on the PACS-CS and MILC Asqtad ensembles are not considered in this work. They imply either a pion or kaon mass larger than 550 MeV.

\begin{table}[t]
\setlength{\tabcolsep}{2.5mm}
\renewcommand{\arraystretch}{1.15}
\begin{center}
\begin{tabular}{l|cccc|| c} 
                                                           &  Fit 1    &  Fit 2  &  Fit 3   &  Fit 4  &  global Fit   \\ \hline

$a^{\beta = 6.76}_{c,\rm HPQCD}\,   \hfill \mathrm{[fm]}$  &  0.1367   &  0.1359 &  0.1336   &  0.1367 &  0.1387(1) \\ 
  $\Delta^{\beta =6.76 }_{c,\rm HPQCD}$                    &  0.1500   &  0.1494 &  0.1184   &  0.1500 &  0.1640(2) \\

$a^{\beta =7.09 }_{c,\rm HPQCD}\,   \hfill \mathrm{[fm]}$  &  0.0953   &  0.0991 &  0.0970   &  0.0992 &  0.0979(0) \\ 
$\Delta^{\beta = 7.09}_{c,\rm HPQCD}$                      &  0.0936   &  0.1336 &  0.1049   &  0.1282 &  0.1160(4) \\  \hline 

$a^{\beta = 1.90 }_{c,\rm ETMC}\,   \hfill \mathrm{[fm]}$  &  0.1018   &  0.0996 &  0.1025  &  0.1027 &  0.1016(1) \\
$\Delta^{\beta = 1.90}_{c,\rm ETMC}$                       &  0.0983   &  0.0747 &  0.1041  &  0.1086 &  0.0978(7) \\ 

$a^{\beta = 1.95}_{\rm ETMC}\,   \hfill \mathrm{[fm]}$     &  0.0934   &  0.0925 &  0.0928  &  0.0943 &  0.0920(0) \\
$\Delta^{\beta =1.95 }_{c,\rm ETMC}$                       &  0.0908   &  0.0817 &  0.0817  &  0.1005 &  0.0781(1) \\ 

$a^{\beta = 2.10}_{\rm ETMC}\,   \hfill \mathrm{[fm]}$     &  0.0695   &  0.0704 &  0.0695  &  0.0699 &  0.0687(0) \\
$\Delta^{\beta = 2.10}_{c,\rm ETMC}$                       &  0.0629   &  0.0728 &  0.0608  &  0.0659 &  0.0554(1) \\ \hline 

$a_{\rm HSC}^{m_\pi \simeq\, 239 \, \rm{MeV}} \,   \hfill \mathrm{[fm]}$                    &  0.1211   &  0.1243 &  0.1242  &  0.1242 &  0.1164(0) \\ 
$\Delta_{c,\rm HSC}$                                       &  0.0050   &  0.0337 &  0.0328  &  0.0343 &  0.0081(1) \\

$a_{\rm HSC}^{m_\pi \simeq \,391\, \rm{MeV} }  \,   \hfill \mathrm{[fm]}$                    &  0.1211   &  0.1243 &  0.1242  &  0.1242 &  0.1244(1) \\ 
$\Delta_{c,\rm HSC}$                                       &  0.0050   &  0.0337 &  0.0328  &  0.0343 &  0.0101(4) \\ 

\end{tabular}

\caption{Results for Fit 1 - Fit 4 from \cite{Guo:2018kno} as compared to our 'global Fit'. The offset parameter $\Delta_c$ is introduced in (\ref{def-Deltac}). The set of lattice data fitted in our global Fit is described in the text. While we characterize the ensembles of HPQCD and ETMC by their $\beta $ values, it is more conventient to use the approximate pion masses for the considered HSC ensembles. They are based on 
an anisotropic framework with temporal $a_t \simeq a/3.5 $ and spatial lattice spacing $a$.}
\label{tab:lattice-scale-Fits}
\end{center}
\end{table}

For technical lattice issues like scale and charm-quark mass settings we refer to our previous work \cite{Guo:2018kno}. Since a $D$ meson mass, $M_D$, depends quite sensitively on the lattice scale, $a$, and the charm-quark mass, in our global Fit the parameter $\Delta_c$ is introduced,  
\begin{eqnarray}
a\,M_D \to a\, M_D + (1 + \epsilon_D )\,\Delta_c\,, \qquad {\rm with }\qquad \epsilon_D \simeq 0\,,
\label{def-Deltac}
\end{eqnarray}
which is supposed to fine-tune the choice of the charm quark mass. In  principle the value of 
$\epsilon_D$ depends on not only the type of $D$ meson but also the $\beta_{\rm QCD} $ value of the 
ensemble considered. If the lattice group provided a sufficient amount of data, we determine the 
parameters $\epsilon_D$, otherwise we use $\epsilon_D = 0$. 
We do not implement explicit discretization effects in our chiral extrapolation approach. Our unconventional scale setting procedure is set up to minimize uncertainties from discretization effects in the open-charm sector. Therefore, we use the empirical isospin averaged $D$ meson masses with $J^P = 0^-$ and $J^P = 1^-$ quantum numbers as an additional constraint in our analysis.
We perform fits at ad hoc values for the systematic error in the $D$ meson masses. 
They reflect a residual uncertainty from the chiral expansion and/or discretization effects. Once this error is sufficiently large the $\chi^2$ per data point should be close to or below one. We confirm our previous estimate of 5-10 MeV, where the values got systematically smaller with larger $\beta_{\rm QCD} $ values.

Assuming that the lattice data can be properly moved to the physical charm quark mass the low-energy constants are obtained by a global Fit to the Lattice QCD data set. 
A faithful reproduction of the 178 = 55 + 123 data points, with 55 charm meson masses and 123 scattering phase shifts, is achieved. 
In Tab. \ref{tab:lattice-scale-Fits} we show the lattice parameters as they result from our fit that considers not only the $D$ meson masses but also the s- and p-wave scattering phase shifts as will be explained in more detail below. An estimate of 1-sigma statistical uncertainties is provided systematically.   
For the readers convenience we recall the four fit scenarios of our previous study that were based on a significantly smaller data set. It is comforting to see values that are in the range of our previous results. We recall that the available data set on the $D$ meson masses is not able to determine a unique parameter set without additional constraints from scattering data. Additional data on the $D$ meson masses, in particular for the $J^P =1^-$ states combined with a larger sample of lattice volumes, may change that situation. In Tab. \ref{tab:lattice-scale} we illustrate the quality of our global Fit, with respect to the $D$ meson masses on the various lattice ensembles. 
Our current results are comparable with previous ones. The somewhat larger $\chi^2$ value on the HSC ensemble with $m_\pi \simeq 233$ MeV is a consequence of tension in the $D^*$ masses \cite{Guo:2021kdo}. We emphasize that here, in contrast to the HPQCD and ETMC ensembles, our result is heavily constrained also by s-wave and p-wave scattering data on two HSC ensembles. A  detailed presentation of the latter follows in the next Section.

\begin{table}[t]
\setlength{\tabcolsep}{2.5mm}
\renewcommand{\arraystretch}{1.18}
\begin{center}
\begin{tabular}{l|c|c} 

                                                      &  $\chi^2/N$ from global Fit  &  estimate of systematic error  \\ \hline

${\rm HPQCD}\,{(N=6)}$                                & 0.9961 &    10 {\rm MeV}   \\
$ \beta = 6.76$                                       & &        \\ 
                                                      
${\rm HPQCD}\,{(N=5)}$                                & 0.7412 &    10 {\rm MeV}   \\
 $ \beta = 7.09$                                      & &        \\ \hline
                                                      
${\rm ETMC}\,{(N=16)}$                                 & 0.7647 &    10 {\rm MeV}   \\ 
$ \beta = 1.90$                                       & &        \\ 
                                                      
${\rm ETMC}\,{(N=12)}$                                 & 0.7253 &    7.5 {\rm MeV}   \\ 
$ \beta = 1.95$                                       & &        \\ 
                                                    
${\rm ETMC}\,{(N=8)}$                                 & 0.8129 &    5 {\rm MeV}   \\ 
$ \beta = 2.10$                                       & & \\ \hline 
                                                      
${\rm HSC }\,{(N=4)}$                                 & 1.6372 &    10 {\rm MeV}   \\ 
$m_\pi \simeq 239$ MeV                                                      & & \\ 
                                                      
${\rm HSC }\,{(N=4)}$                                 & 1.0767 &    10 {\rm MeV}   \\ 
$m_\pi \simeq 391$ MeV                                & & \\   \hline  

\end{tabular}
\vskip0.2cm
\caption{ Quality with which our global Fit reproduced the $D$ meson masses on the various lattice ensembles. Data with $m_\pi < 550$ MeV and $m_K < 550 $ MeV are considered only. The somewhat adhoc estimate of the systematic error was obatined by the request that the chisquare per data point is about 1. }
\label{tab:lattice-scale}
\end{center}
\end{table}

To establish a set of LEC is a computational challenge. The $D, D_s$ and $D^*, D^*_s$ meson masses  are determined from  a set of coupled and nonlinear equations as detailed in \cite{Guo:2018kno}. Such a framework is implied by using on-shell masses in the loop contributions to hadron masses \cite{Semke:2005sn,Semke:2011ez,Lutz:2014oxa,Lutz:2018cqo,Guo:2018kno}. 
We apply the evolutionary algorithm of 
GENEVA \cite{Geneva} with runs of a population size 8000 on 500 parallel CPU cores. 
For any set four coupled nonlinear equations are to be solved on each lattice ensemble considered. This defines the input required for the coupled-channel computation. Then the scattering equations (\ref{def-non-linear}) in all channels, for which there are lattice data available, have to be solved numerically. For given pion and kaon masses we infer the quark masses from the one-loop mass formulae for the 
pseudo-Goldstone bosons to be used in our expressions for the $D$ meson masses. This involves the 
LEC combinations $L_4 - 2\,L_6, L_5 - 2\,L_8$ and $L_8 + 3\,L_7$, where one additional constraint is defined by the requirement to reproduce the empirical $\eta$ meson mass \cite{Gasser:1984gg}.

Our results coined as our 'global Fit' is based on the GPA recalled in this work in some detail. 
The GPA copes with left-hand cut structures as they are implied for instance by long range t- and u-channel exchange processes in a controlled manner. This is of particular importance for the description of p-wave scattering phase shifts in \cite{Moir:2016srx,Cheung:2020mql,Gayer:2021xzv}. Our global Fit treats the effects of chiral order $Q^3$ and $Q^4$ systematically as taking into account the contributions from one-loop bubble and tadpole diagrams to the generalized potential.  
The long-range part of the interaction is considered in terms of conformal expansions that describe the energy dependence of 
the generalized potential in a manner so that asymptotically it is given by a constant. The details of which are charcterized by a universal value of  
\begin{eqnarray}
\delta \Lambda_s = 0.4088(1)\, {\rm GeV }    \,,   
\end{eqnarray}                 
in (\ref{def-Lambda-max}) as determined from our global Fit. In a given sector at fixed isospin ($I$) and strangeness ($S$) its value quantifies up to which energies above the largest considered threshold our scattering phase shifts may be trusted.  In s-wave and p-wave channels we consider 4 and 2 terms respectively in the conformal expansion (\ref{def-U-xipanded}).

\begin{table}[t]
\setlength{\tabcolsep}{2.5mm}
\renewcommand{\arraystretch}{1.2}
\begin{center}
\begin{tabular}{l|rccc||r} 
                             &  Fit 1    &  Fit 2    &  Fit 3   & Fit 4   & global Fit   \\ \hline
$10^3\,(L_4 - 2\,L_6)\, $    & -0.1395   & -0.1112   & -0.1102  & -0.1575 & -0.2778(48) \\
$10^3\,(L_5 - 2\,L_8)\, $    &  0.0406   & -0.0940   & -0.0235  & -0.0370 &  0.0276(24) \\
$10^3\,(L_8 + 3\,L_7)\, $    & -0.5130   & -0.5127   & -0.4950  & -0.5207 & -0.5698(16)  \\
$\mu $ [GeV]                 &  0.77     & 0.77      & 0.77     & 0.77    & 0.8471 \\ \hline
$f$ [MeV]                    &  92.4     &  92.4     &  92.4    &  92.4   & 93.05(44) \\
$m_s/ m $                    &  26.55   &  26.19   &  26.60  &  26.60 & 27.20(9) 

\end{tabular}

\caption{The low-energy constants $L_n$ are at the renormalization scale, $\mu $, as specified in the table for the various fit scenarios. The values of $10^3\,L_4 = -0.7011(93)$ and $10^3\,L_5 = 0.9019(158)$  were
set as to recover the empirical values of the pion and kaon decay constants $f_\pi \simeq 92.1 $ MeV and $f_K \simeq 110$ MeV. }
\label{tab:Ln}
\end{center}
\end{table}

\begin{table}[t]
\setlength{\tabcolsep}{2.5mm}
\renewcommand{\arraystretch}{1.18}
\begin{center}
\begin{tabular}{l|rrrr|| r}
                                                     &  Fit 1     &  Fit 2    & Fit 3     & Fit 4   & global Fit     \\ \hline
                                            
$ M\;\;$ \hfill [GeV]                                &  1.8762    &  1.9382   &  1.9089  & 1.8846  &  1.8478(0)  \\
$\tilde M -M $\hfill [GeV]                                &  0.1873    &  0.1876   &  0.1834  & 0.1882  &  0.1415(3)  \\ \hline

$ c_0$                                               &  0.2270    &  0.3457   &  0.2957  &  0.3002 &  0.2152(3) \\
$ \tilde c_0$                                        &  0.2089    &  0.3080   &  0.2737  &  0.2790 &  0.3438(30) \\
$ c_1$                                               &  0.6703    &  0.9076   &  0.8765  &  0.8880 &  0.6619(19) \\
$ \tilde c_1$                                        &  0.6406    &  0.9473   &  0.8420  &  0.8583 &  0.7728(73) \\
$ c_2 = \tilde c_2$                                  & -0.5625    & -2.1893   & -1.6224  & -1.3046 & -0.6419(0) \\
$ c_3 = \tilde c_3$                                  &  1.1250    &  4.4956   &  3.2448  &  2.9394 &  2.4707(1) \\
$ c_4 = \tilde c_4$                                  &  0.3644    &  2.0012   &  1.2436  &  0.9122 &  1.0368(1) \\    
$ c_5 = \tilde c_5$                                  & -0.7287    & -4.1445   & -2.4873  & -2.1393 & -2.2743(21) \\    
$ c_6 = \tilde c_6$                                  &            &           &          &         & -0.6457(341) \\    \hline

$d^c_1\,\hfill\mathrm{[GeV^{-2}]}$                   & 1.8331    &  1.6937   &  1.6700   & 1.9425  & 1.1333(8)  \\
$\tilde d^c_1\,\hfill\mathrm{[GeV^{-2}]}$            & 1.6356    &  1.6586   &  1.4701   & 1.7426  & 1.3581(18)  \\
$d^c_2= \tilde d^c_2\,\hfill\mathrm{[GeV^{-2}]}$     & 1.0111    &  0.9954   &  0.8684   & 1.0032  & 0.8268(95)  \\
$d^c_3\,\hfill\mathrm{[GeV^{-2}]}$                   & 0.1556    &  0.0679   &  0.1531   & 0.1109  & 0.2669(39)  \\
$\tilde d^c_3\,\hfill\mathrm{[GeV^{-2}]}$            & 0.2571    &  0.1640   &  0.2597   & 0.2143  & 0.1495(39)  \\
$d^c_4= \tilde d^c_4\,\hfill\mathrm{[GeV^{-2}]}$     & 0.8072    &  1.6392   &  0.8607   & 1.1255  & 0.1407(48)  

\end{tabular}
\caption{The low-energy constants from a fit to the pseudoscalar and vector charmed-meson masses based on Lattice QCD ensembles of 
HPQCD, ETMC and HSC as described in the text. }
\label{tab:FitParametersA}
\end{center}
\end{table}

In Tab. \ref{tab:Ln} and  Tab. \ref{tab:FitParametersA} and Tab. \ref{tab:gn} we confront the LEC of our 'global Fit' with four previous phenomenological scenarios, that were based on distinct data sets as explained above. In the first table we collect the LEC that are required to specify the quark masses to be used in our  computation of the charmed meson masses and the various phase shifts at N$^3$LO. 
While in the phenomenological results an ad hoc value for $f$ was used, in the global fit, instead, it was determined by a global adjustment to the lattice data set. All LEC come together with an estimate of their uncertainties. We find that the statistical uncertainty in such LEC are quite small. In particular our prediction of the quark-mass ratio $m_s/m$ suffers from a rather minor statistical error only. Our value is compatible with the current FLAG report value 27.42(12) \cite{Aoki:2021kgd}.

\begin{table}[t]
\begin{tabular}{l|ccll||r} 

              & \; Fit 1 \phantom{xx}   &  Fit 2  \phantom{xx}  &  Fit 3 \phantom{xx}  &  Fit 4 \phantom{xx}  &  \;global Fit \!\!   \\ \hline
                                                      
$g_1$ \,      &  0                    &  0                    &  0.2240              &  0.2338 & \, 0.1248(36)\\
$g_2$         &  0                    &  0                    &  0.5405              &  0.4663 & -0.1506(1)  \\
$g_3$         &  0                    &  0                    &  0.0399              &  0.0299 &   -0.1569(10) \\  
$g_4$         &  0                    &  0                    &  0                   &  0 &  0.0132(104)  \\
$g_5$         &  0                    &  0                    &  0                   &  0 &  0.0786(15)  

\end{tabular}
\caption{The LEC at chiral order $Q^3$ with $\tilde g_n = g_n$ as considered in this work. }
\label{tab:gn}
\end{table}

In Tab. \ref{tab:FitParametersA} the LEC that are needed in the computation of the charm meson masses at N$^3$LO are presented. Our global Fit results are within range of the previous studies, again with rather insignificant statistical uncertainties. 
In all fit scenarios the four low-energy constants $c_{0,1}$ and $\tilde c_{0,1}$ are adjusted to recover the isospin averaged physical $D$ and $D_s$ meson masses with $J^P =0^-$ and $J^P = 1^-$ quantum numbers from the PDG \cite{PDG}. This suggets deviations from the leading order large-$N_c$ relation, $c_1 =2\,c_0$,  and the heavy-quark symmetry sum rules, $\tilde c_n = c_n$ for $n < 2$. Given the current data situation, we deem any attempt to relax our conditions $\tilde c_n = c_n$ for $n > 1$ unreasonable. Here further accurate lattice data, in particular on scattering phase shifts involving 
the $1 ^-$ states would help. While Fit 1 and Fit 3 impose the leading order large-$N_c$ relations (\ref{large-Nc}) the remaining scenarios keep those parameters unrelated. We do not impose the 
heavy quark-symmetry relations $d_n =\tilde d_n $ for all $n= 1,..., 4$. As was pointed out in  
\cite{Guo:2018kno}, our minimal ansatz $\tilde c_{0,1} \neq c_{0,1}$ requires unavoidably also 
$d_n \neq \tilde d_n$. Only in this case the renormalization-scale invariance of the $D$ meson masses 
can be assured in our framework. It is useful to consider suitable linear combinations of the low-energy constants
\begin{eqnarray}
&& d^{c}_1 = - \frac{1}{23} (26 \,d_1 + 9 \,d_2)\,, \qquad \qquad \qquad
d^{c}_3 = \frac{1}{345}\,\Big( 43 \,d_1 + 60 \,d_2 + 69 \,d_3\Big)\,,
\nonumber\\
&& d^{c}_2 = \frac{1}{276}\Big(-132 \,d_1 + 18 \,d_2\Big) \,,\qquad \quad  \;\;
d^{c}_4   = \frac{1}{45}\,\Big(-11 \,d_1 + 15 \,d_2 - 33 \,d_3 + 45 \,d_4\Big) \,.
\label{def-dc}
\end{eqnarray}
Scale invariant expressions request $d^c_1 \neq \tilde d^c_1$ and $d^c_3 \neq \tilde d^c_3$ but permit 
the assumptions $d^c_2 =\tilde d^c_2$ and $d^c_4 = \tilde d^c_4$. 

Like in our previous work we find significant tension of our LEC with those obtained in \cite{Guo:2008gp,Liu:2009uz,Guo:2011dd,Altenbuchinger:2013vwa}. 
The parameters of Fit 2 are reasonably close to the two sets claimed in \cite{Liu:2009uz} with the notable exception of $c_1$ which differs by about a factor 2. 
Despite the considerable variations in the low-energy constants all parameter sets are acceptable from the perspective of describing the $D$ meson masses. We emphasize that while Fit 1-4 did consider the s-wave scattering lengths of \cite{Liu:2012zya} and achieved a reasonable reproduction, this is not the case for our global fit, in which we rejected that data set. In our recent work \cite{Guo:2021kdo} we pointed out significant tension with the more recent results by HSC \cite{Cheung:2020mql,Gayer:2021xzv}.  
Therefore such scattering lengths are not included in our current chisquare function.

A discriminative constraint among the phenomenological Fit 1-4 is provided by the $D\,\pi$ and $D\,\eta$ phase shifts on a HSC ensemble \cite{Moir:2016srx}.  Here Fit 3 and Fit 4 are much superior, with Fit 1 and Fit 2 being at odds, in particular, with the $D\,\eta$ phase shift as suggested in \cite{Moir:2016srx}. 
The scattering lengths and phase shifts are computed in the infinite volume limit based on the LEC of  Tab. \ref{tab:FitParametersA}. In Fit 1-4 the coupled-channel framework 
established in \cite{Kolomeitsev:2003ac,Hofmann:2003je,Lutz:2007sk} was applied. It relies on the on-shell reduction scheme developed in \cite{Lutz:2001yb} which can be justified 
if the interaction is of short-range nature or the long-range part is negligibly small  \cite{Lutz:2011xc,Lutz:2015lca}.  An alternative chain of works based on a somewhat different short-range 
treatment of the coupled-channel effects is \cite{Guo:2011dd,Guo:2008gp,Liu:2012zya,Altenbuchinger:2013vwa,Cleven:2014oka,Du:2016tgp}.  Our previous work considered the one-loop contributions in a rather phenomenological manner, in which the LEC $g_1, g_2$ and $g_3$ were assumed to carry the integral strength of such contributions. This necessarily limits the extrapolation power of our previous results. In this respect a direct comparison of the $g_n$ in Tab. \ref{tab:gn} from our global Fit to our previous phenomenological fits is not justified. This is reflected in the table.

\begin{table}[t]
\setlength{\tabcolsep}{3.5mm}
\renewcommand{\arraystretch}{1.8}
\begin{tabular}{l|r||l|r||l|r} 
$\mu  $ [GeV]        &   0.8471(58)   &                                                       
$d_5 \;  \hfill\mathrm{[GeV^{-2}]}$      &  0.0400(333)   &
$d^{\,t}_3    \hfill\mathrm{[GeV^{-2}]}$       & -0.7646(8)  \\  \hline
$d^{\,t}_4\; \hfill\mathrm{[GeV^{-2}]}$     &   1.7321(134)  &
$d^{su}_3\; \hfill\mathrm{[GeV^{-2}]}$     &   1.8679(192) &
$d^{su}_{4}\; \hfill\mathrm{[GeV^{-2}]}$     &  -0.7961(118)   \\ \hline
$d_{21} \;\hfill\mathrm{[GeV^{-2}]}$     &  1.9973(18)   &
$d_{23} \hfill\mathrm{[GeV^{-2}]}$       &  0.6828(256)  &
$d_{25} \hfill\mathrm{[GeV^{-2}]}$       &  0.0738(321)

\end{tabular}
\caption{The LEC at chiral order $Q^4$ as considered in this work with $\tilde d_n = d_n$ for $n> 4$. We insist on the large-$N_c$ relations $d_{6-8}=0$ and $d_{11-14}=0$ and  $d_{17-20}=0$ and 
$2\,d_{22}=-d_{21}$ and $2\,d_{24}=-d_{23}$ and   $2\,d_{26}=-d_{25}$ at the natural renormalization scale $ \mu = 0.847(6)$ GeV, as determined in our global Fit.  }
\label{tab:dn}
\end{table}

In Tab. \ref{tab:dn} we provide our values for the LEC that are contributing at chiral order $Q^4$ to the scattering phase shifts but not to the charmed meson masses. We find naturally sized values thereof throughout the list of our LEC. 
A priori we encounter 22 such LEC with $d_{5-26}$. In application of leading-order large-$N_c$ sum rules the number of independent LEC is reduced to the 8 LEC of Tab. \ref{tab:dn}. Here it is important to recall that most of the LEC depend on the particular choice of the renormalization scale $\mu$ as was summarized 
in (\ref{d-running-symmetric}-\ref{def-d-inv}). The implications of large-$N_c$ sum rules can be imposed at a given natural choice of $\mu$ only. This is not at odds with our result that the combination of the one-loop contributions with the set of $d_n$ is renormaliaztion-scale independent strictly.  By virtue of the large-$N_c$ sum rules a preferred choice of $\mu$ is mandated. Indeed, it may be used as a parameter to 
optimize the role of the leading order large-$N_c$ sum rules. In this context we arive 
at the value $\mu =  0.8471(53) $ GeV in Tab.  \ref{tab:dn} with an estimate of its one-sigma uncertainty.

Our preferred fit scenario has a $\chi^2/(N= 55) \simeq 0.80$ for the charm meson masses and a $\chi^2/(N = 123) \simeq 1.44$ for the scattering data.  It is noteworthy that our value of the leading order parameter, $f \simeq 93.1 $ MeV, comes out significantly larger than one may have expected from the FLAG report \cite{Aoki:2021kgd}. In the current setup it does not appear possible to find a global Fit with a value for $f$ that is much smaller. After all the one-sigma error estimate for its value is quite small emphasizing the enormous sensitivity of our global Fit on that parameter. 
We observe quite some tension in the data set, which will be illustrated in the next Section in more depth. Removing 3 particular scattering data points for our given set of LEC, the scattering chisquare is significantly reduced with $\chi^2/(N = 120) \simeq 1.17$. This would imply a total chisquare $\chi^2/(N= 55 +120) \simeq 1.06$. While we 
could further improve the quality of our fit by removing the outlier points in the fit process or by invoking further subleading order terms in the $1/N_c$ expansion we refrain from doing so while the fate of our suspected outlier points is not settled. Further improved lattice data would be highly welcome here. 

\begin{table}[t]
\setlength{\tabcolsep}{2.5mm}
\renewcommand{\arraystretch}{1.5}
\begin{center}
\hspace*{+0.25cm}
\begin{tabular}{c|c|c|c}
\hline
            &$(I,S)= (1,1)$      &$(I,S)= (1/2,0)$    &$(I,S)= (0,-1)$     \\ \hline
Fit 1       & $2.542_{-16}^{+15}-0.114_{-9}^{+19}i$  &   $2.471_{-7}^{+8}-0.046_{-3}^{+7}i$   &   $2.360_{-0}^{-1}-0.143_{-14}^{+17}i$ \\ 
Fit 2       & $2.450_{-9}^{+8}-0.297_{-8}^{+10}i$    &   $2.460_{-11}^{+17}-0.152_{+2}^{-5}i$ &   $2.287_{+2}^{-4}-0.124_{-12}^{+14}i$ \\ 
Fit 3       & $2.389_{-9}^{+6}-0.336_{-6}^{+11}i$    &   $2.463_{-27}^{+37}-0.106_{+6}^{-8}i$ &   $2.230_{+3}^{-4}-0.121_{-11}^{+13}i$ \\ 
Fit 4       & $2.382_{-10}^{+10}-0.322_{-10}^{+12}i$ &   $2.439_{-32}^{+42}-0.092_{+3}^{-7}i$ &   $2.229_{+3}^{-4}-0.083_{-11}^{+13}i$ \\ \hline  \hline
global \;Fit & 2.250 - 0.073\,$i$ & 2.379 - 0.025\,$i$ & 2.325 - 0.023\,$i$\\
\end{tabular}
\caption{Pole masses of the $0^+$  meson resonances in the flavour sextet channels, in units of GeV. The $(1,1), (1/2,0), (0,-1)$  poles are located on the $(-,+), (-,-,+), (-)$ sheets respectively in the notation also used in  \cite{Guo:2018gyd}. }
\label{tab:pole}
\end{center}
\end{table}

Before closing this section, a brief discussion of resonance pole masses with $J^P= 0^+ $ as they are implied by our global Fit is given.  Like in our error estimate of the LEC the statistical uncertainties in the pole masses are of rather minor importance with less than 1 MeV. 
A comparison with known results from the PDG or our pole masses from  previous Fit 1-4 is more useful. 
Our  mass for the $D^*_{s0}(2317)$ is larger by about 15 MeV only, as compared to its PDG value, despite the fact that 
it  was not included in our global fit. One may take this 15 MeV as a rough estimate of our systematic uncertainty in our predicted pole masses.  In Tab. \ref{tab:pole} we list the pole masses in the various sectors that constitute the flavor sextet. While there are important quantitative changes in the pole positions the qualitative pattern is quite similar in our previous Fit 1-4 and our current global Fit. The most striking difference is the novel consideration of the left-hand cut contributions in our global Fit as described above. That leads to the additional feature that a pole in the complex plane is characterized by a phase factor, the size of which reflects the importance of the left-hand cut contributions. Indeed, in the $(1,1)$ sector the pole comes with the phases $-21.05 ^{\circ} $ and $85.45^{\circ} $ in the two contributing channels respectively. In the $(0,-1)$ sector the phase is $-58.92^\circ$ and in the $(1/2,0)$ sector the three phases are $-43.43^\circ, -96.55^\circ $ and $87.03^\circ$.
We note that in all sectors there are additional poles on more distant Rieman sheets. Most strikingly, we identify a pole in the $(I,S)= (1/2,0)$  channel on its $(-,-,-)$ sheet at $(2.236 - 0.094\,i )$ GeV, which we would identify as a member of a flavor antitriplet. We did not find an antitriplet pole on the $(-,-,+)$ sheet. 
In the  $(I,S)= (1,1)$  sector we find two poles on the  $ (-,-)$ sheet with $(2.316 - 0.021 \,i)$ GeV and $(2.264 - 0.050 \,i) $ GeV. While the two states have comparable pole masses, their coupling constants to the two considered channels differ strongly. The first one couples to both states with similar strength, the second dominantly to the second channel. Such poles are characterized by sizeable phase factors. The first pole comes with phases of  about  $24.42^{\circ}$ and $90.51^{\circ}$, the second with $1.52^{\circ}$ and  $138.12^{\circ} $ in the two channels respectively.

A few final remarks on our fit strategy may be useful. Our attempts to establish a data description at order $Q^3$ suffered mainly from unphysical p-wave amplitudes, that we had to reject. We take this as a hint of either a shortcoming of our current setup or the possibility that the available lattice data set is not yet sufficiently consistent for a global fit. Our one-loop computation of the generalized potential has not yet triangle and box digrams included that are shown in Fig. \ref{fig:4}. While all such diagrams vanish in the limit $g_P = \tilde g_P\to 0$ it is unclear how important they will be in a global fit. Though in \cite{Yao:2015qia} it was claimed that such diagrams are not important for s-wave scattering lengths, this may not be the case for p-wave phase shifts. In a future work such missing diagrams will be considered in depth. Here we chose to consider effects of order $Q^4$, which bring in a larger flexibility by means of further so far unknown LEC. The latter may in part compensate for the neglect of such triangle and box diagrams. 

Moreover, we also failed to establish a global data description that includes the s- and p-wave phase shifts in application of our previous short-range framework  \cite{Kolomeitsev:2003ac,Hofmann:2003je,Lutz:2007sk,Guo:2018kno}. The chisquare we obtain in our GPA is better typically by one order of magnitude. It turned out quite impossible to obtain a simultaneous fit of s- and p-wave phase shifts, while keeping the constraints from the charmed meson masses in various Lattice QCD boxes.

\section{Scattering phase shifts from Lattice QCD data}

We turn to the scattering phase-shift results from HSC on two of their ensembles, one with a pion mass of about 233 MeV and the other with 383 MeV \cite{Moir:2016srx,Cheung:2020mql,Gayer:2021xzv}. Since we used here a different scale setting scheme, the quoted pion masses differ slightly from the nominal values as given by HSC.  
Such data provide a unique opportunity to scrutinize a given coupled-channel framework to see whether it is able to cope with the flavor SU(3) symmetry from the chiral Lagrangian as broken by finite chiral quark masses. Here it is instrumental that results are generated on a large selection of different isospin (I) and strangeness (S) channels. 

Our global Fit considers all $(I,S)$ channels for which HSC offers results, where we focus on s-wave and p-wave scattering phase shifts. The latter poses a particular challenge, since the LEC that drive such phases are also responsible for finite box effects in the open-charm meson masses. Therefore, only a simultaneous fit of the $D$  meson masses and the available p-wave scattering phase shifts is  significant. In a finite Lattice QCD box, phase shifts are measured by means of the change of the finite-box spectrum as the box size is changed. For sufficiently large boxes the scattering phase shifts can then be extracted. Ultimately, in order to fully control systematic uncertainties, it may be necessary to compute the chiral coupled-channel framework in the same box and compare the energy levels directly. In our explorative work, we follow a more pragmatic approach, in which we take the scattering phase shifts from HSC as already extracted from their finite box spectra.

\begin{figure}[t]
\center{
\includegraphics[keepaspectratio,width=1.0\textwidth]{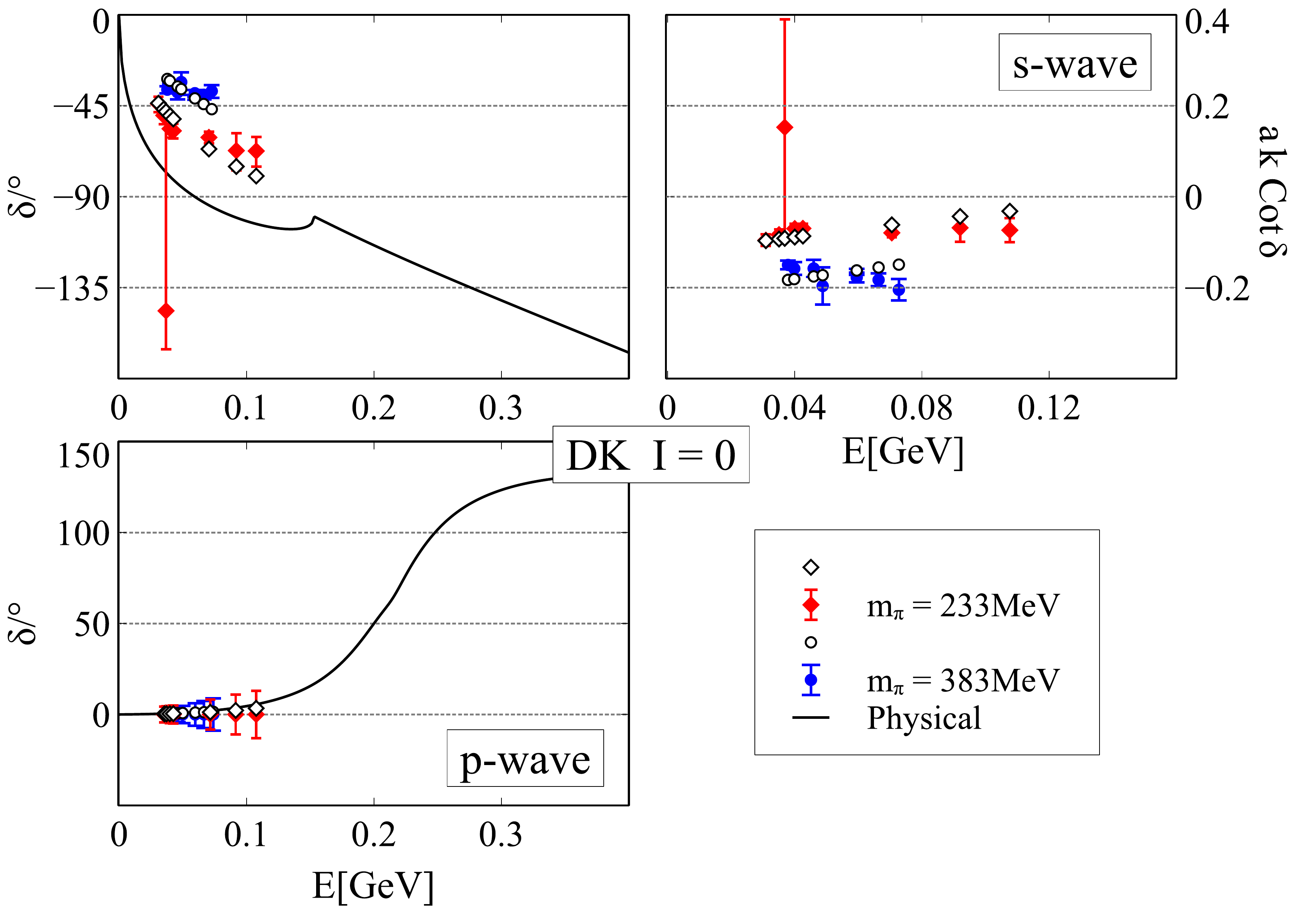} }
\vskip-1.4cm
\caption{\label{fig:ch2} S-wave and p-wave $D K$ phase shifts with $I=0$ and $S=1$ quantum numbers. The red and blue data points are from the ensemble with pion masses of about 233 MeV and 383 MeV respectively. While the solid lines are our predictions at the physical point the open symbols show our global Fit results on the two ensembles.  }
\end{figure}

\begin{figure}[t]
\center{
\includegraphics[keepaspectratio,width=1.0\textwidth]{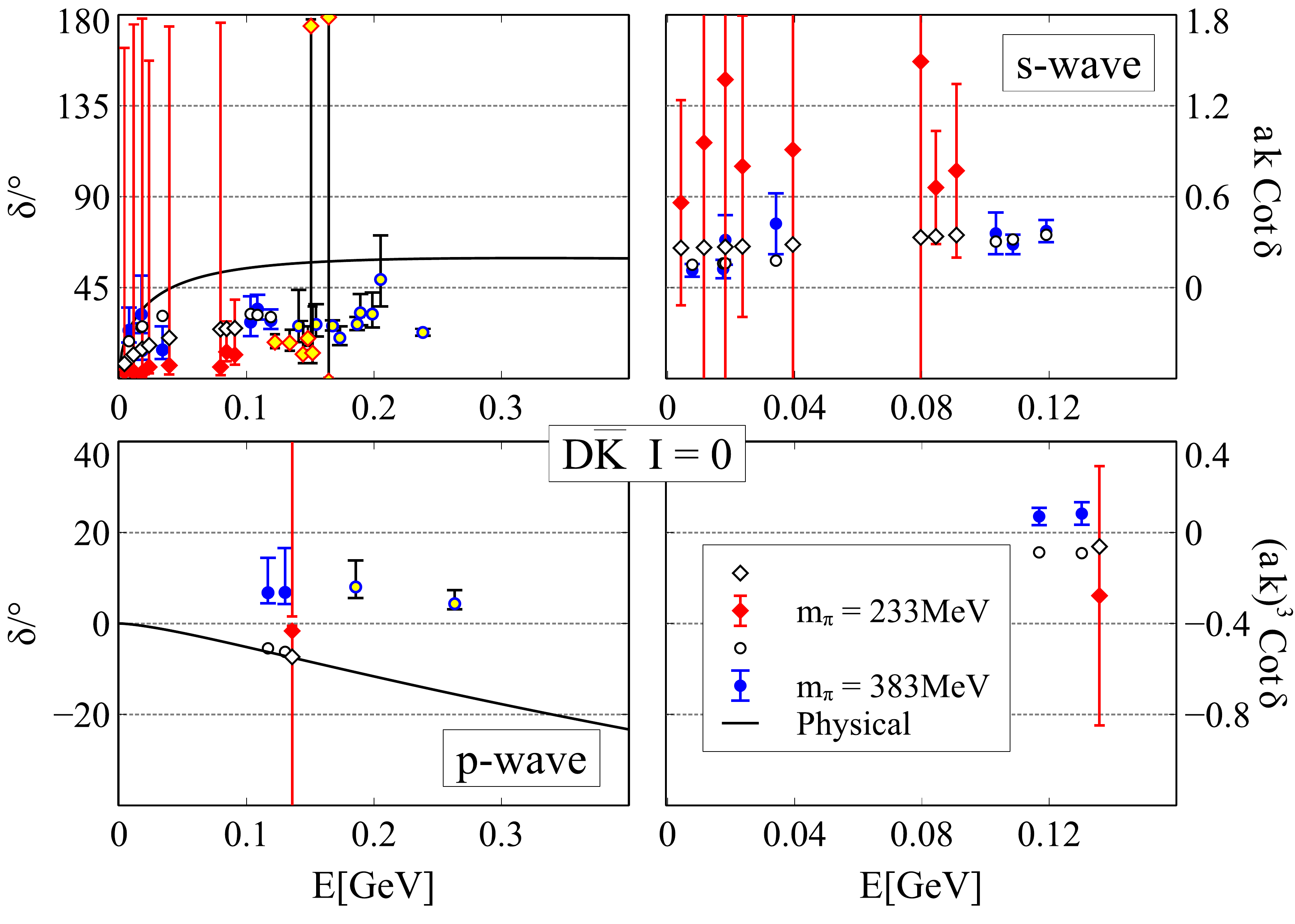} }
\vskip-1.4cm
\caption{\label{fig:ch6} S-wave and p-wave $D \bar K $ phase shifts with $I=0$ and $S=-1$ quantum numbers. While 
the solid lines are our predictions at the physical point the open symbols show our global Fit results. QCD Lattice values are presented by colored symbols, where yellow is used  to mark points that should not be considered. }
\end{figure}

\begin{figure}[t]
\center{
\includegraphics[keepaspectratio,width=1.0\textwidth]{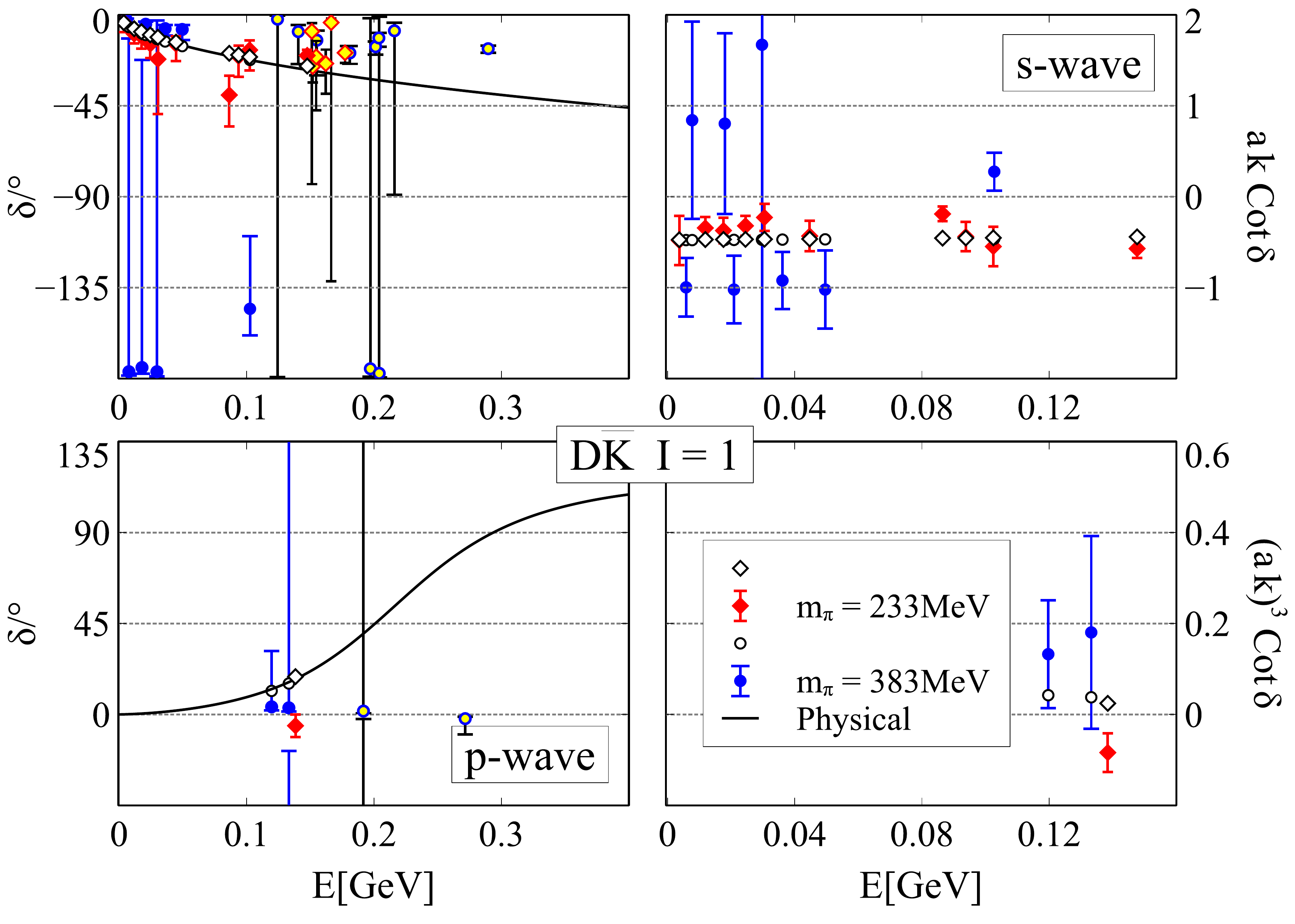} }
\vskip-1.4cm
\caption{\label{fig:ch7} S-wave and p-wave $D \bar K $ phase shifts with $I=1$ and $S=-1$ quantum numbers. Symbols and lines as in Fig. \ref{fig:ch6}.  }
\end{figure}

In all figures of this Section such lattice data are in red and blue colors, where we use red for the ensemble with about 233 MeV pion mass and blue for the one with about 383 MeV pion mass throughout this work. 
Data points which we are reluctant to consider in our global Fit are marked in yellow. 
Our extrapolation to the physical point is provided by the black solid line always. We propagated the statistical uncertainty in the LEC to all such phase shifts. While this is quite a computational challenge, in none of our results we obtained an uncertainty in the phase that was larger than 1 degree, smaller than the width of our phase shift lines. This is a consequence of the large number of data points considered together with the significant nonlinearities of the coupled-channel system considered.  
The energy in all our figures is measured from the threshold value, where we note that the latter does depend on the chosen quark masses. Wherever possible, on the right-hand panel the results of $(a\, k)^{2\,L+ 1} \,\cot \delta$  with $L= 0$ for s-wave and $L=1$ for p-wave channels are shown. Our global Fit is performed using the data set on such right-hand panels, if available. The reason for this lies in the expectation that for the latter a Gaussian-like distribution of the error size is expected. This is so since the latter is quite directly related to the energy levels in the finite Lattice QCD box. Indeed we take such error sizes and translate those into asymmetric error bars in the corresponding phase shifts as shown in the left-hand panels. This is important since a fit to data with asymmetric error bars is ill defined. The asymmetries found in some channels are sizeable.

Consider Fig. \ref{fig:ch2} in which the $D K$ phase shifts with $I= 0$ are shown on the left-hand panels. The s-wave data from HSC are reasonably well recovered. In this case the asymmetry in the phase shift errors is small. We checked that the corresponding subthreshold amplitude shows a pole 
at about 2.33 GeV quite compatible with the empirical mass value of the $J^P= 0^+$ state. In our current setup we assume perfect isospin  symmetry and therefore the state has zero width. Unlike for the s-wave data, unfortunately, HSC does not provide p-wave data on $ \cot \delta $ on either of the two ensembles. A systematic error estimate is provided only for the absolute value of the p-wave phase. This translates into an upper limit on the size of the phase shift as shown in the lower panel of Fig.  \ref{fig:ch2}. 
Nevertheless, we emphasize that this piece of information is an important part of our global chisquare  function. Our global Fit suggests a p-wave resonance in this channel simply because   
the phase shift crosses 90 degrees at  $M_R=2.600$ GeV.  It is interesting to observe that   
in the current PDG \cite{Workman:2022ynf} there is the state $D^*_{s1}(2700)$  with identical quantum numbers. For a sufficently narrow resonance its width would be characterized by the derivative of the phase at $M_R$ with  $\Gamma_R = 2/\delta'(E_R) $. At a more quantitative level the location of the resonance pole in the complex plane is required. Here we provide this rough Breit-Wigner width of $\Gamma_R =115$ MeV for simplicity only, since the presence of left-hand cut lines in the generalized potential reqires a more tedious pole search algorithm. For the s-wave pole masses 
such results are provided in Tab. \ref{tab:pole}.

We proceed with Fig. \ref{fig:ch6} and Fig. \ref{fig:ch7} which show the s- and p-wave phase shifts
for $D \bar K$ in both isospin channels. Here sizeable asymmetries in the errors of the phase shifts 
are seen, that are important for our global Fit. The lattice data show a rather wild distribution of 
points that make it difficult to reach a convincing chisquare value in these channels. 
In our global Fit we consider only such levels that are below the nominal $D^*\bar K$ threshold, despite the fact that additional levels were generated and analyzed in these channels by HSC. The latter are shown in yellow symbols on the left-hand panels with their asymmetric errors as derived from their corresponding $\cot \delta$ representation.
On the right-hand panel we show data points only that were included in our global Fit. 
Though the analysis of the p-wave phases is based on levels which are not affected by the s-wave phase shifts, the presence of the open $D^* \bar K$ brings in possibly an uncontrolled uncertainty in their determination. A two-channel analysis would be required. Concerning the s-wave the following observation may be useful. With a few exceptions such levels beyond the  $D^* \bar K$ threshold couple the s-wave and p-wave channels and should therefore be trusted only if the data set is sufficiently rich such that a simultaneous extraction of s- and p-wave phase shifts is possible.
Moreover we identify some outlier candidates in the p-wave data points. 
In Fig. \ref{fig:ch6} the two blue p-wave points provide a chisquare contribution of about 30 units. Most striking is the single red p-wave point in Fig. \ref{fig:ch7}, that defines a chisquare contribution of about 6 units. We note that there is no evident resonance 
signal from a flavor sextet in the s-wave isospin-zero phase shift, which stays well below 90 degrees. Most striking we find our result of an attractive p-wave isospin-one phase shift, with a rather broad resonance state around 2.656 GeV with width of 354 MeV. For simplicity we again focus on the phase shift where it passes through 90 degrees, with its rough implications on such resonance properties. In the flavor limit such a state would be a member of a 15-plet, 
that cannot be explained in a conventional quark-model picture. 

\begin{figure}[t]
\center{
\includegraphics[keepaspectratio,width=0.6\textwidth]{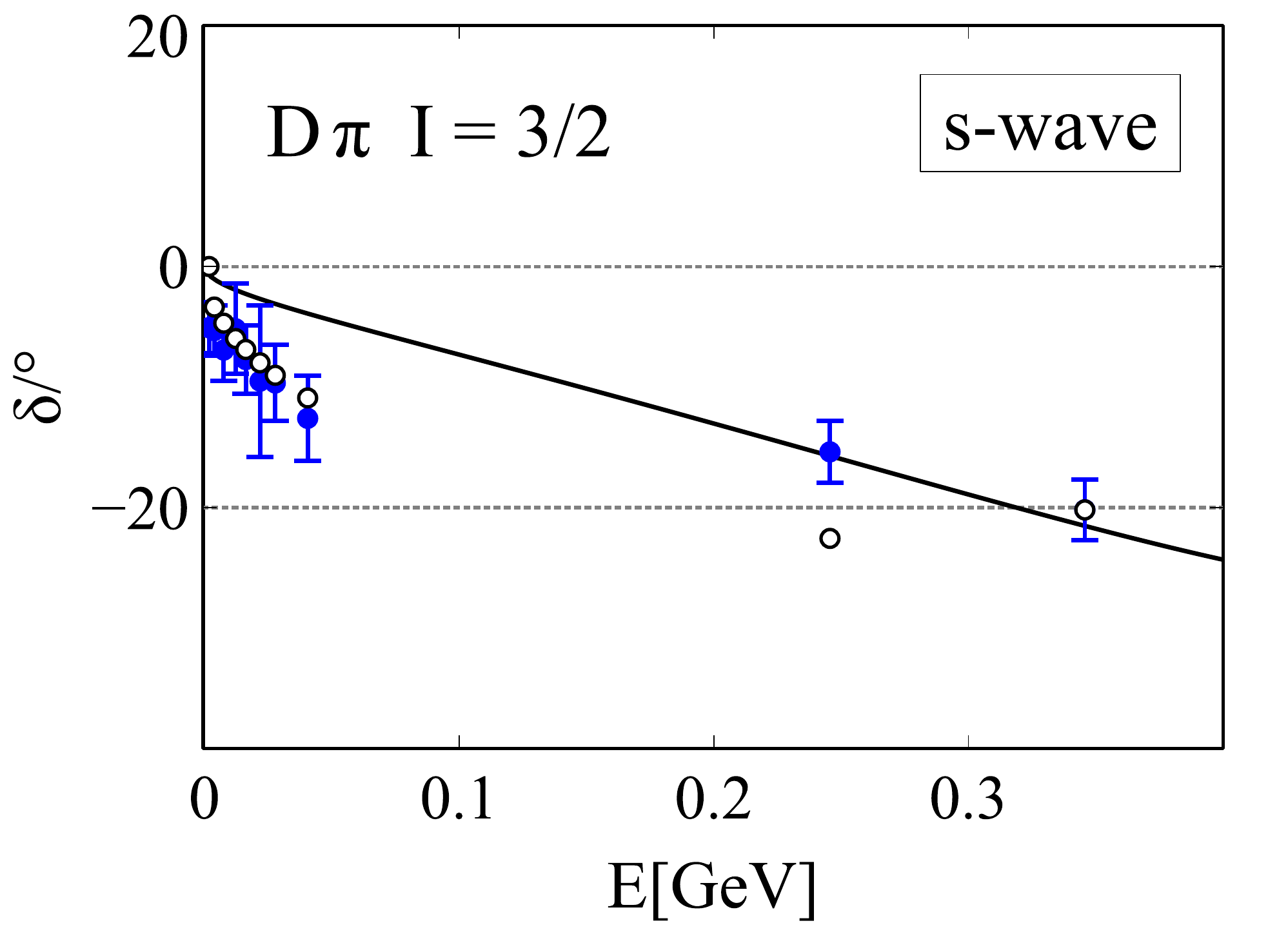} }
\vskip-0.4cm
\caption{\label{fig:ch5} S-wave $D \pi$ phase shift with isospin $I=3/2$. The blue data points are from the ensemble with pion masses of about 389 MeV. While 
the solid line is our prediction at the physical point the open symbols show our global Fit results. QCD Lattice values are presented by blue symbols. }
\end{figure}

\begin{figure}[t]
\center{
\includegraphics[keepaspectratio,width=1.0\textwidth]{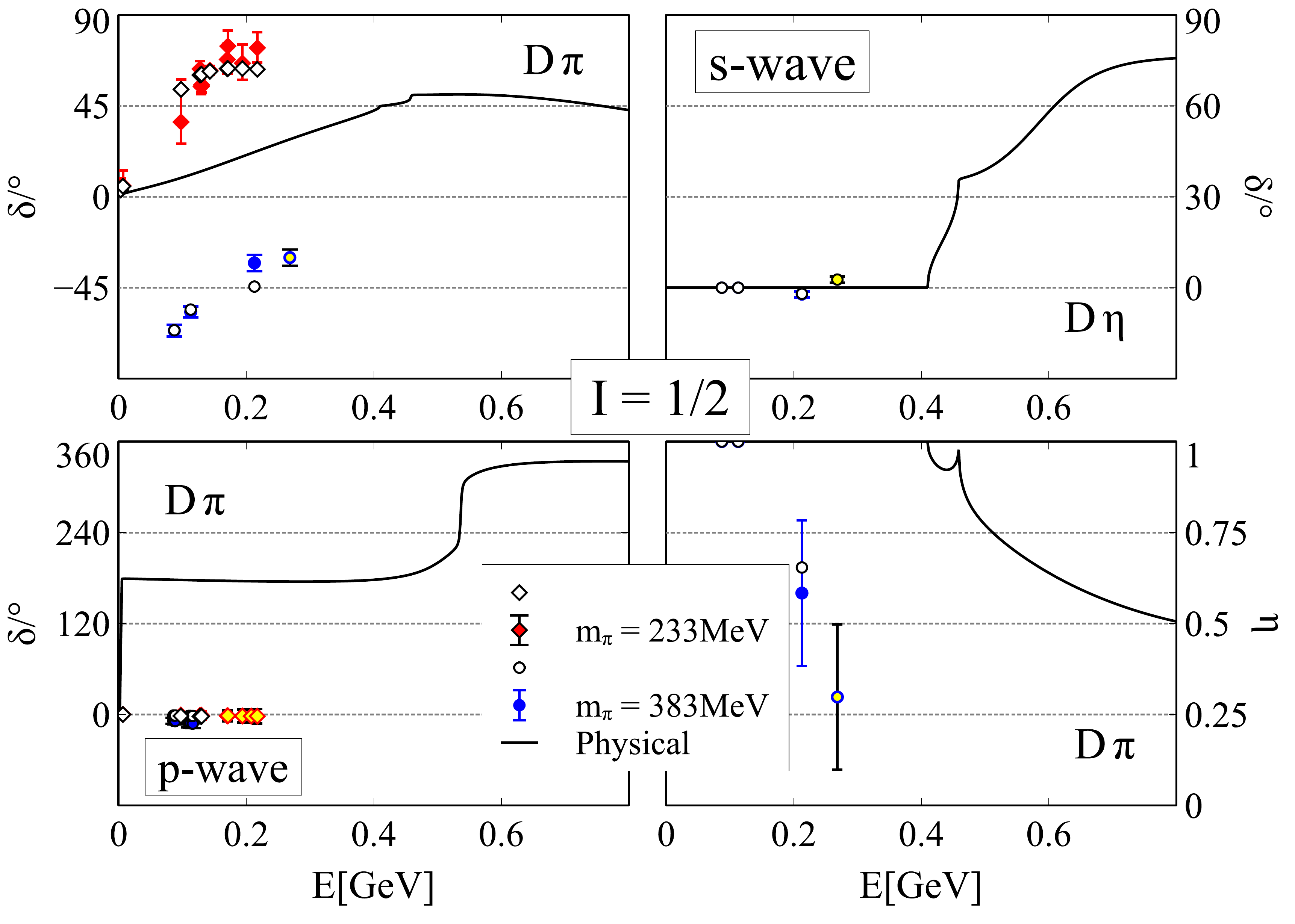} }
\vskip-1.2cm
\caption{\label{fig:ch4} S-wave and p-wave $D\, \pi$ phase shifts with $I=1/2$ quantum numbers. The r.h.p.  shows the s-wave $D\, \eta$ phase shift and inelasticity parameter. While 
the solid lines are our predictions at the physical point the open symbols show our global Fit results. QCD Lattice values are presented by colored symbols, where yellow is used  to mark points that should not be considered. }
\end{figure}

We turn to $D\, \pi$ scattering in the two possible isospin channels. In  Fig. \ref{fig:ch5}
the  s-wave $I=3/2$ phase shift is shown on the heavy pion mass ensemble of HSC. Neither p-wave data nor data on the light pion-mass ensemble is available so far. A fair reproduction of the available
data set is seen. 
In Fig. \ref{fig:ch4} our s- and p-wave results are confronted with data points on the two HSC ensembles in the $I = 1/2$ channel. A striking quark-mass dependence of the s-wave phase shifts is predicted. This confirms the findings of our previous phenomenological approach \cite{Guo:2018kno,Guo:2021kdo}, in which a stunning  extrapolation from the blue data points to the red data points was achieved. Like in our previous works, also our current result predicts still a sizeable step from the red data points to the physical phase shift as shown by the black solid line. However, with our global Fit we predict that the physical phase shift does not cross 90 degrees and no prominent signal of a flavor sextet state is seen in the  s-wave $ D \,\pi $ phase shift. 
The behavior of the p-wave phase is dominated by the fact that at the physical point the $D^*$ meson may decay into its $D \,\pi$ channel, which implies the steep rise of the solid black line. On the two HSC ensembles the decay channel is closed due to the unphysically large pion masses. In this 
case the s-channel $J^P = 1^-$ meson exchange process manisfests itself as a pole on the real axis in the partial-wave scattering amplitude below the $D \,\pi$ threshold. 
We conclude that an accurate direct evaluation of the s- and p-wave $ D\,\pi$  phase shifts would require Lattice QCD ensembles with pion masses smaller than the $233$ MeV so far used by HSC.

We emphasize that our solid line for the p-wave phase shift has to be taken with reservation as it shows a partial computation so far only. In the p-wave channel the $D \,\pi$ and $D^* \pi$ channels mix leading to a 6 channel system all together. This mixing is taken into account in our global Fit systematically. At the physical point the p-wave $D^* \pi$ amplitude picks up an anomalous threshold with its associated anomalous left- and right-hand cut lines. In our current computation such cut lines are dropped for simplicity. It is important to note that our global Fit is not affected by anomalous threshold effects due to the sufficiently large chosen pion masses on the two HSC ensembles. While recently in \cite{Lutz:2018kaz} two of the authors developed a novel framework how to deal with such contributions, the details of such a computation will be documented in a forthcoming work. 
At this stage we may draw a qualitative conclusion nevertheless. The p-wave resonance state seen in the phase shift at about  2.54 GeV would be the flavor antitriplet partner of the $D^*_{s1}(2700)$ state discussed above. So far 
the PDG \cite{Workman:2022ynf} does not claim such a p-wave state. Further studies would be useful to unravel the intricate dynamics of this channel.

\section{Summary and conclusions}

In this work we presented a comprehensive study of open-charm meson systems using lattice QCD data sets from ETMC, HPQCD and HSC. A large class of LEC from the chiral Lagrangian with three light flavors were determined by a global Fit to about 180 data points that include not only charm meson masses but also s- and p-wave scattering data on two ensembles generated by HSC. Despite the larger number of LEC considered, the statistical uncertainties in the LEC of such a global approach lead to scattering phase shifts at the physical point with a statistical uncertainty of less than one degree.   

The charm meson masses were computed at N$^3$LO in finite boxes as set up by corresponding lattice ensembles. 
Our scattering amplitudes were constructed using the generalized potential approach (GPA), in which the potential was constructed at the one-loop level and then properly extrapolated to higher energies by means of conformal variables. In this manner reliable results can be derived that go beyond the applicability domain of conventional chiral perturbation theory ($\chi $PT). The scattering amplitudes are derived in terms of numerical solutions of coupled-channel nonlinear integral equations by means of $N/D$ technology. The largest coupled-channel space with 6 channels was in the isospin one-half p-wave $D\, \pi$ system, where we considered the coupling not only to the $D \,\eta, D_s \bar K$  but also to the $D^* \pi, D^* \eta , D_s^* \bar K$ channels.  

Significant left-hand cut contributions to the generalized potential from  t- and u-channel contributions  were established at  N$^3$LO. Such terms were derived from the chiral Lagrangian using a novel framework, in which loop contributions are derived in terms of on-shell hadron masses while keeping the strict 
renormalization scale invariance of conventional $\chi$PT. All counter terms that are needed to renormalize our one-loop contributions at N$^3$LO were constructed and their scale dependence was derived. 

Our global Fit to the data set leads to predictions of s- and p-wave phase shifts at the physical point, where in some channels a striking quark-mass dependence is found. In particular, the isospin one-half $D\, \pi$ system on  
the HSC ensemble at a nominal pion mass of about 239 MeV is still quite far away from its physical limit. 
While we achieved a reasonable global reproduction of the lattice data set, we observe in part significant tension in the HSC data set that lead us to the identification of some possible outlier p-wave data points. 

Our current set of LEC predicts broad p-wave resonances in the flavor antitriplet channels.  
In the flavor sextet  our LEC do not support a visible resonance effect in the s-wave $D\, \pi$ and $D \,\bar K$ phase shifts, which stay well below 90 degrees. A clear resonance signal, however, is implied in the s-wave $D_s \pi$ phase which crosses 90 degrees at about 2.287 GeV. Its Breit-Wigner width as estimated 
from the derivative of the phase shift is about 169 MeV. Detailed predictions for s-wave pole masses in all flavor sextet sectors are predicted in addition . 
In order to consolidate such predictions further improved Lattice QCD data in particular on p-wave phase shifts would be highly welcome. This should be complemented by a computation of triangle- and box-loop diagrams in the generalized potential, as such contributions were not yet included in our current work. A dedicated study of the so-far poorly understood systematic uncertainties is needed.

\section*{Acknowledgments}

M.F.M. Lutz acknowledges Kilian Schwarz and his team with support on distributed computing issues on the Green Cube at GSI, where Jan Knedlik played an instrumental role at an early stage of the project. Particular thanks go to Daniel Mohler and David Wilson for stimulating discussions on lattice QCD aspects. 
\clearpage

\section*{Appendix A}


In this appendix we detail the form of one-loop contributions to the scattering amplitudes that are of chiral order 4. The set of diagrams is illustrated in Fig. \ref{fig:2}.  
We consider scattering with well defined isospin and strangeness, $I$ and $S$, which implies specific meson masses $(m_b, M_b)$ and $(m_a, M_a)$  in the initial and final states respectively.
Given our renormalization scheme as introduced in Section 3 our results are expressed in terms of a generic tadpole integral $\bar I_Q$ (see (\ref{def-IQ})) together with scalar bubble integrals $\bar I_{PQ}(t), \bar I_{QH}(s)$ and $\bar I_{QH}(u)$ (see (\ref{def-tadpole}, \ref{def-bubble}, \ref{def-bubble-PQ})), where $H$ is a placeholder for our heavy fields, and $P,Q$ for our light fields, i.e. the pion, kaon or eta.  
Here $s,t,u$ are the Mandelstam variables, that are used to describe the kinematics of two-body scattering processes. 

The set of loop diagrams has to be supplemented by suitable tree-level diagrams proportional to the  LEC from (\ref{def-gtilde}, \ref{def-L4}). Only a combined consideration provides renormalization-scale independent expressions as is requested in a consistent EFT approach. 
This was already illustrated for the one-loop diagrams of chiral order $Q^3$ in Section 3. However, since we rely in our work on an uncoventional renormalization scheme, we feel it is useful to make this property explicit also in our results at chiral order $Q^4$. This requires to combine all contributions proportional to the tadpole integrals $\bar I_Q$, being the only contributions that depend on the renormalization scale in our scheme. Typically, this involves a set of specific diagrams, the sum of whose contributions can be displayed as follows. We consider first the contributions that are implied by the symmetry preserving LEC $d_{21-26}$ with
\begin{eqnarray}
&& f^4\,T^{(4-sym)}_{ab} = \frac{(s-u)^4}{64\,M^4}\,\Bigg\{  f^2\,d_{25}\,C_3 + f^2\,d_{26}\,C_2 \Bigg\}+ (\bar q \cdot q)\,\frac{(s-u)^2}{4\,M^2}\,\Bigg\{  f^2\,d_{23}\,C_3 + f^2\,d_{24}\,C_2 \Bigg\} 
\nonumber\\
&& \qquad  \quad +\, 4\,(\bar q \cdot q)^2\,\Bigg\{  f^2\,d_{22}\,C_2
 + \Big( \frac{3}{2}\,c_2 + \frac{1}{4}\,c_3+  \frac{1}{4}\,c_4 + \frac{1}{24}\,c_5 \Big)\,C_2 
 \left( \frac{\bar I_a}{m_a^2}+\frac{\bar I_b}{m_b^2} \right)
\nonumber\\
&& \qquad  \qquad \qquad \quad \;\,\; +\, f^2\,d_{21}\,C_3   +\,\Big( \frac{3}{4}\,c_3 + \frac{1}{8}\,c_5 \Big)\,C_3 \left( \frac{\bar I_a}{m_a^2}+\frac{\bar I_b}{m_b^2} \right)\Bigg\} \,,
\label{res-T4-symmetric}
\end{eqnarray}
where (\ref{res-T4-symmetric}) is renormalization-scale invariant line by line. 
This is a consequence of the relations
\begin{eqnarray}
&& \mu^2\,\frac{d}{d\,\mu^2}\, d^{}_{21} =  \Big(\frac{3}{2}\,c_3+  \frac{1}{4}\,c_5  \Big)\,\frac{1}{(4\,\pi\,f)^2}\,,
\nonumber\\
&& \mu^2\,\frac{d}{d\,\mu^2}\, d^{}_{22} = \Big( 3\,c_2 + \frac{1}{2}\,c_3+  \frac{1}{2}\,c_4 + \frac{1}{12}\,c_5 \Big)\frac{1}{(4\,\pi\,f)^2} \,,\qquad  
\nonumber\\
&& 
\mu^2\,\frac{d}{d\,\mu^2}\, d^{}_n\, =0\qquad \qquad \qquad \;  
\,\qquad  {\rm for}\qquad  n \geq 23  \,,
\label{d-running-symmetric-appendix}
\end{eqnarray}
where our result (\ref{res-T4-symmetric}) includes tree-level contributions of LEC together with their associated tadpole terms. The pertinent Clebsch $C_2$ and $C_3$ were already detailed in Section 3.

We continue with our collection of the tadpole terms that involve either a light-quark mass parameter $m$ and $m_s$ or a mass of the pseudo-Goldstone boson $m_Q$ with
\begin{eqnarray}
&& f^4\,T^{(4-tadp.)}_{ab} = 
\sum_{Q\,\in \{\pi,K,\eta\} } \Bigg\{ 
(\bar q \cdot q)\,C^{(6, t)}_Q  \,\Big(\bar I_Q +f^2\, d^{\,t}_6\,m_Q^2 \Big)
+\frac{(s-u)^2}{16\,M^2}\,C^{(6, su)}_Q \,\Big( \bar I_Q  + f^2\,d^{su}_6 \,m_Q^2\Big) 
\nonumber\\
&& \qquad \qquad \qquad \qquad +\,\sum_{n=0}^5 c_n\,\Big\{ 2\,m\,B_0 \,\bar C^{(n, \pi )}_Q + (m+ m_s)\,B_0\,\bar C^{(n, K)}_Q\Big\} \bar I_Q 
\nonumber\\
&& \qquad \qquad \qquad \qquad +\,\sum_{n=0}^5 c_n\,(\bar q \cdot q)\,\bar C^{(n, t)}_Q  \,\bar I_Q
+ \sum_{n=3}^5 c_n\,\frac{(s-u)^2}{16\,M^2}\,\bar C^{(n, su)}_Q \,\bar I_Q  
\nonumber\\
&& \qquad \qquad \qquad \qquad +\,\sum_{n=0}^1 \,   \Big\{ c_{n}\, \big\{2\,m\,B_0\,\bar L^{(n, \pi)}_Q + (m+m_s)\,B_0\,\bar L^{(n, K )}_Q \big\}
\nonumber\\
&& \qquad \qquad \qquad  \qquad\qquad \qquad +\,c_{n+2}\, (\bar q\cdot q)\,\bar L^{(n+2)}_Q   +\, c_{n+4}\,\frac{(s-u)^2}{16\,M^2}\,\bar L^{(n+4)}_Q   \Big\} \bar I_Q
\nonumber\\
&& \qquad \qquad  \qquad\qquad +\,f^2\,\frac{g_1}{M}\,c_1\,\Big\{2\,m\,B_0\,C^{(0, \pi)}_Q + (m+m_s)\,B_0\,C^{(0, K )}_Q \Big\}\, m_Q^2
\nonumber\\
&& \qquad \qquad \qquad \qquad +\,\sum_{n=1}^5 \Big\{ f^2\, d^{\,\chi }_n\, \Big(  2\,m\,B_0 \,C^{(n, \pi )}_Q + (m+ m_s)\,B_0\,C^{(n, K)}_Q \Big)\Big\} \,m^2_Q 
\nonumber\\
&& \qquad \qquad \qquad \qquad  
+\, \sum_{n=2}^5 f^2\, d^{\,t}_n\, (\bar q \cdot q)\,C^{(n, t)}_Q  \,m^2_Q 
+ \sum_{n=3}^5 f ^2\,d^{su}_n\, \frac{(s-u)^2}{16\,M^2}\,C^{(n, su)}_Q \,m^2_Q   
\nonumber\\
&& \qquad \qquad  \qquad\qquad +\,\sum_{n=0}^1 \,   \Big\{ c_{n}\, \big\{2\,m\,B_0\,L^{(n, \pi)}_Q + (m+m_s)\,B_0\,L^{(n, K )}_Q \big\}
\nonumber\\
&& \qquad \qquad \qquad \qquad  \qquad\qquad +\,c_{n+2}\, (\bar q\cdot q)\,L^{(n+2)}_Q   +\, c_{n+4}\,\frac{(s-u)^2}{16\,M^2}\,L^{(n+4)}_Q   \Big\}\, m_Q^2 
\Bigg\} 
\nonumber\\
&& \qquad \qquad \qquad  +\, (\bar q \cdot q)\, \sum_{n=1}^5\, C^{\chi}_{6+n}\,\Big\{ f^2\,d^{\,t}_n + \frac{c_n}{2}\,
\Big(\frac{\bar I_a}{m_a^2}+  \frac{\bar I_b}{m_b^2}\Big) \Big\} \,,
\label{res-T4-tadpole}
\end{eqnarray}
where our result (\ref{res-T4-tadpole}) includes tree-level contributions of LEC, conveniently combined into $d^{\chi }_n$, $d^{\,t}_n$ and $d^{su}_n$, together with sets of corresponding Clebsch coefficients $C^{(n,.)}_Q, \bar C^{(n,.)}_Q$ and $L^{(n)}_Q, \bar L^{(n)}_Q$ and $C^\chi_n$.  The renormalization-scale dependence in each of the three classes of contributions implied by the sum over $Q$ in (\ref{res-T4-tadpole}) cancel identically. This follows most economically from the identities
\begin{eqnarray}
&& \mu^2\,\frac{d}{d\,\mu^2}\, d^{\,\chi}_n \,=
\mu^2\,\frac{d}{d\,\mu^2}\, d^{\,t}_n = \frac{c_n}{(4\,\pi\,f)^2} \,\qquad  
{\rm for}\qquad n \in \{1,3,5\} \,,
\nonumber\\
&& \mu^2\,\frac{d}{d\,\mu^2}\, d^{\,\chi}_2\, =\frac{c_2 -2\, c_0}{(4\,\pi\,f)^2}  \,,\qquad \;\, \qquad
\mu^2\,\frac{d}{d\,\mu^2}\, d^{\,t}_2 = \frac{c_2 - c_0/2}{(4\,\pi\,f)^2} \,,\qquad  
\nonumber\\
&& \mu^2\,\frac{d}{d\,\mu^2}\, d^{\,\chi}_4\, =\frac{c_4 +12\, c_0}{(4\,\pi\,f)^2}  \,,\qquad \qquad 
\mu^2\,\frac{d}{d\,\mu^2}\, d^{\,t}_4 = \frac{c_4 + 3\,c_0}{(4\,\pi\,f)^2} \,,\qquad  
\nonumber\\
&& 
\mu^2\,\frac{d}{d\,\mu^2}\, d^{su}_n = \frac{c_n}{(4\,\pi\,f)^2} \qquad \qquad \quad \;  \,
\,\qquad  {\rm for}\qquad  n \in \{3,4,5\}  \,,
\nonumber\\
&& \mu^2\,\frac{d}{d\,\mu^2}\, d^{su}_6=
\mu^2\,\frac{d}{d\,\mu^2}\, d^{\,t}_6 = \frac{1}{(4\,\pi\,f)^2}\,,
\label{d-running-appendix}
\end{eqnarray}
together with specific relations among the Clebsch coefficients. The 'bar' Clebsch $\bar C^{(n, .)}_Q$ can always be derived in terms of the 'unbar' Clebsch $C^{
(n, .)}_Q$  provided that  $\bar L^{(n)}_Q$ and $ L^{(n)}_Q$ are known. Note that we introduced those to depend exclusively on $L_4$ or $L_5$ and therefore such terms provide the contributions from  the wave-function renormalization factor of the pion, kaon and eta meson fields.

\begin{table}[]
  \tabcolsep=6.6mm
  \center
    \begin{tabular}{|C||C|C|C|}
      \hline
      (I,\,S)&
      {C}^\chi_7 & {C}^\chi_8 & {C}^\chi_9
      \\
      \hline
      \hline
      (\frac12,\,+2)   &
 -3 \tilde{m}_{K}^2-4 \tilde{m}_\pi^2 & -\frac{4}{3} (9 \tilde{m}_{K}^2+\tilde{m}_\pi^2) & -\frac{2}{9} (15 \tilde{m}_{K}^2+4 \tilde{m}_\pi^2) \\
      \hline
      \hline
      (0,\,+1)         &
 3 \tilde{m}_\pi^2-7 \tilde{m}_{K}^2 & -\frac{4}{3} (9 \tilde{m}_{K}^2+\tilde{m}_\pi^2) & \frac{1}{9} (\tilde{m}_\pi^2-33 \tilde{m}_{K}^2) \\
      \hline
      &
 -\frac{1}{2} \sqrt{3} (\tilde{m}_{K}^2-3 \tilde{m}_\pi^2) & 0 & \frac{2}{3 \sqrt{3}} (3 \tilde{m}_{K}^2-2 \tilde{m}_\pi^2) \\
      \hline
      &
 -6 \tilde{m}_{K}^2 & -\frac{8}{9} (17 \tilde{m}_{K}^2-2 \tilde{m}_\pi^2) & -\frac{2}{27} (43 \tilde{m}_{K}^2+11 \tilde{m}_\pi^2) \\
      \hline
      \hline
      (1,\,+1)         &
 -2 (\tilde{m}_{K}^2+4 \tilde{m}_\pi^2) & -\frac{8}{3} (\tilde{m}_{K}^2+4 \tilde{m}_\pi^2) & -\frac{2}{9} (3 \tilde{m}_{K}^2+19 \tilde{m}_\pi^2) \\
      \hline
      &
 \frac{1}{2} (-5 \tilde{m}_{K}^2-\tilde{m}_\pi^2) & 0 & -\frac{2}{9} (2 \tilde{m}_{K}^2+\tilde{m}_\pi^2) \\
      \hline
      &
 -9 \tilde{m}_{K}^2-\tilde{m}_\pi^2 & -\frac{4}{3} (9 \tilde{m}_{K}^2+\tilde{m}_\pi^2) & \frac{1}{9} (-41 \tilde{m}_{K}^2-3 \tilde{m}_\pi^2) \\
      \hline
      \hline
      (\frac12,\,0)    &
 -3 \tilde{m}_{K}^2-4 \tilde{m}_\pi^2 & -\frac{8}{3} (\tilde{m}_{K}^2+4 \tilde{m}_\pi^2) & -\tilde{m}_{K}^2-\frac{29}{9} \tilde{m}_\pi^2 \\
      \hline
      &
 -3 \tilde{m}_{K}^2 & 0 & \frac{1}{3} (\tilde{m}_{K}^2-3 \tilde{m}_\pi^2) \\
      \hline
      &
 \frac{1}{2} \sqrt{\frac{3}{2}} (5 \tilde{m}_{K}^2+\tilde{m}_\pi^2) & 0 & \frac{1}{3} \sqrt{\frac{2}{3}} (2 \tilde{m}_{K}^2+\tilde{m}_\pi^2) \\
      \hline
      &
 -9 \tilde{m}_{K}^2 & -\frac{8}{9} (17 \tilde{m}_{K}^2-2 \tilde{m}_\pi^2) & -\frac{7}{27} (23 \tilde{m}_{K}^2-5 \tilde{m}_\pi^2) \\
      \hline
      &
 -\frac{1}{2} \sqrt{\frac{3}{2}} (\tilde{m}_{K}^2-3 \tilde{m}_\pi^2) & 0 & \frac{1}{3} \sqrt{\frac{2}{3}} (3 \tilde{m}_{K}^2-2 \tilde{m}_\pi^2) \\
      \hline
      &
 -3 \tilde{m}_{K}^2-4 \tilde{m}_\pi^2 & -\frac{4}{3} (9 \tilde{m}_{K}^2+\tilde{m}_\pi^2) & -\frac{2}{9} (15 \tilde{m}_{K}^2+4 \tilde{m}_\pi^2) \\
      \hline
      \hline
      (\frac32,\,0)    &    
 -3 \tilde{m}_{K}^2-4 \tilde{m}_\pi^2 & -\frac{8}{3} (\tilde{m}_{K}^2+4 \tilde{m}_\pi^2) & -\tilde{m}_{K}^2-\frac{29}{9} \tilde{m}_\pi^2 \\
      \hline
      \hline
      (0,\,-1)         &
 -10 \tilde{m}_{K}^2-3 \tilde{m}_\pi^2 & -\frac{4}{3} (9 \tilde{m}_{K}^2+\tilde{m}_\pi^2) & -\frac{5}{9} (9 \tilde{m}_{K}^2+\tilde{m}_\pi^2) \\
      \hline
      \hline
      (1,\,-1)         &
 \tilde{m}_\pi^2-8 \tilde{m}_{K}^2 & -\frac{4}{3} (9 \tilde{m}_{K}^2+\tilde{m}_\pi^2) & \frac{1}{9} (-37 \tilde{m}_{K}^2-\tilde{m}_\pi^2) \\
      \hline
    \end{tabular}
  \caption{ The coefficients ${C}^{\chi}_7$, ${C}^{\chi }_8$ and ${C}^{\chi}_9$  as introduced in (\ref{res-T4-tadpole}) for
given isospin (I) and strangeness (S). The channel ordering is specified with 
Tab. \ref{tab:states} and Tab. \ref{tab:coeffA}. We use the notation 
$\tilde{m}_\pi^2= 2\,B_0\,m$ and $\tilde{m}_{K}^2= B_0\,(m+m_s)$.
  }
  \label{tab:Clebsch:CChi789}
\end{table}

\begin{table}[]
  \tabcolsep=3.5mm
  \center
    \begin{tabular}{|C||C|C|C||C|C|C||C|C|C|}
      \hline
       (I,\,S)
      & \multicolumn{3}{C||}{
        {C}^{(0,\pi )}_{Q}
      }
      & \multicolumn{3}{C||}{
        {C}^{(1,\pi )}_{Q}
      }
      & \multicolumn{3}{C|}{
        {C}^{(2, \pi)}_{Q}
      }
      \\
      &
      {\pi} & {K} & {\eta} &
      {\pi} & {K} & {\eta} &
      {\pi} & {K} & {\eta} 
      \\
      \hline
      \hline
      (\frac12,\,+2)   &
 0 & 0 & 0 & 0 & \frac{580}{207} & 0 & 0 & -\frac{4}{3} & 0 \\
      \hline
      \hline
      (0,\,+1)         &
 0 & 0 & 0 & 0 & -\frac{4}{5} & 0 & 0 & -\frac{4}{3} & 0 \\
      \hline
      &
 0 & 8 \sqrt{3} & -8 \sqrt{3} & \frac{\sqrt{3}}{2} & \frac{13}{11 \sqrt{3}} & -\frac{1}{\sqrt{3}} & 0 & 0 & 0 \\
      \hline
      &
 0 & 0 & 0 & \frac{46}{13} & -\frac{9292}{4797} & \frac{46}{117} & -8 & \frac{40}{3} & -\frac{8}{9} \\
      \hline
      \hline
      (1,\,+1)         &
 0 & 0 & 0 & \frac{234}{19} & \frac{356}{171} & \frac{18}{19} & -\frac{8}{3} & -\frac{40}{3} & -\frac{8}{3} \\
      \hline
      &
 -8 & 8 & 0 & \frac{11}{15} & \frac{23}{45} & \frac{11}{30} & 0 & 0 & 0 \\
      \hline
      &
 0 & 0 & 0 & 0 & \frac{220}{243} & 0 & 0 & -\frac{4}{3} & 0 \\
      \hline
      \hline
      (\frac12,\,0)    &
 0 & 0 & 0 & \frac{325}{57} & \frac{33974}{8379} & \frac{25}{57} & -\frac{8}{3} & -\frac{40}{3} & -\frac{8}{3} \\
      \hline
      &
 0 & 0 & 0 & \frac{7}{4} & \frac{17}{19} & \frac{7}{12} & 0 & 0 & 0 \\
      \hline
      &
 4 \sqrt{6} & -4 \sqrt{6} & 0 & -\frac{11}{5 \sqrt{6}} & -\frac{23}{15 \sqrt{6}} & -\frac{11}{10 \sqrt{6}} & 0 & 0 & 0 \\
      \hline
      &
 0 & 0 & 0 & \frac{15}{7} & -\frac{49682}{14301} & -\frac{25}{7} & -8 & \frac{40}{3} & -\frac{8}{9} \\
      \hline
      &
 0 & 4 \sqrt{6} & -4 \sqrt{6} & \frac{\sqrt{\frac{3}{2}}}{2} & \frac{13}{11 \sqrt{6}} & -\frac{1}{\sqrt{6}} & 0 & 0 & 0 \\
      \hline
      &
 0 & 0 & 0 & 0 & \frac{580}{207} & 0 & 0 & -\frac{4}{3} & 0 \\
      \hline
      \hline
      (\frac32,\,0)    &    
 0 & 0 & 0 & \frac{325}{57} & \frac{33974}{8379} & \frac{25}{57} & -\frac{8}{3} & -\frac{40}{3} & -\frac{8}{3} \\
      \hline
      \hline
      (0,\,-1)         &
 0 & 0 & 0 & 0 & \frac{72}{43} & 0 & 0 & -\frac{4}{3} & 0 \\
      \hline
      \hline
      (1,\,-1)         &
 0 & 0 & 0 & 0 & \frac{16}{99} & 0 & 0 & -\frac{4}{3} & 0 \\
      \hline
    \end{tabular}
  \caption{The coefficients ${C}^{(0,\pi)}_Q$, ${C}^{(1,\pi)}_Q$, and ${C}^{(2,\pi)}_Q$  as introduced in (\ref{res-T4-tadpole}) for
given isospin (I) and strangeness (S). }
  \label{tab:Clebsch:CQ012Pion}
\end{table}

\begin{table}[]
  \tabcolsep=3.0mm
  \center
    \begin{tabular}{|C||C|C|C||C|C|C||C|C|C|}
      \hline
       (I,\,S)
      & \multicolumn{3}{C||}{
        {C}^{(0,K )}_{Q}
      }
      & \multicolumn{3}{C||}{
        {C}^{(1,K )}_{Q}
      }
      & \multicolumn{3}{C|}{
        {C}^{(2,K)}_{Q}
      }
      \\
      &
      {\pi} & {K} & {\eta} &
      {\pi} & {K} & {\eta} &
      {\pi} & {K} & {\eta} 
      \\
      \hline
      \hline
     (\frac12,\,+2)   &
 0 & 0 & 0 & \frac{5249}{1380} & \frac{29}{5} & \frac{29}{60} & -10 & -\frac{16}{3} & -\frac{14}{3} \\
      \hline
      \hline
      (0,\,+1)         &
 0 & 0 & 0 & -\frac{47}{1260} & \frac{215}{21} & -\frac{43}{84} & -10 & -\frac{16}{3} & -\frac{14}{3} \\
      \hline
      &
 0 & -8 \sqrt{3} & 8 \sqrt{3} & \frac{2500}{2343 \sqrt{3}} & -\frac{875}{142 \sqrt{3}} & -\frac{125}{213 \sqrt{3}} & 0 & 0 & 0 \\
      \hline
      &
 0 & 0 & 0 & \frac{856}{1599} & \frac{160}{39} & \frac{64}{13} & 0 & -32 & \frac{56}{9} \\
      \hline
      \hline
      (1,\,+1)         &
 0 & 0 & 0 & \frac{88}{57} & 0 & 0 & -\frac{8}{3} & 0 & 0 \\
      \hline
      &
 8 & -8 & 0 & \frac{281}{306} & \frac{175}{102} & -\frac{25}{102} & 0 & 0 & 0 \\
      \hline
      &
 0 & 0 & 0 & \frac{18211}{10692} & \frac{455}{33} & \frac{65}{132} & -10 & -\frac{16}{3} & -\frac{14}{3} \\
      \hline
      \hline
      (\frac12,\,0)    &
 0 & 0 & 0 & \frac{132}{49} & 0 & 0 & -\frac{8}{3} & 0 & 0 \\
      \hline
      &
 0 & 0 & 0 & \frac{44}{57} & 0 & 0 & 0 & 0 & 0 \\
      \hline
      &
 -4 \sqrt{6} & 4 \sqrt{6} & 0 & -\frac{281}{102 \sqrt{6}} & -\frac{175}{34 \sqrt{6}} & \frac{25}{34 \sqrt{6}} & 0 & 0 & 0 \\
      \hline
      &
 0 & 0 & 0 & -\frac{438772}{978597} & \frac{3440}{479} & \frac{59168}{4311} & 0 & -32 & \frac{56}{9} \\
      \hline
      &
 0 & -4 \sqrt{6} & 4 \sqrt{6} & \frac{1250 \sqrt{\frac{2}{3}}}{2343} & -\frac{875}{142 \sqrt{6}} & -\frac{125}{213 \sqrt{6}} & 0 & 0 & 0 \\
      \hline
      &
 0 & 0 & 0 & \frac{5249}{1380} & \frac{29}{5} & \frac{29}{60} & -10 & -\frac{16}{3} & -\frac{14}{3} \\
      \hline
      \hline
      (\frac32,\,0)    &    
 0 & 0 & 0 & \frac{132}{49} & 0 & 0 & -\frac{8}{3} & 0 & 0 \\
      \hline
      \hline
      (0,\,-1)         &
 0 & 0 & 0 & \frac{13334}{5031} & \frac{608}{39} & \frac{38}{39} & -10 & -\frac{16}{3} & -\frac{14}{3} \\
      \hline
      \hline
      (1,\,-1)         &
 0 & 0 & 0 & \frac{8}{11} & 12 & 0 & -10 & -\frac{16}{3} & -\frac{14}{3} \\
      \hline
    \end{tabular}
  \caption{Coefficients ${C}^{(0,{K})}_Q$, ${C}^{(1,{K})}_Q$, and ${C}^{(2,{K})}_Q$ as in Tab. \ref{tab:Clebsch:CQ012Pion}.
  }
  \label{tab:Clebsch:CQ012Kaon}
\end{table}

\begin{table}[]
  \tabcolsep=2.7mm
  \center
  \begin{tabular}{|C||C|C|C||C|C|C||C|C|C||C|C|C|}
    \hline
    (I,\,S)
      & \multicolumn{3}{C||}{
        {L}^{(0,\pi )}_{Q,4}
      }
      & \multicolumn{3}{C||}{
        {L}^{(0,K )}_{Q,4}
      }
      & \multicolumn{3}{C||}{
        {L}^{(0,\pi)}_{Q,5}
      }
      & \multicolumn{3}{C|}{
        {L}^{(0,K ) }_{Q,5}
      }
      \\
      &
      {\pi} & {K} & {\eta} &
      {\pi} & {K} & {\eta} &
      {\pi} & {K} & {\eta} &
      {\pi} & {K} & {\eta} 
    \\
    \hline
    \hline
    (\frac12,\,+2)   &
 0 & 0 & 0 & -96 & 0 & -96 & 0 & 0 & 0 & 0 & -64 & 0 \\
    \hline
    \hline
    (0,\,+1)         &
 0 & 0 & 0 & -96 & 0 & -96 & 0 & 0 & 0 & 0 & -64 & 0 \\
    \hline
    &
 0 & 0 & 0 & 0 & 0 & 0 & 0 & 0 & 0 & 0 & 0 & 0 \\
    \hline
    &
 0 & 128 & -64 & 0 & -512 & 256 & 0 & 0 & \frac{64}{3} & 0 & 0 & -\frac{256}{3} \\
    \hline
    \hline
    (1,\,+1)         &
 -64 & -128 & 0 & 0 & 0 & 0 & -64 & 0 & 0 & 0 & 0 & 0 \\
    \hline
    &
 0 & 0 & 0 & 0 & 0 & 0 & 0 & 0 & 0 & 0 & 0 & 0 \\
    \hline
    &
 0 & 0 & 0 & -96 & 0 & -96 & 0 & 0 & 0 & 0 & -64 & 0 \\
    \hline
    \hline
    (\frac12,\,0)    &
 -64 & -128 & 0 & 0 & 0 & 0 & -64 & 0 & 0 & 0 & 0 & 0 \\
    \hline
    &
 0 & 0 & 0 & 0 & 0 & 0 & 0 & 0 & 0 & 0 & 0 & 0 \\
    \hline
    &
 0 & 0 & 0 & 0 & 0 & 0 & 0 & 0 & 0 & 0 & 0 & 0 \\
    \hline
    &
 0 & 128 & -64 & 0 & -512 & 256 & 0 & 0 & \frac{64}{3} & 0 & 0 & -\frac{256}{3} \\
    \hline
    &
 0 & 0 & 0 & 0 & 0 & 0 & 0 & 0 & 0 & 0 & 0 & 0 \\
    \hline
    &
 0 & 0 & 0 & -96 & 0 & -96 & 0 & 0 & 0 & 0 & -64 & 0 \\
    \hline
    \hline
    (\frac32,\,0)    &    
 -64 & -128 & 0 & 0 & 0 & 0 & -64 & 0 & 0 & 0 & 0 & 0 \\
    \hline
    \hline
    (0,\,-1)         &
 0 & 0 & 0 & -96 & 0 & -96 & 0 & 0 & 0 & 0 & -64 & 0 \\
    \hline
    \hline
    (1,\,-1)         &
 0 & 0 & 0 & -96 & 0 & -96 & 0 & 0 & 0 & 0 & -64 & 0 \\
    \hline
  \end{tabular}
  \caption{
    The coefficients $L^{(0,\pi)}_Q= L_4\,L^{(0,\pi)}_{Q,4}+ L_5\,L^{(0,\pi)}_{Q,5}$ and 
    $L^{(0, K)}_Q= L_4\,L^{(0,K)}_{Q,4}+ L_5\,L^{(0,K)}_{Q,5}$ in (\ref{res-T4-tadpole}) for
given isospin (I) and strangeness (S). 
  }
  \label{tab:Clebsch:L0}
\end{table}

\begin{table}[]
  \tabcolsep=0.99mm
  \center
  \begin{tabular}{|C||C|C|C||C|C|C||C|C|C||C|C|C|}
    \hline
     (I,\,S)
      & \multicolumn{3}{C||}{
        {L}^{(1,\pi )}_{Q,4}
      }
      & \multicolumn{3}{C||}{
        {L}^{(1,K )}_{Q,4}
      }
      & \multicolumn{3}{C||}{
        {L}^{(1,\pi)}_{Q,5}
      }
      & \multicolumn{3}{C|}{
        {L}^{(1,K ) }_{Q,5}
      }
      \\
      &
      {\pi} & {K} & {\eta} &
      {\pi} & {K} & {\eta} &
      {\pi} & {K} & {\eta} &
      {\pi} & {K} & {\eta} 
    \\
    \hline
    \hline
    (\frac12,\,+2)   &
 0 & 0 & 0 & 24 & 0 & 24 & 0 & 0 & 0 & 0 & 16 & 0 \\
    \hline
    \hline
    (0,\,+1)         &
 0 & 0 & 0 & 0 & 0 & 0 & 0 & 0 & 0 & 0 & 0 & 0 \\
    \hline
    &
 6 \sqrt{3} & 24 \sqrt{3} & -6 \sqrt{3} & -10 \sqrt{3} & -40 \sqrt{3} & 10 \sqrt{3} & 0 & 4 \sqrt{3} & 4 \sqrt{3} & 0 & -\frac{20}{\sqrt{3}} & -\frac{20}{\sqrt{3}} \\
    \hline
    &
 0 & 64 & -32 & 0 & 0 & 0 & 0 & 0 & \frac{32}{3} & 0 & 0 & 0 \\
    \hline
    \hline
    (1,\,+1)         &
 32 & 64 & 0 & 0 & 0 & 0 & 32 & 0 & 0 & 0 & 0 & 0 \\
    \hline
    &
 10 & 8 & 6 & 10 & 8 & 6 & 4 & 4 & 0 & 4 & 4 & 0 \\
    \hline
    &
 0 & 0 & 0 & 48 & 0 & 48 & 0 & 0 & 0 & 0 & 32 & 0 \\
    \hline
    \hline
    (\frac12,\,0)    &
 16 & 32 & 0 & 0 & 0 & 0 & 16 & 0 & 0 & 0 & 0 & 0 \\
    \hline
    &
 8 & 64 & -24 & 0 & 0 & 0 & 8 & 0 & 8 & 0 & 0 & 0 \\
    \hline
    &
 -5 \sqrt{6} & -4 \sqrt{6} & -3 \sqrt{6} & -5 \sqrt{6} & -4 \sqrt{6} & -3 \sqrt{6} & -2 \sqrt{6} & -2 \sqrt{6} & 0 & -2 \sqrt{6} & -2 \sqrt{6} & 0 \\
    \hline
    &
 0 & -96 & 48 & 0 & 256 & -128 & 0 & 0 & -16 & 0 & 0 & \frac{128}{3} \\
    \hline
    &
 3 \sqrt{6} & 12 \sqrt{6} & -3 \sqrt{6} & -5 \sqrt{6} & -20 \sqrt{6} & 5 \sqrt{6} & 0 & 2 \sqrt{6} & 2 \sqrt{6} & 0 & \frac{-20}{\sqrt{6}} & \frac{-20}{\sqrt{6}} \\
    \hline
    &
 0 & 0 & 0 & 24 & 0 & 24 & 0 & 0 & 0 & 0 & 16 & 0 \\
    \hline
    \hline
    (\frac32,\,0)    &    
 16 & 32 & 0 & 0 & 0 & 0 & 16 & 0 & 0 & 0 & 0 & 0 \\
    \hline
    \hline
    (0,\,-1)         &
 0 & 0 & 0 & 72 & 0 & 72 & 0 & 0 & 0 & 0 & 48 & 0 \\
    \hline
    \hline
    (1,\,-1)         &
 0 & 0 & 0 & 24 & 0 & 24 & 0 & 0 & 0 & 0 & 16 & 0 \\
    \hline
   \end{tabular}
  \caption{The coefficients $L^{(1,\pi)}_Q= L_4\,L^{(1,\pi)}_{Q,4}+ L_5\,L^{(1,\pi)}_{Q,5}$ and 
    $L^{(1, K)}_Q= L_4\,L^{(1,K)}_{Q,4}+ L_5\,L^{(1,K)}_{Q,5}$  in (\ref{res-T4-tadpole}) for
given isospin (I) and strangeness (S).  }
  \label{tab:Clebsch:L1}
\end{table}

\begin{table}[]
  \tabcolsep=3.8mm
  \center
    \begin{tabular}{|C||C|C|C||C|C|C||C|C|C|}
      \hline
       (I,\,S)
      & \multicolumn{3}{C||}{
        {C}^{(2,t )}_{Q}
      }
      & \multicolumn{3}{C||}{
        {C}^{(6,t )}_{Q}
      }
      & \multicolumn{3}{C|}{
        {C}^{(6,su)}_{Q}
      }
      \\
      &
      {\pi} & {K} & {\eta} &
      {\pi} & {K} & {\eta} &
      {\pi} & {K} & {\eta} 
      \\
      \hline
      \hline
      (\frac12,\,+2)   &
 8 & -24 & 8 & \frac{3}{8} & \frac{1}{2} & \frac{3}{8} & -\frac{3}{4} & -1 & -\frac{3}{4} \\
      \hline
      \hline
      (0,\,+1)         &
 8 & -24 & 8 & 0 & \frac{5}{4} & \frac{3}{4} & 0 & -\frac{5}{2} & -\frac{3}{2} \\
      \hline
      &
 0 & 0 & 0 & 0 & \frac{\sqrt{3}}{4} & 0 & 0 & -\frac{\sqrt{3}}{2} & 0 \\
      \hline
      &
 0 & 32 & -40 & 0 & \frac{3}{2} & 0 & 0 & -3 & 0 \\
      \hline
      \hline
      (1,\,+1)         &
 -\frac{56}{3} & \frac{32}{3} & 0 & 0 & \frac{1}{2} & 0 & 0 & -1 & 0 \\
      \hline
      &
 0 & 0 & 0 & -\frac{1}{2} & -\frac{1}{4} & 0 & 1 & \frac{1}{2} & 0 \\
      \hline
      &
 8 & -24 & 8 & \frac{1}{4} & \frac{1}{4} & 0 & -\frac{1}{2} & -\frac{1}{2} & 0 \\
      \hline
      \hline
      (\frac12,\,0)    &
 -\frac{56}{3} & \frac{32}{3} & 0 & 1 & \frac{1}{4} & 0 & -2 & -\frac{1}{2} & 0 \\
      \hline
      &
 0 & 0 & 0 & 0 & -\frac{3}{4} & 0 & 0 & \frac{3}{2} & 0 \\
      \hline
      &
 0 & 0 & 0 & \frac{1}{2} \sqrt{\frac{3}{2}} & \frac{1}{4} \sqrt{\frac{3}{2}} & 0 & -\sqrt{\frac{3}{2}} & -\frac{1}{2} \sqrt{\frac{3}{2}} & 0 \\
      \hline
      &
 0 & 32 & -40 & 0 & \frac{3}{4} & 0 & 0 & -\frac{3}{2} & 0 \\
      \hline
      &
 0 & 0 & 0 & 0 & \frac{1}{4} \sqrt{\frac{3}{2}} & 0 & 0 & -\frac{1}{2} \sqrt{\frac{3}{2}} & 0 \\
      \hline
      &
 8 & -24 & 8 & \frac{3}{8} & \frac{1}{2} & \frac{3}{8} & -\frac{3}{4} & -1 & -\frac{3}{4} \\
      \hline
      \hline
      (\frac32,\,0)    &    
 -\frac{56}{3} & \frac{32}{3} & 0 & 1 & \frac{1}{4} & 0 & -2 & -\frac{1}{2} & 0 \\
      \hline
      \hline
      (0,\,-1)         &
 8 & -24 & 8 & \frac{3}{8} & -\frac{1}{4} & -\frac{3}{8} & -\frac{3}{4} & \frac{1}{2} & \frac{3}{4} \\
      \hline
      \hline
      (1,\,-1)         &
 8 & -24 & 8 & \frac{1}{8} & \frac{3}{4} & \frac{3}{8} & -\frac{1}{4} & -\frac{3}{2} & -\frac{3}{4} \\
      \hline
    \end{tabular}
  \caption{The coefficients $C^{(2,t)}_Q$, $C^{(6,t)}_Q$ and $C^{(6,su)}_Q$  as introduced (\ref{res-T4-tadpole}).}
\label{tab:Clebsch:CQ26}
\end{table}

\begin{table}[]
  \tabcolsep=1.5mm
  \center
    \begin{tabular}{|C||C|C|C||C|C|C||C|C|C||C|C|C|}
      \hline
        (I,\,S)
      & \multicolumn{3}{C||}{
        {C}^{(3,\pi )}_{Q}
      }
      & \multicolumn{3}{C||}{
        {C}^{(3,K )}_{Q}
      }
      & \multicolumn{3}{C||}{
        {C}^{(3,t)}_{Q}
      }
      & \multicolumn{3}{C|}{
        {C}^{(3,su ) }_{Q}
      }
      \\
      &
      {\pi} & {K} & {\eta} &
      {\pi} & {K} & {\eta} &
      {\pi} & {K} & {\eta} &
      {\pi} & {K} & {\eta} 
      \\
      \hline
      \hline
      (\frac12,\,+2)   &
 0 & -\frac{8}{9} & 0 & -5 & \frac{1}{9} & -\frac{7}{9} & \frac{11}{2} & -\frac{23}{3} & \frac{5}{6} & -3 & 0 & 1 \\
      \hline
      \hline
      (0,\,+1)         &
 0 & \frac{1}{9} & 0 & -2 & -\frac{14}{9} & -\frac{13}{9} & 0 & \frac{1}{3} & \frac{7}{3} & 0 & -6 & -2 \\
      \hline
      &
 -\frac{2}{\sqrt{3}} & 0 & \frac{1}{2 \sqrt{3}} & \frac{11}{4 \sqrt{3}} & \frac{5}{6 \sqrt{3}} & -\frac{5}{12 \sqrt{3}} & -\frac{3 \sqrt{3}}{2} & \frac{7}{2 \sqrt{3}} & \frac{5}{\sqrt{3}} & \sqrt{3} & -2 \sqrt{3} & -\sqrt{3} \\
      \hline
      &
 -4 & \frac{10}{3} & \frac{14}{27} & 0 & -8 & \frac{58}{27} & \frac{10}{3} & -\frac{10}{3} & 0 & 0 & -4 & 0 \\
      \hline
      \hline
      (1,\,+1)         &
 -\frac{34}{9} & -\frac{10}{3} & -\frac{4}{9} & -\frac{2}{3} & 0 & 0 & -6 & \frac{2}{3} & 0 & 0 & 4 & 0 \\
      \hline
      &
 -\frac{13}{18} & 0 & 0 & -\frac{1}{36} & -\frac{5}{6} & -\frac{1}{12} & -\frac{7}{3} & -\frac{3}{2} & -\frac{1}{6} & 3 & 2 & 1 \\
      \hline
      &
 0 & -\frac{1}{3} & 0 & -\frac{8}{3} & -\frac{28}{9} & -\frac{19}{9} & \frac{5}{3} & -11 & 4 & 2 & 2 & 0 \\
      \hline
      \hline
      (\frac12,\,0)    &
 \frac{5}{9} & -5 & -\frac{10}{9} & -1 & 0 & 0 & -\frac{19}{3} & 5 & 0 & 0 & -2 & 0 \\
      \hline
      &
 \frac{1}{6} & -\frac{5}{3} & \frac{1}{6} & -\frac{1}{2} & \frac{2}{3} & -\frac{1}{2} & -\frac{9}{2} & 3 & -\frac{5}{2} & 4 & 2 & 0 \\
      \hline
      &
 \frac{13}{6 \sqrt{6}} & 0 & 0 & \frac{1}{12 \sqrt{6}} & \frac{5}{2 \sqrt{6}} & \frac{1}{4 \sqrt{6}} & \frac{7}{\sqrt{6}} & \frac{3}{2} \sqrt{\frac{3}{2}} & \frac{1}{2 \sqrt{6}} & -3 \sqrt{\frac{3}{2}} & -\sqrt{6} & -\sqrt{\frac{3}{2}} \\
      \hline
      &
 -2 & 5 & -\frac{19}{27} & 0 & -12 & \frac{55}{27} & 0 & 11 & -15 & 0 & 2 & 0 \\
      \hline
      &
 -\sqrt{\frac{2}{3}} & 0 & \frac{1}{2 \sqrt{6}} & \frac{11}{4 \sqrt{6}} & \frac{5}{6 \sqrt{6}} & -\frac{5}{12 \sqrt{6}} & -\frac{3}{2} \sqrt{\frac{3}{2}} & \frac{7}{2 \sqrt{6}} & \frac{5}{\sqrt{6}} & \sqrt{\frac{3}{2}} & -\sqrt{6} & -\sqrt{\frac{3}{2}} \\
      \hline
      &
 0 & -\frac{8}{9} & 0 & -5 & \frac{1}{9} & -\frac{7}{9} & \frac{11}{2} & -\frac{23}{3} & \frac{5}{6} & -3 & 0 & 1 \\
      \hline
      \hline
      (\frac32,\,0)    &    
 \frac{5}{9} & -5 & -\frac{10}{9} & -1 & 0 & 0 & -\frac{19}{3} & 5 & 0 & 0 & -2 & 0 \\
      \hline
      \hline
      (0,\,-1)         &
 0 & -\frac{5}{9} & 0 & -3 & -\frac{35}{9} & -\frac{22}{9} & \frac{5}{2} & -\frac{50}{3} & \frac{29}{6} & 3 & 6 & 1 \\
      \hline
      \hline
      (1,\,-1)         &
 0 & -\frac{1}{9} & 0 & -\frac{7}{3} & -\frac{7}{3} & -\frac{16}{9} & \frac{5}{6} & -\frac{16}{3} & \frac{19}{6} & 1 & -2 & -1 \\
      \hline
    \end{tabular}
  \caption{
     The coefficients $C^{(3,\pi)}_Q$, $C^{(3,K)}_Q$ and $C^{(3,t)}_Q$, $C^{(3,su)}_Q$ from  (\ref{res-T4-tadpole}).
  }
  \label{tab:Clebsch:CQ3}
\end{table}


By construction the various Clebsch $C^{(n, .) }_Q$ in (\ref{res-T4-tadpole}) are related to the Clebsch $C^{\chi }_n$ and $C^{\chi\chi }_n$ already introduced in (\ref{res-T4-counter}), which summarized the 
impact of $d_{1-26}$ on the tree-level scattering amplitudes. While we could document the Clebsch 
used in (\ref{res-T4-counter}) it is more convenient to do so directly for the Clebsch as used in (\ref{res-T4-tadpole}). This is so since given the Clebsch $C^{(n,.) }_Q$ 
it is straightforward to read off the values for $C^{\chi }_n$ and $C^{\chi\chi }_n$, however, the inverse procedure is not so immediate. This becomes clear from the set of identities 
\begin{eqnarray}
&& \sum_{n=1}^8 \, d_n\,C^{\chi\chi }_n = \sum_{n=1}^5\,\sum_{ Q \in\{\pi, K, \eta \} }\, d^{\,\chi }_n \,B_0\,\Big(
 2\,m\,C^{(n,\pi) }_Q +  (m + m_s)\,C^{(n, K) }_Q\Big)\,m_Q^2 +  \sum_{n=1}^3 d^{  \,\chi}_{5+n}\, C^{\chi \chi}_{2+2\,n} \,,
\nonumber\\
&& \sum_{n=1}^6 \, d_{n +8}\,C^{ \chi }_n = \sum_{n=2}^6\,\sum_{ Q \in\{\pi, K, \eta \} }\, d^{\, t }_n \,C^{(n, t) }_Q \,m_Q^2  + d^{\,t}_1\,C^\chi_7 +  \Big( d^{\, t }_2+ \frac{1}{6}\,d^{\,t}_4 \Big)\,C^\chi_{8} +
 \Big( d^{\, t }_3+ \frac{1}{6}\,d^{\,t}_5 \Big)\,C^\chi_{9}  
\,,
\nonumber\\
&& \sum_{n=1}^6 \, d_{n +14}\,C^{ \chi }_n = \sum_{n=1}^2 d^{  su }_{n}\,C^\chi_{n+2}+ \sum_{n=3}^6\,\sum_{ Q \in\{\pi, K, \eta \} }\, d^{ su }_n \,C^{(n, su) }_Q \,m_Q^2 \,,
\nonumber\\ \nonumber\\
&& C^\chi_7= -\frac{1}{2}\,C^\chi_{1} -\frac{1}{8}\,C^\chi_{2} +\frac{3}{4}\,C^\chi_{3}  +\frac{1}{2}\,C^\chi_{4} -\frac{3}{4}\,C^\chi_{5} -\frac{3}{4}\,\,C^\chi_{6}   \,,\qquad C^\chi_8 =  6\, C^\chi_{10} = -\frac{7}{3}\,C^\chi_5-\frac{1}{3}\,C^\chi_6  \,,
\nonumber\\
&& C^\chi_9 = 6\, C^\chi_{11} =  \frac{7}{36}\,C^\chi_{1} +\frac{7}{36}\,C^\chi_{2} -\frac{1}{12}\,C^\chi_{3}  -\frac{35}{36}\,C^\chi_{5} -\frac{1}{12}\,\,C^\chi_{6}     \,,   
 \label{res-rewrite-d-Clebsch}
\end{eqnarray}
which hold for $m_\pi^2 \to  2\,m\,B_0$, $m^2_K \to (m+m_s)\,B_0$ and $m_\eta^2 \to 2\, (m+ 2\,m_s)\,B_0/3$, the relations predicted by $\chi$PT at order 2. As already emphasized various times in this manuscript, our result (\ref{res-T4-tadpole}) is derived using strict on-shell masses instead. To this extent, we deem our particular rewrite of the tree-level contributions as is implied by (\ref{res-rewrite-d-Clebsch}) once the on-shell masses, e.g. with  $m^2_\pi \neq 2\,m\,B_0 $ and $m_K^2\neq (m+ m_s)\,B_0$, are used, unambiguous and well motivated. 

It is left to detail the advocated relations among the 'bar' and 'unbar' Clebsch for which we find the set of identities
\begin{eqnarray}
&&\bar C^{(0, \pi )}_Q =  -2\, C^{(2, \pi )}_Q  +12\,C^{(4, \pi )}_Q   +\Delta L^{(0,\pi)}_Q \,,\qquad \;\;\;\,\, 
\nonumber\\
&& \bar C^{(1, \pi )}_Q =  C^{(1, \pi )}_Q - \frac{1}{32}\,C^{(0, \pi )}_Q +\Delta L^{(1,\pi)}_Q \,,
\nonumber\\
&& \bar C^{(0, K )}_Q =   -2\,C^{(2, K )}_Q +12\,C^{(4, K )}_Q+ \Delta L^{(0, K)}_Q \,,\qquad \;\;\;\,\,
\nonumber\\
&& \bar C^{(1, K )}_Q =  C^{(1, K )}_Q- \frac{1}{32}\,C^{(0, K )}_Q + \Delta L^{(1, K)}_Q\,,
\nonumber\\
&&\bar C^{(n, \pi )}_Q =  C^{(n, \pi )}_Q \,,\qquad \qquad  \qquad \quad \;\,\,\bar C^{(n, K )}_Q =  C^{(n, K )}_Q\qquad \qquad  \qquad \;\;{\rm for }\qquad n > 1\,,
\nonumber\\
&& \bar C^{(0, t )}_Q = -\frac{1}{2}\, C^{( 2,t )}_Q + 3\,C^{(4 ,t)}_Q \,, \qquad \; \,
\bar C^{(1, t)}_Q  = 0 \,, 
\nonumber\\
&& \bar C^{(n, t )}_Q =  C^{(n, t )}_Q + \Delta L^{(n)}_Q  \qquad \qquad \;\;\; \, {\rm for }\qquad n = 2 \; {\rm or}\;  n = 3 \,,
\nonumber\\
&& \bar C^{(n, t )}_Q =  C^{(n, t )}_Q \qquad \qquad \qquad \qquad \;\;\, {\rm for }\qquad  n = 4\; {\rm or }\;  n = 5 \,,
\nonumber\\
&& \bar C^{(n, su )}_Q =  C^{(n, su )}_Q \qquad \qquad \qquad  \qquad \! {\rm for }\qquad n  \neq  4 \; {\rm and}\;  n \neq 5 \,,
\nonumber\\
&& \bar C^{(n, su )}_Q =  C^{(n, su )}_Q + \Delta L^{(n)}_Q \qquad  \qquad \, {\rm for }\qquad n  =  4 \; {\rm or}\;  n = 5 \,,
\nonumber\\
&& \quad {\rm where} \ \qquad  \Delta L^{(n,x  )}_Q =  L^{(n,x)}_Q\big|_{3\,L_4 = L_5 =-3/16} -\bar L^{(n,x)}_Q  \qquad {\rm with} \qquad x \in\{\pi, K \} \,,
\nonumber\\
&& \quad {\rm \phantom{ where} }\ \qquad  \Delta L^{(n)}_Q \;\,=  L^{(n)}_Q  \big|_{3\,L_4 = L_5 =-3/16}- \bar L^{(n)}_Q \qquad \quad \,\, {\rm with} \qquad n > 1 \,.
\end{eqnarray}
We claim that it suffices to specify the 'unbar' Clebsch coefficients. This is evident from (\ref{res-T4-tadpole}) by evaluating the three terms in the $Q=\pi, K, \eta$ sum at distinct renormalization scales $\mu = m_\pi$, $\mu = m_K$ and $\mu=m_\eta$. While the LEC need to be  evaluated at the three scales, all terms proportional to $\bar I_Q$ vanish identically in this case.  The number of Clebsch that need to be detailed here is further reduced 
by the relations
\begin{eqnarray}
 L^{(4)}_Q  = L^{(2)}_Q  \,,\qquad  L^{(5)}_Q  = L^{(3)}_Q \,, \qquad 
 \bar L^{(4)}_Q  = \bar L^{(2)}_Q  \,,\qquad  \bar L^{(5)}_Q  = \bar L^{(3)}_Q \,.
\end{eqnarray}
A complete documentation of the Clebsch required in the evaluation of (\ref{res-T4-tadpole}) is provided by Tab. \ref{tab:Clebsch:CChi789} - Tab. \ref{tab:Clebsch:n5}. 

\begin{table}[]
  \tabcolsep=2.0mm
  \center
  \begin{tabular}{|C||C|C|C||C|C|C||C|C|C||C|C|C|}
    \hline
    (I,\,S)
    & \multicolumn{3}{C||}{
        {L}^{(2 )}_{Q,4}
      }
      & \multicolumn{3}{C||}{
        {L}^{(3 )}_{Q,4}
      }
      & \multicolumn{3}{C||}{
        {L}^{(2)}_{Q,5}
      }
      & \multicolumn{3}{C|}{
        {L}^{(3 ) }_{Q,5}
      }
      \\
      &
      {\pi} & {K} & {\eta} &
      {\pi} & {K} & {\eta} &
      {\pi} & {K} & {\eta} &
      {\pi} & {K} & {\eta} 
    \\
    \hline
    \hline
   (\frac12,\,+2)   &
 -96 & 0 & -96 & -24 & 0 & -24 & 0 & -64 & 0 & 0 & -64 & 0 \\
    \hline
    \hline
    (0,\,+1)         &
 -96 & 0 & -96 & 0 & 0 & 0 & 0 & -64 & 0 & 0 & -64 & 0 \\
    \hline
    &
 0 & 0 & 0 & 4 \sqrt{3} & 16 \sqrt{3} & -4 \sqrt{3} & 0 & 0 & 0 & 0 & 0 & 0 \\
    \hline
    &
 0 & -384 & 192 & 0 & -64 & 32 & 0 & 0 & -64 & 0 & 0 & -64 \\
    \hline
    \hline
    (1,\,+1)         &
 -64 & -128 & 0 & -32 & -64 & 0 & -64 & 0 & 0 & -64 & 0 & 0 \\
    \hline
    &
 0 & 0 & 0 & -20 & -16 & -12 & 0 & 0 & 0 & 0 & 0 & 0 \\
    \hline
    &
 -96 & 0 & -96 & -48 & 0 & -48 & 0 & -64 & 0 & 0 & -64 & 0 \\
    \hline
    \hline
    (\frac12,\,0)    &
 -64 & -128 & 0 & -16 & -32 & 0 & -64 & 0 & 0 & -64 & 0 & 0 \\
    \hline
    &
 0 & 0 & 0 & -8 & -64 & 24 & 0 & 0 & 0 & 0 & 0 & 0 \\
    \hline
    &
 0 & 0 & 0 & 10 \sqrt{6} & 8 \sqrt{6} & 6 \sqrt{6} & 0 & 0 & 0 & 0 & 0 & 0 \\
    \hline
    &
 0 & -384 & 192 & 0 & -160 & 80 & 0 & 0 & -64 & 0 & 0 & -64 \\
    \hline
    &
 0 & 0 & 0 & 2 \sqrt{6} & 8 \sqrt{6} & -2 \sqrt{6} & 0 & 0 & 0 & 0 & 0 & 0 \\
    \hline
    &
 -96 & 0 & -96 & -24 & 0 & -24 & 0 & -64 & 0 & 0 & -64 & 0 \\
    \hline
    \hline
    (\frac32,\,0)    &    
 -64 & -128 & 0 & -16 & -32 & 0 & -64 & 0 & 0 & -64 & 0 & 0 \\
    \hline
    \hline
    (0,\,-1)         &
 -96 & 0 & -96 & -72 & 0 & -72 & 0 & -64 & 0 & 0 & -64 & 0 \\
    \hline
    \hline
    (1,\,-1)         &
 -96 & 0 & -96 & -24 & 0 & -24 & 0 & -64 & 0 & 0 & -64 & 0 \\
    \hline
  \end{tabular}
  \caption{ The coeffficients 
     $L^{(n)}_Q=L_4\,L^{(n)}_{Q,4}+L_5\,L^{(n)}_{Q,5}$
    with $n=2,\,3$.
  }
  \label{tab:Clebsch:L23}
\end{table}

\begin{table}[]
  \tabcolsep=2.8mm
  \center
    \begin{tabular}{|C||C|C|C||C|C|C||C|C|C||C|C|C|}
      \hline
       (I,\,S)
      & \multicolumn{3}{C||}{
        {C}^{(4,\pi )}_{Q}
      }
      & \multicolumn{3}{C||}{
        {C}^{(4,K )}_{Q}
      }
      & \multicolumn{3}{C||}{
        {C}^{(4,t)}_{Q}
      }
      & \multicolumn{3}{C|}{
        {C}^{(4,su ) }_{Q}
      }
      \\
      &
      {\pi} & {K} & {\eta} &
      {\pi} & {K} & {\eta} &
      {\pi} & {K} & {\eta} &
      {\pi} & {K} & {\eta} 
      \\
      \hline
      \hline
      (\frac12,\,+2)   &
 0 & -\frac{14}{27} & 0 & -\frac{20}{9} & -\frac{94}{27} & -\frac{8}{9} & \frac{10}{3} & 0 & \frac{10}{3} & -6 & -12 & -6 \\
      \hline
      \hline
      (0,\,+1)         &
 0 & -\frac{14}{27} & 0 & -\frac{20}{9} & -\frac{94}{27} & -\frac{8}{9} & \frac{10}{3} & 0 & \frac{10}{3} & -6 & -12 & -6 \\
      \hline
      &
 0 & 0 & 0 & 0 & 0 & 0 & 0 & 0 & 0 & 0 & 0 & 0 \\
      \hline
      &
 -2 & \frac{80}{27} & \frac{10}{27} & 0 & -\frac{176}{27} & -\frac{52}{27} & 0 & \frac{40}{3} & -\frac{20}{3} & 0 & -24 & 0 \\
      \hline
      \hline
      (1,\,+1)         &
 -\frac{22}{9} & -\frac{80}{27} & -\frac{2}{3} & -\frac{28}{27} & 0 & 0 & \frac{20}{9} & \frac{40}{9} & 0 & -16 & -8 & 0 \\
      \hline
      &
 0 & 0 & 0 & 0 & 0 & 0 & 0 & 0 & 0 & 0 & 0 & 0 \\
      \hline
      &
 0 & -\frac{14}{27} & 0 & -\frac{20}{9} & -\frac{94}{27} & -\frac{8}{9} & \frac{10}{3} & 0 & \frac{10}{3} & -6 & -12 & -6 \\
      \hline
      \hline
      (\frac12,\,0)    &
 -\frac{22}{9} & -\frac{80}{27} & -\frac{2}{3} & -\frac{28}{27} & 0 & 0 & \frac{20}{9} & \frac{40}{9} & 0 & -16 & -8 & 0 \\
      \hline
      &
 0 & 0 & 0 & 0 & 0 & 0 & 0 & 0 & 0 & 0 & 0 & 0 \\
      \hline
      &
 0 & 0 & 0 & 0 & 0 & 0 & 0 & 0 & 0 & 0 & 0 & 0 \\
      \hline
      &
 -2 & \frac{80}{27} & \frac{10}{27} & 0 & -\frac{176}{27} & -\frac{52}{27} & 0 & \frac{40}{3} & -\frac{20}{3} & 0 & -24 & 0 \\
      \hline
      &
 0 & 0 & 0 & 0 & 0 & 0 & 0 & 0 & 0 & 0 & 0 & 0 \\
      \hline
      &
 0 & -\frac{14}{27} & 0 & -\frac{20}{9} & -\frac{94}{27} & -\frac{8}{9} & \frac{10}{3} & 0 & \frac{10}{3} & -6 & -12 & -6 \\
      \hline
      \hline
      (\frac32,\,0)    &    
 -\frac{22}{9} & -\frac{80}{27} & -\frac{2}{3} & -\frac{28}{27} & 0 & 0 & \frac{20}{9} & \frac{40}{9} & 0 & -16 & -8 & 0 \\
      \hline
      \hline
      (0,\,-1)         &
 0 & -\frac{14}{27} & 0 & -\frac{20}{9} & -\frac{94}{27} & -\frac{8}{9} & \frac{10}{3} & 0 & \frac{10}{3} & -6 & -12 & -6 \\
      \hline
      \hline
      (1,\,-1)         &
 0 & -\frac{14}{27} & 0 & -\frac{20}{9} & -\frac{94}{27} & -\frac{8}{9} & \frac{10}{3} & 0 & \frac{10}{3} & -6 & -12 & -6 \\
      \hline
    \end{tabular}
 \caption{The coefficients $C^{(4,\pi)}_Q$, $C^{(4,K)}_Q$ and $C^{(4,t)}_Q$, $C^{(4,su)}_Q$ from  (\ref{res-T4-tadpole}). }
\label{tab:Clebsch:CQ4}
\end{table}

\begin{table}[]
  \tabcolsep=0.9mm
  \center
    \begin{tabular}{|C||C|C|C||C|C|C||C|C|C||C|C|C|}
      \hline
      (I,\,S)
      & \multicolumn{3}{C||}{
        {C}^{(5,\pi )}_{Q}
      }
      & \multicolumn{3}{C||}{
        {C}^{(5,K )}_{Q}
      }
      & \multicolumn{3}{C||}{
        {C}^{(5,t)}_{Q}
      }
      & \multicolumn{3}{C|}{
        {C}^{(5,su ) }_{Q}
      }
      \\
      &
      {\pi} & {K} & {\eta} &
      {\pi} & {K} & {\eta} &
      {\pi} & {K} & {\eta} &
      {\pi} & {K} & {\eta} 
      \\
      \hline
      \hline
      (\frac12,\,+2)   &
 0 & -\frac{28}{81} & 0 & -\frac{10}{9} & -\frac{47}{81} & -\frac{4}{27} & \frac{5}{3} & -\frac{5}{18} & \frac{5}{9} & -\frac{9}{2} & -3 & -\frac{1}{2} \\
      \hline
      \hline
      (0,\,+1)         &
 0 & \frac{7}{162} & 0 & -\frac{1}{2} & -\frac{139}{162} & -\frac{17}{54} & 0 & \frac{5}{9} & \frac{5}{9} & 0 & -6 & -2 \\
      \hline
      &
 -\frac{71}{144 \sqrt{3}} & -\frac{7}{72 \sqrt{3}} & \frac{7}{48 \sqrt{3}} & \frac{83}{144 \sqrt{3}} & \frac{\sqrt{3}}{8} & \frac{7}{144 \sqrt{3}} & -\frac{5}{4 \sqrt{3}} & -\frac{5}{12 \sqrt{3}} & \frac{5}{6 \sqrt{3}} & \frac{3 \sqrt{3}}{2} & -\frac{3 \sqrt{3}}{2} & 0 \\
      \hline
      &
 -1 & \frac{20}{27} & \frac{14}{81} & 0 & -\frac{44}{27} & -\frac{23}{81} & 0 & \frac{10}{3} & -\frac{5}{3} & 0 & -8 & 0 \\
      \hline
      \hline
      (1,\,+1)         &
 -\frac{44}{27} & -\frac{20}{27} & -\frac{1}{9} & -\frac{7}{27} & 0 & 0 & \frac{5}{3} & \frac{10}{9} & 0 & -8 & 0 & 0 \\
      \hline
      &
 -\frac{233}{1296} & -\frac{37}{648} & \frac{1}{432} & -\frac{115}{1296} & -\frac{143}{648} & -\frac{5}{432} & \frac{5}{18} & \frac{5}{12} & \frac{5}{36} & \frac{1}{6} & -\frac{1}{6} & 0 \\
      \hline
      &
 0 & -\frac{7}{54} & 0 & -\frac{31}{54} & -\frac{89}{54} & -\frac{7}{18} & \frac{10}{9} & 0 & \frac{5}{3} & -1 & -4 & -3 \\
      \hline
      \hline
      (\frac12,\,0)    &
 -\frac{11}{27} & -\frac{10}{9} & -\frac{5}{18} & -\frac{7}{18} & 0 & 0 & \frac{5}{18} & \frac{5}{3} & 0 & -4 & -4 & 0 \\
      \hline
      &
 -\frac{1}{36} & -\frac{10}{27} & -\frac{1}{36} & -\frac{7}{36} & \frac{7}{27} & -\frac{7}{36} & -\frac{5}{12} & \frac{5}{3} & -\frac{5}{12} & 2 & -2 & 0 \\
      \hline
      &
 \frac{233}{432 \sqrt{6}} & \frac{37}{216 \sqrt{6}} & -\frac{1}{144 \sqrt{6}} & \frac{115}{432 \sqrt{6}} & \frac{143}{216 \sqrt{6}} & \frac{5}{144 \sqrt{6}} & -\frac{5}{6 \sqrt{6}} & -\frac{5}{4 \sqrt{6}} & -\frac{5}{12 \sqrt{6}} & -\frac{1}{2 \sqrt{6}} & \frac{1}{2 \sqrt{6}} & 0 \\
      \hline
      &
 -\frac{1}{2} & \frac{10}{9} & \frac{8}{81} & 0 & -\frac{22}{9} & -\frac{133}{162} & 0 & 5 & -\frac{5}{2} & 0 & -8 & 0 \\
      \hline
      &
 -\frac{71}{144 \sqrt{6}} & -\frac{7}{72 \sqrt{6}} & \frac{7}{48 \sqrt{6}} & \frac{83}{144 \sqrt{6}} & \frac{1}{8} \sqrt{\frac{3}{2}} & \frac{7}{144 \sqrt{6}} & -\frac{5}{4 \sqrt{6}} & -\frac{5}{12 \sqrt{6}} & \frac{5}{6 \sqrt{6}} & \frac{3}{2} \sqrt{\frac{3}{2}} & -\frac{3}{2} \sqrt{\frac{3}{2}} & 0 \\
      \hline
      &
 0 & -\frac{28}{81} & 0 & -\frac{10}{9} & -\frac{47}{81} & -\frac{4}{27} & \frac{5}{3} & -\frac{5}{18} & \frac{5}{9} & -\frac{9}{2} & -3 & -\frac{1}{2} \\
      \hline
      \hline
      (\frac32,\,0)    &    
 -\frac{11}{27} & -\frac{10}{9} & -\frac{5}{18} & -\frac{7}{18} & 0 & 0 & \frac{5}{18} & \frac{5}{3} & 0 & -4 & -4 & 0 \\
      \hline
      \hline
      (0,\,-1)         &
 0 & -\frac{35}{162} & 0 & -\frac{11}{18} & -\frac{331}{162} & -\frac{23}{54} & \frac{5}{3} & -\frac{5}{18} & \frac{20}{9} & -\frac{3}{2} & -3 & -\frac{7}{2} \\
      \hline
      \hline
      (1,\,-1)         &
 0 & -\frac{7}{162} & 0 & -\frac{29}{54} & -\frac{203}{162} & -\frac{19}{54} & \frac{5}{9} & \frac{5}{18} & \frac{10}{9} & -\frac{1}{2} & -5 & -\frac{5}{2} \\
      \hline
    \end{tabular}
 \caption{The coefficients $C^{(5,\pi)}_Q$, $C^{(5,K)}_Q$ and $C^{(5,t)}_Q$, $C^{(5,su)}_Q$ from  (\ref{res-T4-tadpole}). }
\label{tab:Clebsch:n5}
\end{table}

We continue with contributions from the s- and u-channel exchange diagrams that are proportional to their corresponding scalar bubble integrals 
with \begin{eqnarray}
&& f^2\,T^{(4-su)}_{ab}(s,t,u) = \sum_{QH}\,\frac{s-u}{f^2}\,\Bigg\{ \Big[2\,(m_a^2 +m_b^2) + M_a^2+M_b^2 -2\,M_H^2 - 2\,t  \Big]  \,C_{QH}^{(s)} 
\nonumber\\
&& \qquad \quad  + \,2\,B_0\,m \,\big( c_0 \,C^{(s-0),\pi}_{QH} + c_1 \,C^{(s-1),\pi}_{QH}\big)
+ B_0\,(m+m_s) \,\big( c_0\, C^{(s-0),K}_{QH} +  c_1\,C^{(s-1),K}_{QH}\big)
\nonumber\\
&& \qquad \quad  + \left(\frac{s-u}{4\,M_H} \right)^2 \Big[- 4\,C_{QH}^{(s)} + (c_{2} + c_4 )\,C_{QH}^{(s-2)}  + (c_{3} + c_5)\,C_{QH}^{(s-3)} \Big]  \Bigg\} \,\bar I_{QH}(s)
\nonumber\\
&& \qquad - \sum_{QH}\,\frac{s-u}{f^2}\,\Bigg\{  \Big[2\,(m_a^2 +m_b^2) + M_a^2+M_b^2 -2\,M_H^2 - 2\,t   \Big] \,C_{QH}^{(u)} 
\nonumber\\
&& \qquad \quad  + \,2\,B_0\,m \,\big( c_0 \,C^{(u-0),\pi}_{QH} + c_1 \,C^{(u-1),\pi}_{QH}\big)
+ B_0\,( m+ m_s) \,\big(c_0\, C^{(u-0),K}_{QH} +c_1\,  C^{(u-1),K}_{QH}\big)
\nonumber\\
&& \qquad \quad  + \left( \frac{s-u}{4\,M_H} \right)^2 \Big[- 4\,C_{QH}^{(u)} + (c_{2} + c_4 )\,C_{QH}^{(u-2)}  + (c_{3} + c_5)\,C_{QH}^{(u-3)} \Big] \Bigg\}\,\bar I_{QH}(u)\,,
\label{res-T4-su-channel}
\end{eqnarray}
where we encounter a further set of Clebsch $C^{(s \cdots)}_{QH}$ and $C^{(u \cdots )}_{QH}$. Their specific form can be easily derived from our $C_{\rm WT}$, $C_{0-3}$ coefficients. Therefore we refrain from constructing further tables to display them here.

\begin{table}[]
  \tabcolsep=3.4mm
  \center
    \begin{tabular}{|C||C|C|C||C|C|C||C|C|C|}
      \hline
       (I,\,S)
      & \multicolumn{3}{C||}{
        {C}^{(\chi),\pi}_{PQ}
      }
      & \multicolumn{3}{C||}{
        {C}^{(\chi),{K}}_{PQ}
      }
       & \multicolumn{3}{C|}{
        {C}^{( -)}_{PQ}
      }
      \\
      &
      {\pi{K}} & {\pi\eta} & {{K}\eta} & 
      {\pi{K}} & {\pi\eta} & {{K}\eta} & 
      {\pi{K}} & {\pi\eta} & {{K}\eta} 
      \\
      \hline
      \hline
    (0,\,+1)_{12}      &
 \frac{1}{8 \sqrt{3}} & 0 & \frac{1}{8 \sqrt{3}} & -\frac{1}{8 \sqrt{3}} & 0 & -\frac{\sqrt{3}}{8} & \frac{-\sqrt{3}}{16} & 0 & \frac{\sqrt{3}}{16} \\
      \hline
      \hline
      (1,\,+1)_{12}      &
 -\frac{1}{24} & 0 & \frac{1}{24} & -\frac{1}{24} & 0 & -\frac{1}{24} & -\frac{5}{48} & 0 & -\frac{1}{16} \\
      \hline
      \hline
      (\frac12,\,0)_{12} &
 0 & 0 & 0 & 0 & 0 & 0 & 0 & 0 & 0 \\
      \hline
      (\frac12,\,0)_{13} &
 -\frac{1}{8 \sqrt{6}} & 0 & \frac{1}{8 \sqrt{6}} & -\frac{1}{8 \sqrt{6}} & 0 & -\frac{1}{8 \sqrt{6}} & -\frac{5}{16 \sqrt{6}} & 0 & -\frac{1}{16} \sqrt{\frac{3}{2}} \\
      \hline
      (\frac12,0)_{23} &
 \frac{1}{8 \sqrt{6}} & 0 & \frac{1}{8 \sqrt{6}} & -\frac{1}{8 \sqrt{6}} & 0 & -\frac{1}{8} \sqrt{\frac{3}{2}} & -\frac{1}{16} \sqrt{\frac{3}{2}} & 0 & \frac{1}{16} \sqrt{\frac{3}{2}} \\
      \hline
      \hline
    \end{tabular}
  \caption{The coefficients  $C^{(\chi),\pi}_{PQ}$, $C^{(\chi),{K}}_{PQ}$ and $C^{(-)}_{PQ}$ from  (\ref{res-T4-t-channel}) for given isospin (I) and strangeness (S). Note that $C^{(+)}_{PQ} = 3\,C^{(t)}_{PQ}/4$ with $C^{(t)}_{PQ}$ already specified in Tab. \ref{tab:coeffB}.  }
  \label{tab:Clebsch:tWT}
\end{table}

\begin{table}[]
  \tabcolsep=2.8mm
  \center
  \begin{tabular}{|C||C|C||C|C|C||C|C||C|C|C|C|C|}
    \hline
    (I,\,S)
    & \multicolumn{2}{C||}{
      {C}^{(\chi-0),\pi\pi}_{PQ}
    }
    & \multicolumn{3}{C||}{
      {C}^{(\chi-0),\pi{K}}_{PQ}
    }
    & \multicolumn{2}{C||}{
      {C}^{(\chi-0),{K}{K}}_{PQ}
    }
    & \multicolumn{5}{C|}{
      {C}^{(\chi-1),\pi\pi}_{PQ}
    }
    \\
    &
    {\pi\pi} & {\eta\eta} &
    {\pi\pi} & {{K}{K}} & {\eta\eta} &
    {{K}{K}} & {\eta\eta} &
    {\pi\pi} & {\pi{K}} & {\pi\eta} &
    {{K}\eta} & {\eta\eta}
    \\
    \hline
    \hline
    (\frac12,\,+2)   &
 2 & \frac{2}{9} & 2 & 0 & -\frac{14}{9} & 8 & \frac{8}{3} & -1 & 0 & 0 & 0 & \frac{1}{9} \\
    \hline
    \hline
    (0,\,+1)         &
 2 & \frac{2}{9} & 2 & 0 & -\frac{14}{9} & 8 & \frac{8}{3} & -\frac{1}{2} & 0 & \frac{1}{3} & 0 & -\frac{1}{6} \\
    \hline
    &
 0 & 0 & 0 & 0 & 0 & 0 & 0 & 0 & -\frac{1}{2 \sqrt{3}} & 0 & \frac{1}{2 \sqrt{3}} & 0 \\
    \hline
    &
 4 & \frac{28}{27} & 0 & -\frac{8}{3} & -\frac{176}{27} & 8 & \frac{256}{27} & -2 & 0 & 0 & 0 & \frac{14}{27} \\
    \hline
    \hline
    (1,\,+1)         &
 \frac{20}{3} & -\frac{4}{9} & 0 & \frac{8}{3} & \frac{16}{9} & \frac{8}{3} & 0 & -\frac{10}{3} & 0 & 0 & 0 & -\frac{2}{9} \\
    \hline
    &
 0 & 0 & 0 & 0 & 0 & 0 & 0 & 0 & -\frac{1}{6} & 0 & -\frac{1}{6} & 0 \\
    \hline
    &
 2 & \frac{2}{9} & 2 & 0 & -\frac{14}{9} & 8 & \frac{8}{3} & -\frac{1}{2} & 0 & -\frac{1}{9} & 0 & -\frac{1}{6} \\
    \hline
    \hline
    (\frac12,\,0)    &
 \frac{20}{3} & -\frac{4}{9} & 0 & \frac{8}{3} & \frac{16}{9} & \frac{8}{3} & 0 & -\frac{5}{3} & 0 & 0 & 0 & \frac{1}{3} \\
    \hline
    &
 0 & 0 & 0 & 0 & 0 & 0 & 0 & 0 & 0 & -\frac{2}{3} & 0 & 0 \\
    \hline
    &
 0 & 0 & 0 & 0 & 0 & 0 & 0 & 0 & \frac{1}{2 \sqrt{6}} & 0 & \frac{1}{2 \sqrt{6}} & 0 \\
    \hline
    &
 4 & \frac{28}{27} & 0 & -\frac{8}{3} & -\frac{176}{27} & 8 & \frac{256}{27} & -1 & 0 & 0 & 0 & -\frac{7}{9} \\
    \hline
    &
 0 & 0 & 0 & 0 & 0 & 0 & 0 & 0 & -\frac{1}{2 \sqrt{6}} & 0 & \frac{1}{2 \sqrt{6}} & 0 \\
    \hline
    &
 2 & \frac{2}{9} & 2 & 0 & -\frac{14}{9} & 8 & \frac{8}{3} & -1 & 0 & 0 & 0 & \frac{1}{9} \\
    \hline
    \hline
    (\frac32,\,0)    &    
 \frac{20}{3} & -\frac{4}{9} & 0 & \frac{8}{3} & \frac{16}{9} & \frac{8}{3} & 0 & -\frac{5}{3} & 0 & 0 & 0 & \frac{1}{3} \\
    \hline
    \hline
    (0,\,-1)         &
 2 & \frac{2}{9} & 2 & 0 & -\frac{14}{9} & 8 & \frac{8}{3} & -\frac{1}{2} & 0 & -\frac{1}{3} & 0 & -\frac{1}{6} \\
    \hline
    \hline
    (1,\,-1)         &
 2 & \frac{2}{9} & 2 & 0 & -\frac{14}{9} & 8 & \frac{8}{3} & -\frac{1}{2} & 0 & \frac{1}{9} & 0 & -\frac{1}{6} \\
    \hline
  \end{tabular}
  \caption{
   The coefficients $C^{(\chi-0),\pi\pi}_{PQ}$, $C^{(\chi-0),\pi{K}}_{PQ}$, $C^{(\chi-0),{K}{K}}_{PQ}$ and $C^{(\chi-1),{\pi}{\pi}}_{PQ}$
    from (\ref{res-T4-t-channel}). Coefficients that are zero are not shown always.
  }
  \label{tab:Clebsch:t0xx12}
\end{table}

\begin{table}[]
  \tabcolsep=2.6mm
  \center
  \begin{tabular}{|C||C|C|C|C|C|C||C|C|C|C|C|C|}
    \hline
    (I,\,S)
    & \multicolumn{6}{C||}{
      {C}^{(\chi-1),\pi{K}}_{PQ}
    }
    & \multicolumn{6}{C|}{
      {C}^{(\chi-1),{K}{K}}_{PQ}
    }
    \\
    &
    {\pi\pi} & {\pi{K}} & {\pi\eta} &
    {{K}{K}} & {{K}\eta} & {\eta\eta} &
    {\pi\pi} & {\pi{K}} & {\pi\eta} &
    {{K}{K}} & {{K}\eta} & {\eta\eta}
    \\
    \hline
    \hline
    (\frac12,\,+2)   &
 -1 & 0 & 0 & 0 & 0 & -\frac{1}{3} & 0 & 0 & 0 & -2 & 0 & 0 \\
    \hline
    \hline
    (0,\,+1)         &
 -\frac{1}{2} & 0 & -\frac{1}{3} & 0 & 0 & \frac{17}{18} & 0 & 0 & 0 & -2 & 0 & -\frac{4}{3} \\
    \hline
    &
 0 & 0 & 0 & 0 & -\frac{7}{3 \sqrt{3}} & 0 & 0 & \frac{1}{2 \sqrt{3}} & 0 & 0 & \frac{5}{2 \sqrt{3}} & 0 \\
    \hline
    &
 0 & 0 & 0 & \frac{2}{3} & 0 & -\frac{32}{27} & 0 & 0 & 0 & -2 & 0 & 0 \\
    \hline
    \hline
    (1,\,+1)         &
 0 & 0 & 0 & -\frac{2}{3} & 0 & 0 & 0 & 0 & 0 & -\frac{2}{3} & 0 & 0 \\
    \hline
    &
 0 & -\frac{1}{3} & 0 & 0 & \frac{4}{9} & 0 & 0 & -\frac{1}{6} & 0 & 0 & -\frac{5}{18} & 0 \\
    \hline
    &
 -\frac{1}{2} & 0 & \frac{1}{9} & 0 & 0 & \frac{17}{18} & 0 & 0 & 0 & -\frac{10}{3} & 0 & -\frac{4}{3} \\
    \hline
    \hline
    (\frac12,\,0)    &
 0 & 0 & 0 & -1 & 0 & -\frac{8}{9} & 0 & 0 & 0 & -1 & 0 & 0 \\
    \hline
    &
 0 & 0 & 0 & -\frac{1}{3} & 0 & 0 & 0 & 0 & 0 & \frac{1}{3} & 0 & 0 \\
    \hline
    &
 0 & \frac{1}{\sqrt{6}} & 0 & 0 & -\frac{4}{3 \sqrt{6}} & 0 & 0 & \frac{1}{2 \sqrt{6}} & 0 & 0 & \frac{5}{6 \sqrt{6}} & 0 \\
    \hline
    &
 0 & 0 & 0 & 1 & 0 & \frac{104}{27} & 0 & 0 & 0 & -3 & 0 & -\frac{128}{27} \\
    \hline
    &
 0 & 0 & 0 & 0 & -\frac{7}{3 \sqrt{6}} & 0 & 0 & \frac{1}{2 \sqrt{6}} & 0 & 0 & \frac{5}{2 \sqrt{6}} & 0 \\
    \hline
    &
 -1 & 0 & 0 & 0 & 0 & -\frac{1}{3} & 0 & 0 & 0 & -2 & 0 & 0 \\
    \hline
    \hline
    (\frac32,\,0)    &    
 0 & 0 & 0 & -1 & 0 & -\frac{8}{9} & 0 & 0 & 0 & -1 & 0 & 0 \\
    \hline
    \hline
    (0,\,-1)         &
 -\frac{1}{2} & 0 & \frac{1}{3} & 0 & 0 & \frac{17}{18} & 0 & 0 & 0 & -4 & 0 & -\frac{4}{3} \\
    \hline
    \hline
    (1,\,-1)         &
 -\frac{1}{2} & 0 & -\frac{1}{9} & 0 & 0 & \frac{17}{18} & 0 & 0 & 0 & -\frac{8}{3} & 0 & -\frac{4}{3} \\
    \hline
  \end{tabular}
  \caption{
  The coefficients $C^{(\chi-1),\pi{K}}_{PQ}$ and $C^{(\chi-1),{K}{K}}_{PQ}$
    from (\ref{res-T4-t-channel}).
  }
  \label{tab:Clebsch:t1xx12}
\end{table}

\begin{table}[]
  \tabcolsep=3.1mm
  \center
    \begin{tabular}{|C||C|C|C|C|C|C||C|C|C|C|C|C|}
      \hline
      (I,\,S)
      & \multicolumn{6}{C||}{
        {C}^{(\chi-2),\pi}_{PQ}
      }
      & \multicolumn{6}{C|}{
        {C}^{(\chi-2),{K}}_{PQ}
      }
      \\
      &
      {\pi\pi} & {\pi{K}} & {\pi\eta} &
      {{K}{K}} & {{K}\eta} & {\eta\eta} &
      {\pi\pi} & {\pi{K}} & {\pi\eta} &
      {{K}{K}} & {{K}\eta} & {\eta\eta}
      \\
      \hline
      \hline
      (\frac12,\,+2)   &
 1 & 0 & 0 & 0 & 0 & -\frac{1}{3} & 1 & 0 & 0 & 4 & 0 & 1 \\
      \hline
      \hline
      (0,\,+1)         &
 1 & 0 & 0 & 0 & 0 & -\frac{1}{3} & 1 & 0 & 0 & 4 & 0 & 1 \\
      \hline
      &
 0 & 0 & 0 & 0 & 0 & 0 & 0 & 0 & 0 & 0 & 0 & 0 \\
      \hline
      &
 2 & 0 & 0 & -\frac{4}{3} & 0 & -\frac{14}{9} & 0 & 0 & 0 & 4 & 0 & \frac{32}{9} \\
      \hline
      \hline
      (1,\,+1)         &
 \frac{10}{3} & 0 & 0 & \frac{4}{3} & 0 & \frac{2}{3} & 0 & 0 & 0 & \frac{4}{3} & 0 & 0 \\
      \hline
      &
 0 & 0 & 0 & 0 & 0 & 0 & 0 & 0 & 0 & 0 & 0 & 0 \\
      \hline
      &
 1 & 0 & 0 & 0 & 0 & -\frac{1}{3} & 1 & 0 & 0 & 4 & 0 & 1 \\
      \hline
      \hline
      (\frac12,\,0)    &
 \frac{10}{3} & 0 & 0 & \frac{4}{3} & 0 & \frac{2}{3} & 0 & 0 & 0 & \frac{4}{3} & 0 & 0 \\
      \hline
      &
 0 & 0 & 0 & 0 & 0 & 0 & 0 & 0 & 0 & 0 & 0 & 0 \\
      \hline
      &
 0 & 0 & 0 & 0 & 0 & 0 & 0 & 0 & 0 & 0 & 0 & 0 \\
      \hline
      &
 2 & 0 & 0 & -\frac{4}{3} & 0 & -\frac{14}{9} & 0 & 0 & 0 & 4 & 0 & \frac{32}{9} \\
      \hline
      &
 0 & 0 & 0 & 0 & 0 & 0 & 0 & 0 & 0 & 0 & 0 & 0 \\
      \hline
      &
 1 & 0 & 0 & 0 & 0 & -\frac{1}{3} & 1 & 0 & 0 & 4 & 0 & 1 \\
      \hline
      \hline
      (\frac32,\,0)    &    
 \frac{10}{3} & 0 & 0 & \frac{4}{3} & 0 & \frac{2}{3} & 0 & 0 & 0 & \frac{4}{3} & 0 & 0 \\
      \hline
      \hline
      (0,\,-1)         &
 1 & 0 & 0 & 0 & 0 & -\frac{1}{3} & 1 & 0 & 0 & 4 & 0 & 1 \\
      \hline
      \hline
      (1,\,-1)         &
 1 & 0 & 0 & 0 & 0 & -\frac{1}{3} & 1 & 0 & 0 & 4 & 0 & 1 \\
      \hline
    \end{tabular}
  \caption{The coefficients  $C^{(\chi-2),\pi}_{PQ}$ and $C^{(\chi-2),{K}}_{PQ}$ from  (\ref{res-T4-t-channel}).  }
  \label{tab:Clebsch:t2x}
\end{table}

\begin{table}[]
  \tabcolsep=2.5mm
  \center
    \begin{tabular}{|C||C|C|C|C|C|C||C|C|C|C|C|C|}
      \hline
      (I,\,S)
      & \multicolumn{6}{C||}{
        {C}^{(\chi-3),\pi}_{PQ}
      }
      & \multicolumn{6}{C|}{
        {C}^{(\chi-3),{K}}_{PQ}
      }
      \\
      &
      {\pi\pi} & {\pi{K}} & {\pi\eta} &
      {{K}{K}} & {{K}\eta} & {\eta\eta} &
      {\pi\pi} & {\pi{K}} & {\pi\eta} &
      {{K}{K}} & {{K}\eta} & {\eta\eta}
      \\
      \hline
      \hline
      (\frac12,\,+2)   &
 \frac{1}{2} & 0 & 0 & 0 & 0 & -\frac{1}{18} & \frac{1}{2} & 0 & 0 & 1 & 0 & \frac{1}{6} \\
      \hline
      \hline
      (0,\,+1)         &
 \frac{1}{4} & 0 & -\frac{1}{6} & 0 & 0 & -\frac{5}{36} & \frac{1}{4} & 0 & \frac{1}{6} & 1 & 0 & \frac{5}{12} \\
      \hline
      &
 0 & \frac{1}{2 \sqrt{3}} & 0 & 0 & \frac{1}{6 \sqrt{3}} & 0 & 0 & -\frac{1}{2 \sqrt{3}} & 0 & 0 & -\frac{1}{2 \sqrt{3}} & 0 \\
      \hline
      &
 1 & 0 & 0 & -\frac{1}{3} & 0 & -\frac{7}{27} & 0 & 0 & 0 & 1 & 0 & \frac{16}{27} \\
      \hline
      \hline
      (1,\,+1)         &
 \frac{5}{3} & 0 & 0 & \frac{1}{3} & 0 & \frac{1}{9} & 0 & 0 & 0 & \frac{1}{3} & 0 & 0 \\
      \hline
      &
 0 & \frac{1}{6} & 0 & 0 & -\frac{1}{18} & 0 & 0 & \frac{1}{6} & 0 & 0 & \frac{1}{18} & 0 \\
      \hline
      &
 \frac{1}{4} & 0 & \frac{1}{18} & 0 & 0 & -\frac{5}{36} & \frac{1}{4} & 0 & -\frac{1}{18} & \frac{5}{3} & 0 & \frac{5}{12} \\
      \hline
      \hline
      (\frac12,\,0)    &
 \frac{5}{6} & 0 & 0 & \frac{1}{2} & 0 & \frac{5}{18} & 0 & 0 & 0 & \frac{1}{2} & 0 & 0 \\
      \hline
      &
 0 & 0 & \frac{1}{3} & \frac{1}{6} & 0 & 0 & 0 & 0 & 0 & -\frac{1}{6} & 0 & 0 \\
      \hline
      &
 0 & -\frac{1}{2 \sqrt{6}} & 0 & 0 & \frac{1}{6 \sqrt{6}} & 0 & 0 & -\frac{1}{2 \sqrt{6}} & 0 & 0 & -\frac{1}{6 \sqrt{6}} & 0 \\
      \hline
      &
 \frac{1}{2} & 0 & 0 & -\frac{1}{2} & 0 & -\frac{35}{54} & 0 & 0 & 0 & \frac{3}{2} & 0 & \frac{40}{27} \\
      \hline
      &
 0 & \frac{1}{2 \sqrt{6}} & 0 & 0 & \frac{1}{6 \sqrt{6}} & 0 & 0 & -\frac{1}{2 \sqrt{6}} & 0 & 0 & -\frac{1}{2 \sqrt{6}} & 0 \\
      \hline
      &
 \frac{1}{2} & 0 & 0 & 0 & 0 & -\frac{1}{18} & \frac{1}{2} & 0 & 0 & 1 & 0 & \frac{1}{6} \\
      \hline
      \hline
      (\frac32,\,0)    &    
 \frac{5}{6} & 0 & 0 & \frac{1}{2} & 0 & \frac{5}{18} & 0 & 0 & 0 & \frac{1}{2} & 0 & 0 \\
      \hline
      \hline
      (0,\,-1)         &
 \frac{1}{4} & 0 & \frac{1}{6} & 0 & 0 & -\frac{5}{36} & \frac{1}{4} & 0 & -\frac{1}{6} & 2 & 0 & \frac{5}{12} \\
      \hline
      \hline
      (1,\,-1)         &
 \frac{1}{4} & 0 & -\frac{1}{18} & 0 & 0 & -\frac{5}{36} & \frac{1}{4} & 0 & \frac{1}{18} & \frac{4}{3} & 0 & \frac{5}{12} \\
      \hline
    \end{tabular}
  \caption{ The coefficients  $C^{(\chi-3),\pi}_{PQ}$ and $C^{(\chi-3),{K}}_{PQ}$ from  (\ref{res-T4-t-channel}).  }
  \label{tab:Clebsch:t3x}
\end{table}

\begin{table}[]
  \tabcolsep=3.0mm
  \center
    \begin{tabular}{|C||C|C|C|C|C|C||C|C|C|C|C|C|}
      \hline
      (I,\,S)
      & \multicolumn{6}{C||}{
        {C}^{(t+0),\pi}_{PQ}
      }
      & \multicolumn{6}{C|}{
        {C}^{(t+0),{K}}_{PQ}
      }
      \\
      &
      {\pi\pi} & {\pi{K}} & {\pi\eta} &
      {{K}{K}} & {{K}\eta} & {\eta\eta} &
      {\pi\pi} & {\pi{K}} & {\pi\eta} &
      {{K}{K}} & {{K}\eta} & {\eta\eta}
      \\
      \hline
      \hline
      (\frac12,\,+2)   &
 2 & 0 & 0 & 0 & 0 & -\frac{2}{3} & 0 & 0 & 0 & 4 & 0 & \frac{8}{3} \\
      \hline
      \hline
      (0,\,+1)         &
 2 & 0 & 0 & 0 & 0 & -\frac{2}{3} & 0 & 0 & 0 & 4 & 0 & \frac{8}{3} \\
      \hline
      &
 0 & 0 & 0 & 0 & 0 & 0 & 0 & 0 & 0 & 0 & 0 & 0 \\
      \hline
      &
 0 & 0 & 0 & 0 & 0 & 0 & 0 & 0 & 0 & 8 & 0 & 0 \\
      \hline
      \hline
      (1,\,+1)         &
 \frac{16}{3} & 0 & 0 & 0 & 0 & 0 & 0 & 0 & 0 & \frac{8}{3} & 0 & 0 \\
      \hline
      &
 0 & 0 & 0 & 0 & 0 & 0 & 0 & 0 & 0 & 0 & 0 & 0 \\
      \hline
      &
 2 & 0 & 0 & 0 & 0 & -\frac{2}{3} & 0 & 0 & 0 & 4 & 0 & \frac{8}{3} \\
      \hline
      \hline
      (\frac12,\,0)    &
 \frac{16}{3} & 0 & 0 & 0 & 0 & 0 & 0 & 0 & 0 & \frac{8}{3} & 0 & 0 \\
      \hline
      &
 0 & 0 & 0 & 0 & 0 & 0 & 0 & 0 & 0 & 0 & 0 & 0 \\
      \hline
      &
 0 & 0 & 0 & 0 & 0 & 0 & 0 & 0 & 0 & 0 & 0 & 0 \\
      \hline
      &
 0 & 0 & 0 & 0 & 0 & 0 & 0 & 0 & 0 & 8 & 0 & 0 \\
      \hline
      &
 0 & 0 & 0 & 0 & 0 & 0 & 0 & 0 & 0 & 0 & 0 & 0 \\
      \hline
      &
 2 & 0 & 0 & 0 & 0 & -\frac{2}{3} & 0 & 0 & 0 & 4 & 0 & \frac{8}{3} \\
      \hline
      \hline
      (\frac32,\,0)    &    
 \frac{16}{3} & 0 & 0 & 0 & 0 & 0 & 0 & 0 & 0 & \frac{8}{3} & 0 & 0 \\
      \hline
      \hline
      (0,\,-1)         &
 2 & 0 & 0 & 0 & 0 & -\frac{2}{3} & 0 & 0 & 0 & 4 & 0 & \frac{8}{3} \\
      \hline
      \hline
      (1,\,-1)         &
 2 & 0 & 0 & 0 & 0 & -\frac{2}{3} & 0 & 0 & 0 & 4 & 0 & \frac{8}{3} \\
      \hline
      \hline
      (I,\,S)
      & \multicolumn{6}{C||}{
        {C}^{(t-0),\pi}_{PQ}
      }
      & \multicolumn{6}{C|}{
        {C}^{(t-0),{K}}_{PQ}
      }
      \\
         &
      {\pi\pi} & {\pi{K}} & {\pi\eta} &
      {{K}{K}} & {{K}\eta} & {\eta\eta} &
      {\pi\pi} & {\pi{K}} & {\pi\eta} &
      {{K}{K}} & {{K}\eta} & {\eta\eta}
      \\
      \hline
      \hline
      (0,\,+1)_{12}      &
      0 & 0 & 0 & 0 & 0 & 0 &
      0 & 0 & 0 & 0 & 0 & 0 \\
      \hline
      \hline
      (1,\,+1)_{12}      &
      0 & 0 & 0 & 0 & 0 & 0 &
      0 & 0 & 0 & 0 & 0 & 0 \\
      \hline
      \hline
      (\frac12,\,0)_{12} &
      0 & 0 & 0 & 0 & 0 & 0 &
      0 & 0 & 0 & 0 & 0 & 0 \\
      \hline
      (\frac12,\,0)_{13} &
      0 & 0 & 0 & 0 & 0 & 0 &
      0 & 0 & 0 & 0 & 0 & 0 \\
      \hline
      (\frac12,0)_{23} &
      0 & 0 & 0 & 0 & 0 & 0 &
      0 & 0 & 0 & 0 & 0 & 0 \\
      \hline
      \hline
    \end{tabular}
   \caption{The coefficients $C^{(t\pm 0 ), \pi}_{QP}$ and $C^{(t\pm 0 ), K}_{QP}$ from  (\ref{res-T4-t-channel}).  }
\label{tab:Clebsch:t0}
\end{table}

\begin{table}[]
  \tabcolsep=2.1mm
  \center
    \begin{tabular}{|C||C|C|C|C|C|C||C|C|C|C|C|C|}
      \hline
      (I,\,S)
      & \multicolumn{6}{C||}{
        {C}^{(t+1),\pi}_{PQ}
      }
      & \multicolumn{6}{C|}{
        {C}^{(t+1),{K}}_{PQ}
      }
      \\
      &
      {\pi\pi} & {\pi{K}} & {\pi\eta} &
      {{K}{K}} & {{K}\eta} & {\eta\eta} &
      {\pi\pi} & {\pi{K}} & {\pi\eta} &
      {{K}{K}} & {{K}\eta} & {\eta\eta}
      \\
      \hline
      \hline
      (\frac12,\,+2)   &
 -1 & 0 & 0 & 0 & 0 & -\frac{1}{3} & 0 & 0 & 0 & -1 & 0 & 0 \\
      \hline
      \hline
      (0,\,+1)         &
 -\frac{1}{2} & 0 & 1 & 0 & 0 & \frac{1}{2} & 0 & 0 & 0 & -1 & 0 & -\frac{4}{3} \\
      \hline
      &
 0 & \frac{\sqrt{3}}{4} & 0 & 0 & \frac{\sqrt{3}}{4} & 0 & 0 & \frac{\sqrt{3}}{4} & 0 & 0 & -\frac{5}{4 \sqrt{3}} & 0 \\
      \hline
      &
 0 & 0 & 0 & 0 & 0 & 0 & 0 & 0 & 0 & -2 & 0 & 0 \\
      \hline
      \hline
      (1,\,+1)         &
 -\frac{8}{3} & 0 & 0 & 0 & 0 & 0 & 0 & 0 & 0 & -\frac{2}{3} & 0 & 0 \\
      \hline
      &
 0 & -\frac{5}{12} & 0 & 0 & \frac{1}{4} & 0 & 0 & -\frac{5}{12} & 0 & 0 & -\frac{5}{12} & 0 \\
      \hline
      &
 -\frac{1}{2} & 0 & -\frac{1}{3} & 0 & 0 & \frac{1}{2} & 0 & 0 & 0 & -\frac{5}{3} & 0 & -\frac{4}{3} \\
      \hline
      \hline
      (\frac12,\,0)    &
 -\frac{4}{3} & 0 & 0 & 0 & 0 & 0 & 0 & 0 & 0 & -1 & 0 & 0 \\
      \hline
      &
 0 & 0 & 0 & 0 & 0 & 0 & 0 & 0 & 0 & -1 & 0 & 0 \\
      \hline
      &
 0 & \frac{5}{4 \sqrt{6}} & 0 & 0 & -\frac{1}{4} \sqrt{\frac{3}{2}} & 0 & 0 & \frac{5}{4 \sqrt{6}} & 0 & 0 & \frac{5}{4 \sqrt{6}} & 0 \\
      \hline
      &
 0 & 0 & 0 & 0 & 0 & 0 & 0 & 0 & 0 & -3 & 0 & 0 \\
      \hline
      &
 0 & \frac{1}{4} \sqrt{\frac{3}{2}} & 0 & 0 & \frac{1}{4} \sqrt{\frac{3}{2}} & 0 & 0 & \frac{1}{4} \sqrt{\frac{3}{2}} & 0 & 0 & -\frac{5}{4 \sqrt{6}} & 0 \\
      \hline
      &
 -1 & 0 & 0 & 0 & 0 & -\frac{1}{3} & 0 & 0 & 0 & -1 & 0 & 0 \\
      \hline
      \hline
      (\frac32,\,0)    &    
 -\frac{4}{3} & 0 & 0 & 0 & 0 & 0 & 0 & 0 & 0 & -1 & 0 & 0 \\
      \hline
      \hline
      (0,\,-1)         &
 -\frac{1}{2} & 0 & -1 & 0 & 0 & \frac{1}{2} & 0 & 0 & 0 & -2 & 0 & -\frac{4}{3} \\
      \hline
      \hline
      (1,\,-1)         &
 -\frac{1}{2} & 0 & \frac{1}{3} & 0 & 0 & \frac{1}{2} & 0 & 0 & 0 & -\frac{4}{3} & 0 & -\frac{4}{3} \\
      \hline
      \hline
      (I,\,S)
      & \multicolumn{6}{C||}{
        {C}^{(t-1),\pi}_{PQ}
      }
      & \multicolumn{6}{C|}{
        {C}^{(t-1),{K}}_{PQ}
      }
      \\
      &
      {\pi\pi} & {\pi{K}} & {\pi\eta} &
      {{K}{K}} & {{K}\eta} & {\eta\eta} &
      {\pi\pi} & {\pi{K}} & {\pi\eta} &
      {{K}{K}} & {{K}\eta} & {\eta\eta}
      \\
      \hline
      \hline
      (0,\,+1)_{12}      &
 0 & -\frac{3 \sqrt{3}}{8} & 0 & 0 & \frac{3 \sqrt{3}}{8} & 0 &
 0 & -\frac{3 \sqrt{3}}{8} & 0 & 0 & -\frac{5 \sqrt{3}}{8} & 0 \\
      \hline
      \hline
      (1,\,+1)_{12}      &
 0 & \frac{3}{8} & 0 & 0 & -\frac{3}{8} & 0 &
 0 & \frac{3}{8} & 0 & 0 & \frac{5}{8} & 0 \\
      \hline
      \hline
      (\frac12,\,0)_{12} &
      0 & 0 & 0 & 0 & 0 & 0 &
      0 & 0 & 0 & 0 & 0 & 0 \\
      \hline
      (\frac12,\,0)_{13} &
 0 & -\frac{3}{8} \sqrt{\frac{3}{2}} & 0 & 0 & \frac{3}{8} \sqrt{\frac{3}{2}} & 0 &
 0 & -\frac{3}{8} \sqrt{\frac{3}{2}} & 0 & 0 & -\frac{5}{8} \sqrt{\frac{3}{2}} & 0 \\
      \hline
      (\frac12,0)_{23} &
 0 & \frac{3}{8} \sqrt{\frac{3}{2}} & 0 & 0 & -\frac{3}{8} \sqrt{\frac{3}{2}} & 0 &
 0 & \frac{3}{8} \sqrt{\frac{3}{2}} & 0 & 0 & \frac{5}{8} \sqrt{\frac{3}{2}} & 0 \\
      \hline
      \hline
    \end{tabular}
 \caption{The coefficients $C^{(t\pm 1 ), \pi}_{QP}$ and $C^{(t\pm 1 ), K}_{QP}$ from  (\ref{res-T4-t-channel}).  }
\label{tab:Clebsch:t1}
\end{table}

\begin{table}[]
  \tabcolsep=2.6mm
  \center
    \begin{tabular}{|C||C|C|C|C|C|C||C|C|C|C|C|C|}
      \hline
      (I,\,S)
      & \multicolumn{6}{C||}{
        {C}^{(t+2)}_{PQ}
      }
      & \multicolumn{6}{C|}{
        {C}^{(t+3)}_{PQ}
      }
      \\
      &
      {\pi\pi} & {\pi{K}} & {\pi\eta} &
      {{K}{K}} & {{K}\eta} & {\eta\eta} &
      {\pi\pi} & {\pi{K}} & {\pi\eta} &
      {{K}{K}} & {{K}\eta} & {\eta\eta}
      \\
      \hline
      \hline
      (\frac12,\,+2)   &
 1 & 0 & 0 & 2 & 0 & 1 & \frac{1}{2} & 0 & 0 & \frac{1}{2} & 0 & \frac{1}{6} \\
      \hline
      \hline
      (0,\,+1)         &
 1 & 0 & 0 & 2 & 0 & 1 & \frac{1}{4} & 0 & -\frac{1}{2} & \frac{1}{2} & 0 & \frac{5}{12} \\
      \hline
      &
 0 & 0 & 0 & 0 & 0 & 0 & 0 & -\frac{\sqrt{3}}{4} & 0 & 0 & \frac{1}{4 \sqrt{3}} & 0 \\
      \hline
      &
 0 & 0 & 0 & 4 & 0 & 0 & 0 & 0 & 0 & 1 & 0 & 0 \\
      \hline
      \hline
      (1,\,+1)         &
 \frac{8}{3} & 0 & 0 & \frac{4}{3} & 0 & 0 & \frac{4}{3} & 0 & 0 & \frac{1}{3} & 0 & 0 \\
      \hline
      &
 0 & 0 & 0 & 0 & 0 & 0 & 0 & \frac{5}{12} & 0 & 0 & \frac{1}{12} & 0 \\
      \hline
      &
 1 & 0 & 0 & 2 & 0 & 1 & \frac{1}{4} & 0 & \frac{1}{6} & \frac{5}{6} & 0 & \frac{5}{12} \\
      \hline
      \hline
      (\frac12,\,0)    &
 \frac{8}{3} & 0 & 0 & \frac{4}{3} & 0 & 0 & \frac{2}{3} & 0 & 0 & \frac{1}{2} & 0 & 0 \\
      \hline
      &
 0 & 0 & 0 & 0 & 0 & 0 & 0 & 0 & 0 & \frac{1}{2} & 0 & 0 \\
      \hline
      &
 0 & 0 & 0 & 0 & 0 & 0 & 0 & -\frac{5}{4 \sqrt{6}} & 0 & 0 & -\frac{1}{4 \sqrt{6}} & 0 \\
      \hline
      &
 0 & 0 & 0 & 4 & 0 & 0 & 0 & 0 & 0 & \frac{3}{2} & 0 & 0 \\
      \hline
      &
 0 & 0 & 0 & 0 & 0 & 0 & 0 & -\frac{\sqrt{\frac{3}{2}}}{4} & 0 & 0 & \frac{1}{4 \sqrt{6}} & 0 \\
      \hline
      &
 1 & 0 & 0 & 2 & 0 & 1 & \frac{1}{2} & 0 & 0 & \frac{1}{2} & 0 & \frac{1}{6} \\
      \hline
      \hline
      (\frac32,\,0)    &    
 \frac{8}{3} & 0 & 0 & \frac{4}{3} & 0 & 0 & \frac{2}{3} & 0 & 0 & \frac{1}{2} & 0 & 0 \\
      \hline
      \hline
      (0,\,-1)         &
 1 & 0 & 0 & 2 & 0 & 1 & \frac{1}{4} & 0 & \frac{1}{2} & 1 & 0 & \frac{5}{12} \\
      \hline
      \hline
      (1,\,-1)         &
 1 & 0 & 0 & 2 & 0 & 1 & \frac{1}{4} & 0 & -\frac{1}{6} & \frac{2}{3} & 0 & \frac{5}{12} \\
      \hline
      \hline
       (I,\,S)
      & \multicolumn{6}{C||}{
        {C}^{(t-2)}_{PQ}
      }
      & \multicolumn{6}{C|}{
        {C}^{(t-3)}_{PQ}
      }
      \\
      &
      {\pi\pi} & {\pi{K}} & {\pi\eta} &
      {{K}{K}} & {{K}\eta} & {\eta\eta} &
      {\pi\pi} & {\pi{K}} & {\pi\eta} &
      {{K}{K}} & {{K}\eta} & {\eta\eta}
      \\
      \hline
      \hline
      (0,\,+1)_{12}      &
      0 & 0 & 0 & 0 & 0 & 0 &
 0 & \frac{3 \sqrt{3}}{8} & 0 & 0 & \frac{\sqrt{3}}{8} & 0 \\
      \hline
      \hline
      (1,\,+1)_{12}      &
      0 & 0 & 0 & 0 & 0 & 0 &
 0 & -\frac{3}{8} & 0 & 0 & -\frac{1}{8} & 0 \\
      \hline
      \hline
      (\frac12,\,0)_{12} &
      0 & 0 & 0 & 0 & 0 & 0 &
 0 & 0 & 0 & 0 & 0 & 0 \\
      \hline
      (\frac12,\,0)_{13} &
      0 & 0 & 0 & 0 & 0 & 0 &
 0 & \frac{3}{8} \sqrt{\frac{3}{2}} & 0 & 0 & \frac{1}{8} \sqrt{\frac{3}{2}} & 0 \\
      \hline
      (\frac12,0)_{23} &
      0 & 0 & 0 & 0 & 0 & 0 &
 0 & -\frac{3}{8} \sqrt{\frac{3}{2}} & 0 & 0 & -\frac{1}{8} \sqrt{\frac{3}{2}} & 0 \\
      \hline
      \hline
    \end{tabular}
 \caption{The coefficients $C^{(t\pm 2 )}_{QP}$ and $C^{(t\pm 3 )}_{QP}$  from  (\ref{res-T4-t-channel}).  }
\label{tab:Clebsch:t23}
\end{table}

We close this appendix with a discussion of contributions form t-channel processes. Again we systematically split such terms into tadpole-type terms proportional to $\bar I_Q$, that are already considered in (\ref{res-T4-tadpole}), and bubble-type terms proportional to $\bar I_{PQ}(t)$. However, the particular and instrumental form of the tadpole terms in (\ref{res-T4-tadpole})  requires to do so in a specific manner. A direct evaluation of the t-channel loops  
leads to structures of the form $m_P^2\,\bar I_Q/m_Q^2$ or $m_Q^2\,\bar I_P/m_P^2$, that are not part of (\ref{res-T4-tadpole}). Nevertheless, they can be transformed into the generic tadpoles used in (\ref{res-T4-tadpole}), however, with extra finite scale-invariant terms generated. It is therefore convenient to express our results in terms of more general bubble integrals $J_{ab,PQ}^{(\cdots)}(t)$ and 
associated Clebsch coefficients $C_{ab,PQ}^{(\cdots)}$. We find
\begin{eqnarray}
&& f^4\,T^{(4-t)}_{ab}(s,t,u) = \sum_{PQ} \,\Big\{ 
\Big(2 \,B_0\,m\,C_{PQ}^{(\chi),\pi} + B_0\,(m + m_s)\,C_{PQ}^{(\chi ) ,K} \Big) \,J^{(\chi )}_{ab,PQ}(t)
\nonumber\\
&& \qquad \qquad \qquad   +\, C_{PQ}^{( +)}\,J^{(+)}_{ab,PQ}(t) +\, C_{PQ}^{(-)}\,J^{(-)}_{ab,PQ}(t)
\Big\}
\nonumber\\
&& \quad +\,\sum_{n=0}^1 c_n\,\sum_{PQ} \,
\Big\{ \Big( 2\,B_0\,m\,C_{PQ}^{(t+n),\pi} + B_0\,(m + m_s)\,C_{PQ}^{(t+n),K} \Big)\,J^{(n+)}_{ab,PQ}(t)
\nonumber\\
&& \qquad \qquad \qquad \;\; +\, \Big(2 \,B_0\,m\,C_{PQ}^{(t-n),\pi} + B_0\,(m + m_s)\,C_{PQ}^{(t-n),K} \Big)\, J^{(n-)}_{ab,PQ}(t)
\nonumber\\
&& \qquad \qquad   \qquad \;\; +\, 
 \Big( 4\,B_0^2\,m^2\,C_{PQ}^{(\chi-n),\pi \pi} + 2\,B_0^2\,m\,(m + m_s)\,C_{PQ}^{(\chi-n),\pi K }
 \nonumber\\
&& \qquad \qquad   \qquad \qquad \qquad \qquad \;\; \;\; +\, 
B_0^2\,(m + m_s)^2\,C_{PQ}^{(\chi-n),K K }\Big) \,J^{(\chi-n )}_{ab,PQ}(t)
\Big\} 
\nonumber\\
&& \quad +\,\sum_{n=2}^5 c_n\,\sum_{PQ} \,\Big\{
 \Big(2 \,B_0\,m\,C_{PQ}^{(\chi-n),\pi} + B_0\,(m + m_s)\,C_{PQ}^{(\chi-n),K} \Big)\,  J^{(\chi-n )}_{ab,PQ}(t)
\nonumber\\
&& \qquad \qquad +\,
 C_{PQ}^{(t+n)}\,J^{(n+)}_{ab,PQ}(t) + C_{PQ}^{(t-n)}\,J^{(n-)}_{ab,PQ}(t)
\Big\} \,,
\label{res-T4-t-channel}
\end{eqnarray}
with
\begin{eqnarray}
&&  C_{PQ}^{(t\pm 4)}=C_{PQ}^{(t\pm 2)}\,, \qquad C_{PQ}^{(t\pm 5)}=C_{PQ}^{(t\pm 3)} \,,\qquad 
C_{PQ}^{(\chi- 4)}=C_{PQ}^{(\chi- 2)}\,,\qquad  
C_{PQ}^{(\chi-5)}=C_{PQ}^{(\chi- 3)}\,,
\nonumber\\
&& J^{(n \pm )}_{ab,PQ}(t) = \pm  J^{(n \pm)}_{ba,PQ}(t) \,, \qquad  \qquad 
 \big[ C_{PQ}^{(t\pm n)} \big]_{ab} = \pm \, \big[ C_{PQ}^{(t\pm n)} \big]_{ba} \,,
\end{eqnarray} 
and 
\begin{eqnarray}
&&  J^{(\chi )}_{ab,PQ}(t) =   \Big[ (M^2_a-M^2_b)\,(m_P^2-m_Q^2)\,\frac{\bar I_{PQ}(t)}{t} \Big]_{\rm reg} \,,
\nonumber\\
&& J^{(+)}_{ab,PQ}(t) =\Big[ (M^2_a-M^2_b)\,(m_a^2-m_b^2)\, (4\,(m_P^2-m_Q^2)^2 -2\,t\,(m_P^2+m_Q^2)+ t^2 )\,\frac{ \bar I_{PQ}(t)}{3\,t^2} \Big]_{\rm reg } 
\nonumber\\
&& \qquad \qquad - \,
(M_a^2 -M_b^2)\,\frac{m_a^2}{6\,(4\pi)^2}\,\log \frac{m_P^2\,m_Q^2}{m_a^4} 
+(M_a^2 -M_b^2)\,\frac{m_b^2}{6\,(4\pi)^2}\,\log \frac{m_P^2\,m_Q^2}{m_b^4} 
\nonumber\\
&& \qquad \qquad  +\, \frac{2}{9}\,\frac{(M_a^2 -M_b^2)\,(m_a^2 -m_b^2)}{(4\pi)^2}\,,
\nonumber\\
&& J^{(-)}_{ab,PQ}(t) = \Big[ (M^2_a-M^2_b)\, (m_P^2-m_Q^2)\,(3\,t -m_P^2-m_Q^2-m_a^2 -m_b^2)\,\frac{ \bar I_{PQ}(t)}{2\,t} \Big]_{\rm reg }
\nonumber\\
&& \qquad \qquad + \,
3\,(M_a^2 -M_b^2)\,\frac{m_P^2+m_Q^2}{4\,(4\pi)^2}\,\log \frac{m_P^2}{m_Q^2}\,,
\nonumber\\ \nonumber\\
&&  J^{(\chi-1 )}_{ab,PQ}(t) = J^{(\chi-0 )}_{ab,PQ}(t) =   \bar I_{PQ}(t) \,,
\nonumber\\
&&  J^{(1+)}_{ab,PQ}(t) =  J^{(0+)}_{ab,PQ}(t) = \frac{1}{2}\,\big(3\, t-  m_P^2-m_Q^2 -m_a^2- m_b^2  \big)\, \bar I_{PQ}(t) - \frac{m_P^2-m_Q^2}{4\,(4 \pi)^2}\,\log \frac{m_P^2}{m_Q^2}
\nonumber\\
&& \qquad \qquad -\,\frac{m_a^2}{2\,(4\pi)^2}\,\log \frac{m_P^2\,m_Q^2}{m_a^4}-\frac{m_b^2}{2\,(4\pi)^2}\,\log \frac{m_P^2\,m_Q^2}{m_b^4} \,,
\nonumber\\
&&  J^{(1-)}_{ab,PQ}(t) = J^{(0-)}_{ab,PQ}(t) =  \Big[ (m_a^2-m_b^2)\, ( m_P^2-m_Q^2 )\, \frac{ \bar I_{PQ}(t)}{t} \Big]_{\rm reg } \,,
\nonumber\\  \nonumber\\
&&  J^{(\chi-3 )}_{ab,PQ}(t) = J^{(\chi-2 )}_{ab,PQ}(t) = ( m_P^2+m_Q^2 -t )\, \bar I_{PQ}(t)
+ \frac{m_P^2-m_Q^2}{2\,(4 \pi)^2}\,\log \frac{m_P^2}{m_Q^2}
\nonumber\\
&& \qquad \qquad +\,\frac{m_a^2}{2\,(4\pi)^2}\,\log \frac{m_P^2\,m_Q^2}{m_a^4}+\frac{m_b^2}{2\,(4\pi)^2}\,\log \frac{m_P^2\,m_Q^2}{m_b^4} - \frac{(\bar q  \cdot q)}{(4\pi)^2}\,\log \frac{m_P^2\,m_Q^2}{m_a^2\,m_b^2}
\,,
\nonumber\\
&&  J^{(3+)}_{ab,PQ}(t) = J^{(2+)}_{ab,PQ}(t) = \frac{1}{2}\,\big(3\, t-  m_P^2-m_Q^2 -m_a^2- m_b^2  \big)\,\big( m_P^2+m_Q^2 - t\big)\, \bar I_{PQ}(t) 
\nonumber\\
&& \qquad \qquad -\,  \frac{m_P^4-m_Q^4}{4\,(4 \pi)^2}\,\log \frac{m_P^2}{m_Q^2}
+\,\frac{m_a^4}{2\,(4\pi)^2}\,\log \frac{m_P^2\,m_Q^2}{m_a^4}+\frac{m_b^4}{2\,(4\pi)^2}\,\log \frac{m_P^2\,m_Q^2}{m_b^4}
\nonumber\\
&& \qquad \qquad 
+\,\frac{m_a^2\,m_b^2}{(4\pi)^2}\,\log \frac{m_P^2\,m_Q^2}{m_a^2\,m_b^2}
-\frac{m_P^2\,m_a^2}{4\,(4 \pi)^2}\,\log \frac{m_P^2\,m_Q^6}{m_a^8} -  \frac{m_P ^2\,m_b^2}{4\,(4 \pi)^2}\,\log \frac{m_P^2\,m_Q^6}{m_b^8}
\nonumber\\
&& \qquad \qquad -\,
\frac{m_Q^2\,m_b^2}{4\,(4 \pi)^2}\,\log \frac{m_P^6\,m_Q^2}{m_b^8} -  \frac{m_Q^2\,m_a^2}{4\,(4 \pi)^2}\,\log \frac{m_P^6\,m_Q^2}{m_a^8} 
-\, 2\,(\bar q\cdot q)\, \frac{m_P^2-m_Q^2}{(4 \pi)^2}\,\log \frac{m_P^2}{m_Q^2}
\nonumber\\
&& \qquad \qquad 
-\,\frac{5}{2}\,\frac{(\bar q \cdot q)\,m_a^2}{(4\pi)^2}\,\log \frac{m_P^2\,m_Q^2}{m_a^4}-\frac{5}{2}\,\frac{(\bar q \cdot q)\,m_b^2}{(4\pi)^2}\,\log \frac{m_P^2\,m_Q^2}{m_b^4}
\nonumber\\
&& \qquad \qquad +\,3\,\frac{(\bar q\cdot q)^2}{(4\pi)^2}\, \log \frac{m_P^2 \,m_Q^2}{m_a^2\,m_b^2}
\,,
\nonumber\\
&&  J^{(3-)}_{ab,PQ}(t) =  J^{(2-)}_{ab,PQ}(t) =\Big[(m_a^2-m_b^2)\,  \big(m_P^2-m_Q^2\big)\,\big(m_P^2+m_Q^2- t  \big) \, \frac{\bar I_{PQ}(t)}{t}  \Big]_{\rm reg} 
\nonumber\\
&& \qquad \qquad +\,\frac{m_P^2\,m_a^2}{6\,(4 \pi)^2}\,\log \frac{m_P^2\,m_Q^6}{m_a^8} -  \frac{m_P ^2\,m_b^2}{6\,(4 \pi)^2}\,\log \frac{m_P^2\,m_Q^6}{m_b^8}
\nonumber\\
&& \qquad \qquad +\,
\frac{m_Q^2\,m_b^2}{6\,(4 \pi)^2}\,\log \frac{m_P^6\,m_Q^2}{m_b^8} -  \frac{m_Q^2\,m_a^2}{6\,(4 \pi)^2}\,\log \frac{m_P^6\,m_Q^2}{m_a^8}
\,,
\nonumber\\ \nonumber\\
&& J^{(\chi-5)}_{ab,PQ}(t)=J^{(\chi-4)}_{ab,PQ}(t)  = -\frac{2}{3}\,\Big[ p^2_{PQ}(t)  \bar I_{PQ}(t) \Big]_{\rm reg }
+ \frac{m_P^2-m_Q^2}{6\,(4 \pi)^2}\,\log \frac{m_P^2}{m_Q^2}
\nonumber\\
&& \qquad \qquad +\,\frac{m_a^2}{12\,(4\pi)^2}\,\log \frac{m_P^2\,m_Q^2}{m_a^4}+\frac{m_b^2}{12\,(4\pi)^2}\,\log \frac{m_P^2\,m_Q^2}{m_b^4} - 
\frac{( \bar q\cdot q)}{6\,(4\pi)^2}\,\log \frac{m_P^2\,m_Q^2}{m_a^2\,m_b^2}
\nonumber\\
&& \qquad \qquad
+ \, \frac{2\,(\bar q\cdot q) -m_a^2-m_b^2+3\,(m_P^2+m_Q^2)  }{9\,(4\pi)^2}\,,
\nonumber\\
&&  J^{(5+)}_{ab,PQ}(t) = J^{(4+)}_{ab,PQ}(t) = -\frac{1}{3}\,\Big[\big(3\,t- m_P^2-m_Q^2-m_a^2-m_b^2  \big)\, p^2_{PQ}(t)\,\bar I_{PQ}(t) \Big]_{\rm reg }
\nonumber\\
&& \qquad \qquad -\, 5\, \frac{m_P^4-m_Q^4}{24\,(4 \pi)^2}\,\log \frac{m_P^2}{m_Q^2}
+\,\frac{m_a^4}{12\,(4\pi)^2}\,\log \frac{m_P^2\,m_Q^2}{m_a^4}+\frac{m_b^4}{12\,(4\pi)^2}\,\log \frac{m_P^2\,m_Q^2}{m_b^4}
\nonumber\\
&& \qquad \qquad 
+\,\frac{m_a^2\,m_b^2}{6\,(4\pi)^2}\,\log \frac{m_P^2\,m_Q^2}{m_a^2\,m_b^2}
-5\,\frac{m_P^2\,m_a^2}{72\,(4 \pi)^2}\,\log \frac{m_P^2\,m_Q^6}{m_a^8} -  5\,\frac{m_P ^2\,m_b^2}{72\,(4 \pi)^2}\,\log \frac{m_P^2\,m_Q^6}{m_b^8} 
\nonumber\\
&& \qquad \qquad -\,5\,
\frac{m_Q^2\,m_b^2}{72\,(4 \pi)^2}\,\log \frac{m_P^6\,m_Q^2}{m_b^8} -  5\,\frac{m_Q^2\,m_a^2}{72\,(4 \pi)^2}\,\log \frac{m_P^6\,m_Q^2}{m_a^8}
\nonumber\\
&& \qquad \qquad - \,\frac{7}{24}\,\frac{m_P^4+m_Q^4}{(4\pi)^2}+
\frac{7}{18}\,\frac{(m_P^2+m_Q^2)\,(m_a^2+ m_b^2)}{(4\pi)^2} - \frac{1}{9}\,
\frac{(m_a^2+m_b^2)^2 +3\, m_P^2\,m_Q^2}{(4\pi)^2}
\nonumber\\
&& \qquad \qquad 
-\,\frac{5}{9}\,\big( 2\,(m_P^2+m_Q^2) -m_a^2-m_b^2\big)\,\frac{(\bar q \cdot q)}{(4\pi)^2}
-\, \frac{7}{12}\,(\bar q\cdot q)\, \frac{m_P^2-m_Q^2}{(4 \pi)^2}\,\log \frac{m_P^2}{m_Q^2}
\nonumber\\
&& \qquad \qquad 
-\,\frac{5}{12}\,\frac{(\bar q \cdot q)\,m_a^2}{(4\pi)^2}\,\log \frac{m_P^2\,m_Q^2}{m_a^4}-\frac{5}{12}\,\frac{(\bar q \cdot q)\,m_b^2}{(4\pi)^2}\,\log \frac{m_P^2\,m_Q^2}{m_b^4}
\nonumber\\
&& \qquad \qquad 
-\,\frac{2}{3}\,\frac{(\bar q \cdot q)^2}{(4\pi)^2} + \frac{(\bar q\cdot q)^2}{2\,(4\pi)^2}\, \log \frac{m_P^2 \,m_Q^2}{m_a^2\,m_b^2}\,,
\nonumber\\
&&  J^{(5-)}_{ab,PQ}(t) =J^{(4-)}_{ab,PQ}(t) =-\frac{2}{3}\, \Big[(m_a^2-m_b^2)\,(m_P^2-m_Q^2)\,p_{PQ}^2(t) \frac{\bar I_{PQ}(t)}{t}  \Big]_{\rm reg} 
\nonumber\\
&& \qquad \qquad +\,
\frac{m_P^2\,m_a^2}{36\,(4 \pi)^2}\,\log \frac{m_P^2\,m_Q^6}{m_a^8}- \, \frac{m_P ^2\,m_b^2}{36\,(4 \pi)^2}\,\log \frac{m_P^2\,m_Q^6}{m_b^8}
\nonumber\\
&& \qquad \qquad +\,
\frac{m_Q^2\,m_b^2}{36\,(4 \pi)^2}\,\log \frac{m_P^6\,m_Q^2}{m_b^8} -  \frac{m_Q^2\,m_a^2}{36\,(4 \pi)^2}\,\log \frac{m_P^6\,m_Q^2}{m_a^8} - \frac{(m_a^2-m_b^2)\,(m_P^2-m_Q^2)}{9\,(4\pi)^2}\,,
\nonumber\\ \nonumber\\
&&{\rm where} \qquad  \Big[ \frac{h(t)}{t^n} \Big]_{\rm reg } =  \frac{1}{t^n}\,\Big\{ 1 - \sum_{k=0}^{n-1}\,\frac{t^k}{ k !}\,\left( \frac{d }{d t} \right)^k \bigg|_{t=0} \Big\}\,h(t)\,.
\end{eqnarray}

\clearpage

\bibliography{literature}

\begin{thebibliography}{75}%
\makeatletter
\providecommand \@ifxundefined [1]{%
 \@ifx{#1\undefined}
}%
\providecommand \@ifnum [1]{%
 \ifnum #1\expandafter \@firstoftwo
 \else \expandafter \@secondoftwo
 \fi
}%
\providecommand \@ifx [1]{%
 \ifx #1\expandafter \@firstoftwo
 \else \expandafter \@secondoftwo
 \fi
}%
\providecommand \natexlab [1]{#1}%
\providecommand \enquote  [1]{``#1''}%
\providecommand \bibnamefont  [1]{#1}%
\providecommand \bibfnamefont [1]{#1}%
\providecommand \citenamefont [1]{#1}%
\providecommand \href@noop [0]{\@secondoftwo}%
\providecommand \href [0]{\begingroup \@sanitize@url \@href}%
\providecommand \@href[1]{\@@startlink{#1}\@@href}%
\providecommand \@@href[1]{\endgroup#1\@@endlink}%
\providecommand \@sanitize@url [0]{\catcode `\\12\catcode `\$12\catcode
  `\&12\catcode `\#12\catcode `\^12\catcode `\_12\catcode `\%12\relax}%
\providecommand \@@startlink[1]{}%
\providecommand \@@endlink[0]{}%
\providecommand \url  [0]{\begingroup\@sanitize@url \@url }%
\providecommand \@url [1]{\endgroup\@href {#1}{\urlprefix }}%
\providecommand \urlprefix  [0]{URL }%
\providecommand \Eprint [0]{\href }%
\providecommand \doibase [0]{http://dx.doi.org/}%
\providecommand \selectlanguage [0]{\@gobble}%
\providecommand \bibinfo  [0]{\@secondoftwo}%
\providecommand \bibfield  [0]{\@secondoftwo}%
\providecommand \translation [1]{[#1]}%
\providecommand \BibitemOpen [0]{}%
\providecommand \bibitemStop [0]{}%
\providecommand \bibitemNoStop [0]{.\EOS\space}%
\providecommand \EOS [0]{\spacefactor3000\relax}%
\providecommand \BibitemShut  [1]{\csname bibitem#1\endcsname}%
\let\auto@bib@innerbib\@empty
\bibitem [{\citenamefont {Casalbuoni}\ \emph {et~al.}(1997)\citenamefont
  {Casalbuoni}, \citenamefont {Deandrea}, \citenamefont {Di~Bartolomeo},
  \citenamefont {Gatto}, \citenamefont {Feruglio} \emph
  {et~al.}}]{Casalbuoni:1996pg}%
  \BibitemOpen
  \bibfield  {author} {\bibinfo {author} {\bibfnamefont {R.}~\bibnamefont
  {Casalbuoni}}, \bibinfo {author} {\bibfnamefont {A.}~\bibnamefont
  {Deandrea}}, \bibinfo {author} {\bibfnamefont {N.}~\bibnamefont
  {Di~Bartolomeo}}, \bibinfo {author} {\bibfnamefont {R.}~\bibnamefont
  {Gatto}}, \bibinfo {author} {\bibfnamefont {F.}~\bibnamefont {Feruglio}},
  \emph {et~al.},\ }\href {\doibase 10.1016/S0370-1573(96)00027-0} {\bibfield
  {journal} {\bibinfo  {journal} {Phys.Rept.}\ }\textbf {\bibinfo {volume}
  {281}},\ \bibinfo {pages} {145} (\bibinfo {year} {1997})},\ \Eprint
  {http://arxiv.org/abs/hep-ph/9605342} {arXiv:hep-ph/9605342 [hep-ph]}
  \BibitemShut {NoStop}%
\bibitem [{\citenamefont {Lutz}\ \emph {et~al.}(2016)\citenamefont {Lutz} \emph
  {et~al.}}]{Lutz:2015ejy}%
  \BibitemOpen
  \bibfield  {author} {\bibinfo {author} {\bibfnamefont {M.~F.~M.}\
  \bibnamefont {Lutz}} \emph {et~al.},\ }\href {\doibase
  10.1016/j.nuclphysa.2016.01.070} {\bibfield  {journal} {\bibinfo  {journal}
  {Nucl. Phys.}\ }\textbf {\bibinfo {volume} {A948}},\ \bibinfo {pages} {93}
  (\bibinfo {year} {2016})},\ \Eprint {http://arxiv.org/abs/1511.09353}
  {arXiv:1511.09353 [hep-ph]} \BibitemShut {NoStop}%
\bibitem [{\citenamefont {Chen}\ \emph {et~al.}(2017)\citenamefont {Chen},
  \citenamefont {Chen}, \citenamefont {Liu}, \citenamefont {Liu},\ and\
  \citenamefont {Zhu}}]{Chen:2016spr}%
  \BibitemOpen
  \bibfield  {author} {\bibinfo {author} {\bibfnamefont {H.-X.}\ \bibnamefont
  {Chen}}, \bibinfo {author} {\bibfnamefont {W.}~\bibnamefont {Chen}}, \bibinfo
  {author} {\bibfnamefont {X.}~\bibnamefont {Liu}}, \bibinfo {author}
  {\bibfnamefont {Y.-R.}\ \bibnamefont {Liu}}, \ and\ \bibinfo {author}
  {\bibfnamefont {S.-L.}\ \bibnamefont {Zhu}},\ }\href {\doibase
  10.1088/1361-6633/aa6420} {\bibfield  {journal} {\bibinfo  {journal} {Rept.
  Prog. Phys.}\ }\textbf {\bibinfo {volume} {80}},\ \bibinfo {pages} {076201}
  (\bibinfo {year} {2017})},\ \Eprint {http://arxiv.org/abs/1609.08928}
  {arXiv:1609.08928 [hep-ph]} \BibitemShut {NoStop}%
\bibitem [{\citenamefont {Kolomeitsev}\ and\ \citenamefont
  {Lutz}(2004)}]{Kolomeitsev:2003ac}%
  \BibitemOpen
  \bibfield  {author} {\bibinfo {author} {\bibfnamefont {E.}~\bibnamefont
  {Kolomeitsev}}\ and\ \bibinfo {author} {\bibfnamefont {M.~F.~M.}\
  \bibnamefont {Lutz}},\ }\href {\doibase 10.1016/j.physletb.2003.10.118}
  {\bibfield  {journal} {\bibinfo  {journal} {Phys.Lett.}\ }\textbf {\bibinfo
  {volume} {B582}},\ \bibinfo {pages} {39} (\bibinfo {year} {2004})},\ \Eprint
  {http://arxiv.org/abs/hep-ph/0307133} {arXiv:hep-ph/0307133 [hep-ph]}
  \BibitemShut {NoStop}%
\bibitem [{\citenamefont {Hofmann}\ and\ \citenamefont
  {Lutz}(2004)}]{Hofmann:2003je}%
  \BibitemOpen
  \bibfield  {author} {\bibinfo {author} {\bibfnamefont {J.}~\bibnamefont
  {Hofmann}}\ and\ \bibinfo {author} {\bibfnamefont {M.~F.~M.}\ \bibnamefont
  {Lutz}},\ }\href {\doibase 10.1016/j.nuclphysa.2003.12.013} {\bibfield
  {journal} {\bibinfo  {journal} {Nucl. Phys.}\ }\textbf {\bibinfo {volume}
  {A733}},\ \bibinfo {pages} {142} (\bibinfo {year} {2004})},\ \Eprint
  {http://arxiv.org/abs/hep-ph/0308263} {arXiv:hep-ph/0308263 [hep-ph]}
  \BibitemShut {NoStop}%
\bibitem [{\citenamefont {Lutz}\ and\ \citenamefont
  {Soyeur}(2008)}]{Lutz:2007sk}%
  \BibitemOpen
  \bibfield  {author} {\bibinfo {author} {\bibfnamefont {M.~F.~M.}\
  \bibnamefont {Lutz}}\ and\ \bibinfo {author} {\bibfnamefont {M.}~\bibnamefont
  {Soyeur}},\ }\href {\doibase 10.1016/j.nuclphysa.2008.09.003} {\bibfield
  {journal} {\bibinfo  {journal} {Nucl.Phys.}\ }\textbf {\bibinfo {volume}
  {A813}},\ \bibinfo {pages} {14} (\bibinfo {year} {2008})},\ \Eprint
  {http://arxiv.org/abs/0710.1545} {arXiv:0710.1545 [hep-ph]} \BibitemShut
  {NoStop}%
\bibitem [{\citenamefont {Liu}\ \emph {et~al.}(2013)\citenamefont {Liu},
  \citenamefont {Orginos}, \citenamefont {Guo}, \citenamefont {Hanhart},\ and\
  \citenamefont {Mei{\ss}ner}}]{Liu:2012zya}%
  \BibitemOpen
  \bibfield  {author} {\bibinfo {author} {\bibfnamefont {L.}~\bibnamefont
  {Liu}}, \bibinfo {author} {\bibfnamefont {K.}~\bibnamefont {Orginos}},
  \bibinfo {author} {\bibfnamefont {F.-K.}\ \bibnamefont {Guo}}, \bibinfo
  {author} {\bibfnamefont {C.}~\bibnamefont {Hanhart}}, \ and\ \bibinfo
  {author} {\bibfnamefont {U.-G.}\ \bibnamefont {Mei{\ss}ner}},\ }\href
  {\doibase 10.1103/PhysRevD.87.014508} {\bibfield  {journal} {\bibinfo
  {journal} {Phys.Rev.}\ }\textbf {\bibinfo {volume} {D87}},\ \bibinfo {pages}
  {014508} (\bibinfo {year} {2013})},\ \Eprint {http://arxiv.org/abs/1208.4535}
  {arXiv:1208.4535 [hep-lat]} \BibitemShut {NoStop}%
\bibitem [{\citenamefont {Altenbuchinger}\ \emph {et~al.}(2014)\citenamefont
  {Altenbuchinger}, \citenamefont {Geng},\ and\ \citenamefont
  {Weise}}]{Altenbuchinger:2013vwa}%
  \BibitemOpen
  \bibfield  {author} {\bibinfo {author} {\bibfnamefont {M.}~\bibnamefont
  {Altenbuchinger}}, \bibinfo {author} {\bibfnamefont {L.~S.}\ \bibnamefont
  {Geng}}, \ and\ \bibinfo {author} {\bibfnamefont {W.}~\bibnamefont {Weise}},\
  }\href {\doibase 10.1103/PhysRevD.89.014026} {\bibfield  {journal} {\bibinfo
  {journal} {Phys. Rev.}\ }\textbf {\bibinfo {volume} {D89}},\ \bibinfo {pages}
  {014026} (\bibinfo {year} {2014})},\ \Eprint {http://arxiv.org/abs/1309.4743}
  {arXiv:1309.4743 [hep-ph]} \BibitemShut {NoStop}%
\bibitem [{\citenamefont {Cleven}\ \emph {et~al.}(2014)\citenamefont {Cleven},
  \citenamefont {Grie{\ss}hammer}, \citenamefont {Guo}, \citenamefont
  {Hanhart},\ and\ \citenamefont {Mei{\ss}ner}}]{Cleven:2014oka}%
  \BibitemOpen
  \bibfield  {author} {\bibinfo {author} {\bibfnamefont {M.}~\bibnamefont
  {Cleven}}, \bibinfo {author} {\bibfnamefont {H.~W.}\ \bibnamefont
  {Grie{\ss}hammer}}, \bibinfo {author} {\bibfnamefont {F.-K.}\ \bibnamefont
  {Guo}}, \bibinfo {author} {\bibfnamefont {C.}~\bibnamefont {Hanhart}}, \ and\
  \bibinfo {author} {\bibfnamefont {U.-G.}\ \bibnamefont {Mei{\ss}ner}},\
  }\href {\doibase 10.1140/epja/i2014-14149-y} {\bibfield  {journal} {\bibinfo
  {journal} {Eur. Phys. J.}\ }\textbf {\bibinfo {volume} {A50}},\ \bibinfo
  {pages} {149} (\bibinfo {year} {2014})},\ \Eprint
  {http://arxiv.org/abs/1405.2242} {arXiv:1405.2242 [hep-ph]} \BibitemShut
  {NoStop}%
\bibitem [{\citenamefont {Du}\ \emph {et~al.}(2016)\citenamefont {Du},
  \citenamefont {Guo}, \citenamefont {Mei{\ss}ner},\ and\ \citenamefont
  {Yao}}]{Du:2016tgp}%
  \BibitemOpen
  \bibfield  {author} {\bibinfo {author} {\bibfnamefont {M.-L.}\ \bibnamefont
  {Du}}, \bibinfo {author} {\bibfnamefont {F.-K.}\ \bibnamefont {Guo}},
  \bibinfo {author} {\bibfnamefont {U.-G.}\ \bibnamefont {Mei{\ss}ner}}, \ and\
  \bibinfo {author} {\bibfnamefont {D.-L.}\ \bibnamefont {Yao}},\ }\href
  {\doibase 10.1103/PhysRevD.94.094037} {\bibfield  {journal} {\bibinfo
  {journal} {Phys. Rev.}\ }\textbf {\bibinfo {volume} {D94}},\ \bibinfo {pages}
  {094037} (\bibinfo {year} {2016})},\ \Eprint
  {http://arxiv.org/abs/1610.02963} {arXiv:1610.02963 [hep-ph]} \BibitemShut
  {NoStop}%
\bibitem [{\citenamefont {Huang}\ \emph {et~al.}(2022)\citenamefont {Huang},
  \citenamefont {Lin}, \citenamefont {Chen},\ and\ \citenamefont
  {Zhu}}]{Huang:2022cag}%
  \BibitemOpen
  \bibfield  {author} {\bibinfo {author} {\bibfnamefont {B.-L.}\ \bibnamefont
  {Huang}}, \bibinfo {author} {\bibfnamefont {Z.-Y.}\ \bibnamefont {Lin}},
  \bibinfo {author} {\bibfnamefont {K.}~\bibnamefont {Chen}}, \ and\ \bibinfo
  {author} {\bibfnamefont {S.-L.}\ \bibnamefont {Zhu}},\ }\href@noop {} {\
  (\bibinfo {year} {2022})},\ \Eprint {http://arxiv.org/abs/2205.02619}
  {arXiv:2205.02619 [hep-ph]} \BibitemShut {NoStop}%
\bibitem [{\citenamefont {Guo}\ \emph {et~al.}(2018)\citenamefont {Guo},
  \citenamefont {Heo},\ and\ \citenamefont {Lutz}}]{Guo:2018kno}%
  \BibitemOpen
  \bibfield  {author} {\bibinfo {author} {\bibfnamefont {X.-Y.}\ \bibnamefont
  {Guo}}, \bibinfo {author} {\bibfnamefont {Y.}~\bibnamefont {Heo}}, \ and\
  \bibinfo {author} {\bibfnamefont {M.~F.~M.}\ \bibnamefont {Lutz}},\ }\href
  {\doibase 10.1103/PhysRevD.98.014510} {\bibfield  {journal} {\bibinfo
  {journal} {Phys. Rev.}\ }\textbf {\bibinfo {volume} {D98}},\ \bibinfo {pages}
  {014510} (\bibinfo {year} {2018})},\ \Eprint
  {http://arxiv.org/abs/1801.10122} {arXiv:1801.10122 [hep-lat]} \BibitemShut
  {NoStop}%
\bibitem [{\citenamefont {Aoki}\ \emph {et~al.}(2009)\citenamefont {Aoki} \emph
  {et~al.}}]{Aoki:2008sm}%
  \BibitemOpen
  \bibfield  {author} {\bibinfo {author} {\bibfnamefont {S.}~\bibnamefont
  {Aoki}} \emph {et~al.} (\bibinfo {collaboration} {PACS-CS}),\ }\href
  {\doibase 10.1103/PhysRevD.79.034503} {\bibfield  {journal} {\bibinfo
  {journal} {Phys.Rev.}\ }\textbf {\bibinfo {volume} {D79}},\ \bibinfo {pages}
  {034503} (\bibinfo {year} {2009})},\ \Eprint {http://arxiv.org/abs/0807.1661}
  {arXiv:0807.1661 [hep-lat]} \BibitemShut {NoStop}%
\bibitem [{\citenamefont {Mohler}\ and\ \citenamefont
  {Woloshyn}(2011)}]{Mohler:2011ke}%
  \BibitemOpen
  \bibfield  {author} {\bibinfo {author} {\bibfnamefont {D.}~\bibnamefont
  {Mohler}}\ and\ \bibinfo {author} {\bibfnamefont {R.}~\bibnamefont
  {Woloshyn}},\ }\href {\doibase 10.1103/PhysRevD.84.054505} {\bibfield
  {journal} {\bibinfo  {journal} {Phys.Rev.}\ }\textbf {\bibinfo {volume}
  {D84}},\ \bibinfo {pages} {054505} (\bibinfo {year} {2011})},\ \Eprint
  {http://arxiv.org/abs/1103.5506} {arXiv:1103.5506 [hep-lat]} \BibitemShut
  {NoStop}%
\bibitem [{\citenamefont {Na}\ \emph {et~al.}(2012)\citenamefont {Na},
  \citenamefont {Davies}, \citenamefont {Follana}, \citenamefont {Lepage},\
  and\ \citenamefont {Shigemitsu}}]{Na:2012iu}%
  \BibitemOpen
  \bibfield  {author} {\bibinfo {author} {\bibfnamefont {H.}~\bibnamefont
  {Na}}, \bibinfo {author} {\bibfnamefont {C.~T.}\ \bibnamefont {Davies}},
  \bibinfo {author} {\bibfnamefont {E.}~\bibnamefont {Follana}}, \bibinfo
  {author} {\bibfnamefont {G.~P.}\ \bibnamefont {Lepage}}, \ and\ \bibinfo
  {author} {\bibfnamefont {J.}~\bibnamefont {Shigemitsu}},\ }\href {\doibase
  10.1103/PhysRevD.86.054510} {\bibfield  {journal} {\bibinfo  {journal}
  {Phys.Rev.}\ }\textbf {\bibinfo {volume} {D86}},\ \bibinfo {pages} {054510}
  (\bibinfo {year} {2012})},\ \Eprint {http://arxiv.org/abs/1206.4936}
  {arXiv:1206.4936 [hep-lat]} \BibitemShut {NoStop}%
\bibitem [{\citenamefont {Kalinowski}\ and\ \citenamefont
  {Wagner}(2015)}]{Kalinowski:2015bwa}%
  \BibitemOpen
  \bibfield  {author} {\bibinfo {author} {\bibfnamefont {M.}~\bibnamefont
  {Kalinowski}}\ and\ \bibinfo {author} {\bibfnamefont {M.}~\bibnamefont
  {Wagner}},\ }\href {\doibase 10.1103/PhysRevD.92.094508} {\bibfield
  {journal} {\bibinfo  {journal} {Phys. Rev.}\ }\textbf {\bibinfo {volume}
  {D92}},\ \bibinfo {pages} {094508} (\bibinfo {year} {2015})},\ \Eprint
  {http://arxiv.org/abs/1509.02396} {arXiv:1509.02396 [hep-lat]} \BibitemShut
  {NoStop}%
\bibitem [{\citenamefont {Cichy}\ \emph {et~al.}(2016)\citenamefont {Cichy},
  \citenamefont {Kalinowski},\ and\ \citenamefont {Wagner}}]{Cichy:2016bci}%
  \BibitemOpen
  \bibfield  {author} {\bibinfo {author} {\bibfnamefont {K.}~\bibnamefont
  {Cichy}}, \bibinfo {author} {\bibfnamefont {M.}~\bibnamefont {Kalinowski}}, \
  and\ \bibinfo {author} {\bibfnamefont {M.}~\bibnamefont {Wagner}},\ }\href
  {\doibase 10.1103/PhysRevD.94.094503} {\bibfield  {journal} {\bibinfo
  {journal} {Phys. Rev.}\ }\textbf {\bibinfo {volume} {D94}},\ \bibinfo {pages}
  {094503} (\bibinfo {year} {2016})},\ \Eprint
  {http://arxiv.org/abs/1603.06467} {arXiv:1603.06467 [hep-lat]} \BibitemShut
  {NoStop}%
\bibitem [{\citenamefont {Cheung}\ \emph {et~al.}(2016)\citenamefont {Cheung},
  \citenamefont {O'Hara}, \citenamefont {Moir}, \citenamefont {Peardon},
  \citenamefont {Ryan}, \citenamefont {Thomas},\ and\ \citenamefont
  {Tims}}]{Cheung:2016bym}%
  \BibitemOpen
  \bibfield  {author} {\bibinfo {author} {\bibfnamefont {G.~K.~C.}\
  \bibnamefont {Cheung}}, \bibinfo {author} {\bibfnamefont {C.}~\bibnamefont
  {O'Hara}}, \bibinfo {author} {\bibfnamefont {G.}~\bibnamefont {Moir}},
  \bibinfo {author} {\bibfnamefont {M.}~\bibnamefont {Peardon}}, \bibinfo
  {author} {\bibfnamefont {S.~M.}\ \bibnamefont {Ryan}}, \bibinfo {author}
  {\bibfnamefont {C.~E.}\ \bibnamefont {Thomas}}, \ and\ \bibinfo {author}
  {\bibfnamefont {D.}~\bibnamefont {Tims}} (\bibinfo {collaboration} {Hadron
  Spectrum}),\ }\href {\doibase 10.1007/JHEP12(2016)089} {\bibfield  {journal}
  {\bibinfo  {journal} {JHEP}\ }\textbf {\bibinfo {volume} {12}},\ \bibinfo
  {pages} {089} (\bibinfo {year} {2016})},\ \Eprint
  {http://arxiv.org/abs/1610.01073} {arXiv:1610.01073 [hep-lat]} \BibitemShut
  {NoStop}%
\bibitem [{\citenamefont {Moir}\ \emph {et~al.}(2016)\citenamefont {Moir},
  \citenamefont {Peardon}, \citenamefont {Ryan}, \citenamefont {Thomas},\ and\
  \citenamefont {Wilson}}]{Moir:2016srx}%
  \BibitemOpen
  \bibfield  {author} {\bibinfo {author} {\bibfnamefont {G.}~\bibnamefont
  {Moir}}, \bibinfo {author} {\bibfnamefont {M.}~\bibnamefont {Peardon}},
  \bibinfo {author} {\bibfnamefont {S.~M.}\ \bibnamefont {Ryan}}, \bibinfo
  {author} {\bibfnamefont {C.~E.}\ \bibnamefont {Thomas}}, \ and\ \bibinfo
  {author} {\bibfnamefont {D.~J.}\ \bibnamefont {Wilson}},\ }\href {\doibase
  10.1007/JHEP10(2016)011} {\bibfield  {journal} {\bibinfo  {journal} {JHEP}\
  }\textbf {\bibinfo {volume} {10}},\ \bibinfo {pages} {011} (\bibinfo {year}
  {2016})},\ \Eprint {http://arxiv.org/abs/1607.07093} {arXiv:1607.07093
  [hep-lat]} \BibitemShut {NoStop}%
\bibitem [{\citenamefont {Guo}\ \emph {et~al.}(2021)\citenamefont {Guo},
  \citenamefont {Heo},\ and\ \citenamefont {Lutz}}]{Guo:2021kdo}%
  \BibitemOpen
  \bibfield  {author} {\bibinfo {author} {\bibfnamefont {X.-Y.}\ \bibnamefont
  {Guo}}, \bibinfo {author} {\bibfnamefont {Y.}~\bibnamefont {Heo}}, \ and\
  \bibinfo {author} {\bibfnamefont {M.~F.~M.}\ \bibnamefont {Lutz}},\ }in\
  \href@noop {} {\emph {\bibinfo {booktitle} {{38th International Symposium on
  Lattice Field Theory}}}}\ (\bibinfo {year} {2021})\ \Eprint
  {http://arxiv.org/abs/2107.12284} {arXiv:2107.12284 [hep-lat]} \BibitemShut
  {NoStop}%
\bibitem [{\citenamefont {Cheung}\ \emph {et~al.}(2021)\citenamefont {Cheung},
  \citenamefont {Thomas}, \citenamefont {Wilson}, \citenamefont {Moir},
  \citenamefont {Peardon},\ and\ \citenamefont {Ryan}}]{Cheung:2020mql}%
  \BibitemOpen
  \bibfield  {author} {\bibinfo {author} {\bibfnamefont {G.~K.~C.}\
  \bibnamefont {Cheung}}, \bibinfo {author} {\bibfnamefont {C.~E.}\
  \bibnamefont {Thomas}}, \bibinfo {author} {\bibfnamefont {D.~J.}\
  \bibnamefont {Wilson}}, \bibinfo {author} {\bibfnamefont {G.}~\bibnamefont
  {Moir}}, \bibinfo {author} {\bibfnamefont {M.}~\bibnamefont {Peardon}}, \
  and\ \bibinfo {author} {\bibfnamefont {S.~M.}\ \bibnamefont {Ryan}} (\bibinfo
  {collaboration} {Hadron Spectrum}),\ }\href {\doibase
  10.1007/JHEP02(2021)100} {\bibfield  {journal} {\bibinfo  {journal} {JHEP}\
  }\textbf {\bibinfo {volume} {02}},\ \bibinfo {pages} {100} (\bibinfo {year}
  {2021})},\ \Eprint {http://arxiv.org/abs/2008.06432} {arXiv:2008.06432
  [hep-lat]} \BibitemShut {NoStop}%
\bibitem [{\citenamefont {Gayer}\ \emph {et~al.}(2021)\citenamefont {Gayer},
  \citenamefont {Lang}, \citenamefont {Ryan}, \citenamefont {Tims},
  \citenamefont {Thomas},\ and\ \citenamefont {Wilson}}]{Gayer:2021xzv}%
  \BibitemOpen
  \bibfield  {author} {\bibinfo {author} {\bibfnamefont {L.}~\bibnamefont
  {Gayer}}, \bibinfo {author} {\bibfnamefont {N.}~\bibnamefont {Lang}},
  \bibinfo {author} {\bibfnamefont {S.~M.}\ \bibnamefont {Ryan}}, \bibinfo
  {author} {\bibfnamefont {D.}~\bibnamefont {Tims}}, \bibinfo {author}
  {\bibfnamefont {C.~E.}\ \bibnamefont {Thomas}}, \ and\ \bibinfo {author}
  {\bibfnamefont {D.~J.}\ \bibnamefont {Wilson}} (\bibinfo {collaboration}
  {Hadron Spectrum}),\ }\href {\doibase 10.1007/JHEP07(2021)123} {\bibfield
  {journal} {\bibinfo  {journal} {JHEP}\ }\textbf {\bibinfo {volume} {07}},\
  \bibinfo {pages} {123} (\bibinfo {year} {2021})},\ \Eprint
  {http://arxiv.org/abs/2102.04973} {arXiv:2102.04973 [hep-lat]} \BibitemShut
  {NoStop}%
\bibitem [{\citenamefont {Gasparyan}\ and\ \citenamefont
  {Lutz}(2010)}]{Gasparyan:2010xz}%
  \BibitemOpen
  \bibfield  {author} {\bibinfo {author} {\bibfnamefont {A.}~\bibnamefont
  {Gasparyan}}\ and\ \bibinfo {author} {\bibfnamefont {M.~F.~M.}\ \bibnamefont
  {Lutz}},\ }\href {\doibase 10.1016/j.nuclphysa.2010.08.006} {\bibfield
  {journal} {\bibinfo  {journal} {Nucl.Phys.}\ }\textbf {\bibinfo {volume}
  {A848}},\ \bibinfo {pages} {126} (\bibinfo {year} {2010})},\ \Eprint
  {http://arxiv.org/abs/1003.3426} {arXiv:1003.3426 [hep-ph]} \BibitemShut
  {NoStop}%
\bibitem [{\citenamefont {Danilkin}\ \emph
  {et~al.}(2011{\natexlab{a}})\citenamefont {Danilkin}, \citenamefont
  {Gasparyan},\ and\ \citenamefont {Lutz}}]{Danilkin:2010xd}%
  \BibitemOpen
  \bibfield  {author} {\bibinfo {author} {\bibfnamefont {I.}~\bibnamefont
  {Danilkin}}, \bibinfo {author} {\bibfnamefont {A.}~\bibnamefont {Gasparyan}},
  \ and\ \bibinfo {author} {\bibfnamefont {M.~F.~M.}\ \bibnamefont {Lutz}},\
  }\href {\doibase 10.1016/j.physletb.2011.01.036} {\bibfield  {journal}
  {\bibinfo  {journal} {Phys.Lett.}\ }\textbf {\bibinfo {volume} {B697}},\
  \bibinfo {pages} {147} (\bibinfo {year} {2011}{\natexlab{a}})},\ \Eprint
  {http://arxiv.org/abs/1009.5928} {arXiv:1009.5928 [hep-ph]} \BibitemShut
  {NoStop}%
\bibitem [{\citenamefont {Danilkin}\ \emph
  {et~al.}(2011{\natexlab{b}})\citenamefont {Danilkin}, \citenamefont {Gil},\
  and\ \citenamefont {Lutz}}]{Danilkin:2011fz}%
  \BibitemOpen
  \bibfield  {author} {\bibinfo {author} {\bibfnamefont {I.~V.}\ \bibnamefont
  {Danilkin}}, \bibinfo {author} {\bibfnamefont {L.~I.~R.}\ \bibnamefont
  {Gil}}, \ and\ \bibinfo {author} {\bibfnamefont {M.~F.~M.}\ \bibnamefont
  {Lutz}},\ }\href {\doibase 10.1016/j.physletb.2011.08.001} {\bibfield
  {journal} {\bibinfo  {journal} {Phys. Lett. B}\ }\textbf {\bibinfo {volume}
  {703}},\ \bibinfo {pages} {504} (\bibinfo {year} {2011}{\natexlab{b}})},\
  \Eprint {http://arxiv.org/abs/1106.2230} {arXiv:1106.2230 [hep-ph]}
  \BibitemShut {NoStop}%
\bibitem [{\citenamefont {Gasparyan}\ \emph {et~al.}(2012)\citenamefont
  {Gasparyan}, \citenamefont {Lutz},\ and\ \citenamefont
  {Epelbaum}}]{Gasparyan:2012km}%
  \BibitemOpen
  \bibfield  {author} {\bibinfo {author} {\bibfnamefont {A.~M.}\ \bibnamefont
  {Gasparyan}}, \bibinfo {author} {\bibfnamefont {M.~F.~M.}\ \bibnamefont
  {Lutz}}, \ and\ \bibinfo {author} {\bibfnamefont {E.}~\bibnamefont
  {Epelbaum}},\ }\href {\doibase 10.1140/epja/i2013-13115-7} {\bibfield
  {journal} {\bibinfo  {journal} {Eur. Phys. J. A}\ }\textbf {\bibinfo {volume}
  {49}},\ \bibinfo {pages} {115} (\bibinfo {year} {2012})},\ \Eprint
  {http://arxiv.org/abs/1212.3057} {arXiv:1212.3057 [nucl-th]} \BibitemShut
  {NoStop}%
\bibitem [{\citenamefont {Yan}\ \emph {et~al.}(1992)\citenamefont {Yan},
  \citenamefont {Cheng}, \citenamefont {Cheung}, \citenamefont {Lin},
  \citenamefont {Lin},\ and\ \citenamefont {Yu}}]{Yan:1992gz}%
  \BibitemOpen
  \bibfield  {author} {\bibinfo {author} {\bibfnamefont {T.-M.}\ \bibnamefont
  {Yan}}, \bibinfo {author} {\bibfnamefont {H.-Y.}\ \bibnamefont {Cheng}},
  \bibinfo {author} {\bibfnamefont {C.-Y.}\ \bibnamefont {Cheung}}, \bibinfo
  {author} {\bibfnamefont {G.-L.}\ \bibnamefont {Lin}}, \bibinfo {author}
  {\bibfnamefont {Y.~C.}\ \bibnamefont {Lin}}, \ and\ \bibinfo {author}
  {\bibfnamefont {H.-L.}\ \bibnamefont {Yu}},\ }\href {\doibase
  10.1103/PhysRevD.46.1148, 10.1103/PhysRevD.55.5851} {\bibfield  {journal}
  {\bibinfo  {journal} {Phys. Rev.}\ }\textbf {\bibinfo {volume} {D46}},\
  \bibinfo {pages} {1148} (\bibinfo {year} {1992})},\ \bibinfo {note}
  {[Erratum: Phys. Rev.D55,5851(1997)]}\BibitemShut {NoStop}%
\bibitem [{\citenamefont {Guo}\ \emph {et~al.}(2008)\citenamefont {Guo},
  \citenamefont {Hanhart}, \citenamefont {Krewald},\ and\ \citenamefont
  {Mei{\ss}ner}}]{Guo:2008gp}%
  \BibitemOpen
  \bibfield  {author} {\bibinfo {author} {\bibfnamefont {F.-K.}\ \bibnamefont
  {Guo}}, \bibinfo {author} {\bibfnamefont {C.}~\bibnamefont {Hanhart}},
  \bibinfo {author} {\bibfnamefont {S.}~\bibnamefont {Krewald}}, \ and\
  \bibinfo {author} {\bibfnamefont {U.-G.}\ \bibnamefont {Mei{\ss}ner}},\
  }\href {\doibase 10.1016/j.physletb.2008.07.060} {\bibfield  {journal}
  {\bibinfo  {journal} {Phys. Lett.}\ }\textbf {\bibinfo {volume} {B666}},\
  \bibinfo {pages} {251} (\bibinfo {year} {2008})},\ \Eprint
  {http://arxiv.org/abs/0806.3374} {arXiv:0806.3374 [hep-ph]} \BibitemShut
  {NoStop}%
\bibitem [{\citenamefont {Geng}\ \emph {et~al.}(2010)\citenamefont {Geng},
  \citenamefont {Kaiser}, \citenamefont {Martin-Camalich},\ and\ \citenamefont
  {Weise}}]{Geng:2010vw}%
  \BibitemOpen
  \bibfield  {author} {\bibinfo {author} {\bibfnamefont {L.~S.}\ \bibnamefont
  {Geng}}, \bibinfo {author} {\bibfnamefont {N.}~\bibnamefont {Kaiser}},
  \bibinfo {author} {\bibfnamefont {J.}~\bibnamefont {Martin-Camalich}}, \ and\
  \bibinfo {author} {\bibfnamefont {W.}~\bibnamefont {Weise}},\ }\href
  {\doibase 10.1103/PhysRevD.82.054022} {\bibfield  {journal} {\bibinfo
  {journal} {Phys. Rev. D}\ }\textbf {\bibinfo {volume} {82}},\ \bibinfo
  {pages} {054022} (\bibinfo {year} {2010})},\ \Eprint
  {http://arxiv.org/abs/1008.0383} {arXiv:1008.0383 [hep-ph]} \BibitemShut
  {NoStop}%
\bibitem [{\citenamefont {Yao}\ \emph {et~al.}(2015)\citenamefont {Yao},
  \citenamefont {Du}, \citenamefont {Guo},\ and\ \citenamefont
  {Mei{\ss}ner}}]{Yao:2015qia}%
  \BibitemOpen
  \bibfield  {author} {\bibinfo {author} {\bibfnamefont {D.-L.}\ \bibnamefont
  {Yao}}, \bibinfo {author} {\bibfnamefont {M.-L.}\ \bibnamefont {Du}},
  \bibinfo {author} {\bibfnamefont {F.-K.}\ \bibnamefont {Guo}}, \ and\
  \bibinfo {author} {\bibfnamefont {U.-G.}\ \bibnamefont {Mei{\ss}ner}},\
  }\href {\doibase 10.1007/JHEP11(2015)058} {\bibfield  {journal} {\bibinfo
  {journal} {JHEP}\ }\textbf {\bibinfo {volume} {11}},\ \bibinfo {pages} {058}
  (\bibinfo {year} {2015})},\ \Eprint {http://arxiv.org/abs/1502.05981}
  {arXiv:1502.05981 [hep-ph]} \BibitemShut {NoStop}%
\bibitem [{\citenamefont {Du}\ \emph {et~al.}(2017)\citenamefont {Du},
  \citenamefont {Guo}, \citenamefont {Mei{\ss}ner},\ and\ \citenamefont
  {Yao}}]{Du:2017ttu}%
  \BibitemOpen
  \bibfield  {author} {\bibinfo {author} {\bibfnamefont {M.-L.}\ \bibnamefont
  {Du}}, \bibinfo {author} {\bibfnamefont {F.-K.}\ \bibnamefont {Guo}},
  \bibinfo {author} {\bibfnamefont {U.-G.}\ \bibnamefont {Mei{\ss}ner}}, \ and\
  \bibinfo {author} {\bibfnamefont {D.-L.}\ \bibnamefont {Yao}},\ }\href
  {\doibase 10.1140/epjc/s10052-017-5287-6} {\bibfield  {journal} {\bibinfo
  {journal} {Eur. Phys. J.}\ }\textbf {\bibinfo {volume} {C77}},\ \bibinfo
  {pages} {728} (\bibinfo {year} {2017})},\ \Eprint
  {http://arxiv.org/abs/1703.10836} {arXiv:1703.10836 [hep-ph]} \BibitemShut
  {NoStop}%
\bibitem [{\citenamefont {Jiang}\ \emph {et~al.}(2019)\citenamefont {Jiang},
  \citenamefont {Liu},\ and\ \citenamefont {Yang}}]{Jiang:2019hgs}%
  \BibitemOpen
  \bibfield  {author} {\bibinfo {author} {\bibfnamefont {S.-Z.}\ \bibnamefont
  {Jiang}}, \bibinfo {author} {\bibfnamefont {Y.-R.}\ \bibnamefont {Liu}}, \
  and\ \bibinfo {author} {\bibfnamefont {Q.-H.}\ \bibnamefont {Yang}},\ }\href
  {\doibase 10.1103/PhysRevD.99.074018} {\bibfield  {journal} {\bibinfo
  {journal} {Phys. Rev. D}\ }\textbf {\bibinfo {volume} {99}},\ \bibinfo
  {pages} {074018} (\bibinfo {year} {2019})},\ \Eprint
  {http://arxiv.org/abs/1901.09479} {arXiv:1901.09479 [hep-ph]} \BibitemShut
  {NoStop}%
\bibitem [{\citenamefont {Gasser}\ and\ \citenamefont
  {Leutwyler}(1985)}]{Gasser:1984gg}%
  \BibitemOpen
  \bibfield  {author} {\bibinfo {author} {\bibfnamefont {J.}~\bibnamefont
  {Gasser}}\ and\ \bibinfo {author} {\bibfnamefont {H.}~\bibnamefont
  {Leutwyler}},\ }\href {\doibase 10.1016/0550-3213(85)90492-4} {\bibfield
  {journal} {\bibinfo  {journal} {Nucl. Phys.}\ }\textbf {\bibinfo {volume}
  {B250}},\ \bibinfo {pages} {465} (\bibinfo {year} {1985})}\BibitemShut
  {NoStop}%
\bibitem [{\citenamefont {Lutz}\ \emph {et~al.}(2018)\citenamefont {Lutz},
  \citenamefont {Heo},\ and\ \citenamefont {Guo}}]{Lutz:2018cqo}%
  \BibitemOpen
  \bibfield  {author} {\bibinfo {author} {\bibfnamefont {M.~F.~M.}\
  \bibnamefont {Lutz}}, \bibinfo {author} {\bibfnamefont {Y.}~\bibnamefont
  {Heo}}, \ and\ \bibinfo {author} {\bibfnamefont {X.-Y.}\ \bibnamefont
  {Guo}},\ }\href {\doibase 10.1016/j.nuclphysa.2018.05.007} {\bibfield
  {journal} {\bibinfo  {journal} {Nucl. Phys. A}\ }\textbf {\bibinfo {volume}
  {977}},\ \bibinfo {pages} {146} (\bibinfo {year} {2018})},\ \Eprint
  {http://arxiv.org/abs/1801.06417} {arXiv:1801.06417 [hep-lat]} \BibitemShut
  {NoStop}%
\bibitem [{\citenamefont {Bavontaweepanya}\ \emph {et~al.}(2018)\citenamefont
  {Bavontaweepanya}, \citenamefont {Guo},\ and\ \citenamefont
  {Lutz}}]{Bavontaweepanya:2018yds}%
  \BibitemOpen
  \bibfield  {author} {\bibinfo {author} {\bibfnamefont {R.}~\bibnamefont
  {Bavontaweepanya}}, \bibinfo {author} {\bibfnamefont {X.-Y.}\ \bibnamefont
  {Guo}}, \ and\ \bibinfo {author} {\bibfnamefont {M.~F.~M.}\ \bibnamefont
  {Lutz}},\ }\href {\doibase 10.1103/PhysRevD.98.056005} {\bibfield  {journal}
  {\bibinfo  {journal} {Phys. Rev. D}\ }\textbf {\bibinfo {volume} {98}},\
  \bibinfo {pages} {056005} (\bibinfo {year} {2018})},\ \Eprint
  {http://arxiv.org/abs/1801.10522} {arXiv:1801.10522 [hep-ph]} \BibitemShut
  {NoStop}%
\bibitem [{\citenamefont {Guo}\ and\ \citenamefont {Lutz}(2019)}]{Guo:2018zvl}%
  \BibitemOpen
  \bibfield  {author} {\bibinfo {author} {\bibfnamefont {X.-Y.}\ \bibnamefont
  {Guo}}\ and\ \bibinfo {author} {\bibfnamefont {M.~F.~M.}\ \bibnamefont
  {Lutz}},\ }\href {\doibase 10.1016/j.nuclphysa.2019.02.007} {\bibfield
  {journal} {\bibinfo  {journal} {Nucl. Phys. A}\ }\textbf {\bibinfo {volume}
  {988}},\ \bibinfo {pages} {48} (\bibinfo {year} {2019})},\ \Eprint
  {http://arxiv.org/abs/1810.07078} {arXiv:1810.07078 [hep-lat]} \BibitemShut
  {NoStop}%
\bibitem [{\citenamefont {Passarino}\ and\ \citenamefont
  {Veltman}(1979)}]{Passarino:1978jh}%
  \BibitemOpen
  \bibfield  {author} {\bibinfo {author} {\bibfnamefont {G.}~\bibnamefont
  {Passarino}}\ and\ \bibinfo {author} {\bibfnamefont {M.~J.~G.}\ \bibnamefont
  {Veltman}},\ }\href {\doibase 10.1016/0550-3213(79)90234-7} {\bibfield
  {journal} {\bibinfo  {journal} {Nucl. Phys.}\ }\textbf {\bibinfo {volume}
  {B160}},\ \bibinfo {pages} {151} (\bibinfo {year} {1979})}\BibitemShut
  {NoStop}%
\bibitem [{\citenamefont {Lutz}\ \emph {et~al.}(2020)\citenamefont {Lutz},
  \citenamefont {Sauerwein},\ and\ \citenamefont {Timmermans}}]{Lutz:2020dfi}%
  \BibitemOpen
  \bibfield  {author} {\bibinfo {author} {\bibfnamefont {M.~F.~M.}\
  \bibnamefont {Lutz}}, \bibinfo {author} {\bibfnamefont {U.}~\bibnamefont
  {Sauerwein}}, \ and\ \bibinfo {author} {\bibfnamefont {R.~G.~E.}\
  \bibnamefont {Timmermans}},\ }\href {\doibase 10.1140/epjc/s10052-020-8417-5}
  {\bibfield  {journal} {\bibinfo  {journal} {Eur. Phys. J. C}\ }\textbf
  {\bibinfo {volume} {80}},\ \bibinfo {pages} {844} (\bibinfo {year} {2020})},\
  \Eprint {http://arxiv.org/abs/2003.10158} {arXiv:2003.10158 [hep-lat]}
  \BibitemShut {NoStop}%
\bibitem [{\citenamefont {Sauerwein}\ \emph {et~al.}(2021)\citenamefont
  {Sauerwein}, \citenamefont {Lutz},\ and\ \citenamefont
  {Timmermans}}]{Sauerwein:2021jxb}%
  \BibitemOpen
  \bibfield  {author} {\bibinfo {author} {\bibfnamefont {U.}~\bibnamefont
  {Sauerwein}}, \bibinfo {author} {\bibfnamefont {M.~F.~M.}\ \bibnamefont
  {Lutz}}, \ and\ \bibinfo {author} {\bibfnamefont {R.~G.~E.}\ \bibnamefont
  {Timmermans}},\ }\href@noop {} {\  (\bibinfo {year} {2021})},\ \Eprint
  {http://arxiv.org/abs/2105.06755} {arXiv:2105.06755 [hep-ph]} \BibitemShut
  {NoStop}%
\bibitem [{\citenamefont {Lutz}(2000)}]{Lutz:1999yr}%
  \BibitemOpen
  \bibfield  {author} {\bibinfo {author} {\bibfnamefont {M.}~\bibnamefont
  {Lutz}},\ }\href {\doibase 10.1016/S0375-9474(00)00206-2} {\bibfield
  {journal} {\bibinfo  {journal} {Nucl. Phys. A}\ }\textbf {\bibinfo {volume}
  {677}},\ \bibinfo {pages} {241} (\bibinfo {year} {2000})},\ \Eprint
  {http://arxiv.org/abs/nucl-th/9906028} {arXiv:nucl-th/9906028} \BibitemShut
  {NoStop}%
\bibitem [{\citenamefont {Semke}\ and\ \citenamefont
  {Lutz}(2006)}]{Semke:2005sn}%
  \BibitemOpen
  \bibfield  {author} {\bibinfo {author} {\bibfnamefont {A.}~\bibnamefont
  {Semke}}\ and\ \bibinfo {author} {\bibfnamefont {M.~F.~M.}\ \bibnamefont
  {Lutz}},\ }\href {\doibase 10.1016/j.nuclphysa.2006.07.043} {\bibfield
  {journal} {\bibinfo  {journal} {Nucl. Phys. A}\ }\textbf {\bibinfo {volume}
  {778}},\ \bibinfo {pages} {153} (\bibinfo {year} {2006})},\ \Eprint
  {http://arxiv.org/abs/nucl-th/0511061} {arXiv:nucl-th/0511061} \BibitemShut
  {NoStop}%
\bibitem [{\citenamefont {Lutz}\ \emph {et~al.}(2014)\citenamefont {Lutz},
  \citenamefont {Bavontaweepanya}, \citenamefont {Kobdaj},\ and\ \citenamefont
  {Schwarz}}]{Lutz:2014oxa}%
  \BibitemOpen
  \bibfield  {author} {\bibinfo {author} {\bibfnamefont {M.~F.~M.}\
  \bibnamefont {Lutz}}, \bibinfo {author} {\bibfnamefont {R.}~\bibnamefont
  {Bavontaweepanya}}, \bibinfo {author} {\bibfnamefont {C.}~\bibnamefont
  {Kobdaj}}, \ and\ \bibinfo {author} {\bibfnamefont {K.}~\bibnamefont
  {Schwarz}},\ }\href {\doibase 10.1103/PhysRevD.90.054505} {\bibfield
  {journal} {\bibinfo  {journal} {Phys. Rev.}\ }\textbf {\bibinfo {volume}
  {D90}},\ \bibinfo {pages} {054505} (\bibinfo {year} {2014})},\ \Eprint
  {http://arxiv.org/abs/1401.7805} {arXiv:1401.7805 [hep-lat]} \BibitemShut
  {NoStop}%
\bibitem [{\citenamefont {Lutz}\ and\ \citenamefont
  {Vidana}(2012)}]{Lutz:2011xc}%
  \BibitemOpen
  \bibfield  {author} {\bibinfo {author} {\bibfnamefont {M.~F.~M.}\
  \bibnamefont {Lutz}}\ and\ \bibinfo {author} {\bibfnamefont {I.}~\bibnamefont
  {Vidana}},\ }\href {\doibase 10.1140/epja/i2012-12124-4} {\bibfield
  {journal} {\bibinfo  {journal} {Eur. Phys. J. A}\ }\textbf {\bibinfo {volume}
  {48}},\ \bibinfo {pages} {124} (\bibinfo {year} {2012})},\ \Eprint
  {http://arxiv.org/abs/1111.1838} {arXiv:1111.1838 [hep-ph]} \BibitemShut
  {NoStop}%
\bibitem [{\citenamefont {Jacob}\ and\ \citenamefont
  {Wick}(1959)}]{Jacob:1959at}%
  \BibitemOpen
  \bibfield  {author} {\bibinfo {author} {\bibfnamefont {M.}~\bibnamefont
  {Jacob}}\ and\ \bibinfo {author} {\bibfnamefont {G.~C.}\ \bibnamefont
  {Wick}},\ }\href {\doibase 10.1016/0003-4916(59)90051-X} {\bibfield
  {journal} {\bibinfo  {journal} {Annals Phys.}\ }\textbf {\bibinfo {volume}
  {7}},\ \bibinfo {pages} {404} (\bibinfo {year} {1959})}\BibitemShut {NoStop}%
\bibitem [{\citenamefont {Lutz}\ and\ \citenamefont
  {Kolomeitsev}(2004)}]{Lutz:2003fm}%
  \BibitemOpen
  \bibfield  {author} {\bibinfo {author} {\bibfnamefont {M.~F.~M.}\
  \bibnamefont {Lutz}}\ and\ \bibinfo {author} {\bibfnamefont {E.~E.}\
  \bibnamefont {Kolomeitsev}},\ }\href {\doibase
  10.1016/j.nuclphysa.2003.11.009} {\bibfield  {journal} {\bibinfo  {journal}
  {Nucl. Phys. A}\ }\textbf {\bibinfo {volume} {730}},\ \bibinfo {pages} {392}
  (\bibinfo {year} {2004})},\ \Eprint {http://arxiv.org/abs/nucl-th/0307039}
  {arXiv:nucl-th/0307039} \BibitemShut {NoStop}%
\bibitem [{\citenamefont {Jackson}\ and\ \citenamefont
  {Hite}(1968)}]{Jackson:1968rfn}%
  \BibitemOpen
  \bibfield  {author} {\bibinfo {author} {\bibfnamefont {J.~D.}\ \bibnamefont
  {Jackson}}\ and\ \bibinfo {author} {\bibfnamefont {G.~E.}\ \bibnamefont
  {Hite}},\ }\href {\doibase 10.1103/PhysRev.169.1248} {\bibfield  {journal}
  {\bibinfo  {journal} {Phys. Rev.}\ }\textbf {\bibinfo {volume} {169}},\
  \bibinfo {pages} {1248} (\bibinfo {year} {1968})}\BibitemShut {NoStop}%
\bibitem [{\citenamefont {Hara}(1964)}]{Hara:1964zza}%
  \BibitemOpen
  \bibfield  {author} {\bibinfo {author} {\bibfnamefont {Y.}~\bibnamefont
  {Hara}},\ }\href {\doibase 10.1103/PhysRev.136.B507} {\bibfield  {journal}
  {\bibinfo  {journal} {Phys. Rev.}\ }\textbf {\bibinfo {volume} {136}},\
  \bibinfo {pages} {B507} (\bibinfo {year} {1964})}\BibitemShut {NoStop}%
\bibitem [{\citenamefont {Lutz}\ and\ \citenamefont
  {Kolomeitsev}()}]{Lutz:2001yb}%
  \BibitemOpen
  \bibfield  {author} {\bibinfo {author} {\bibfnamefont {M.~F.~M.}\
  \bibnamefont {Lutz}}\ and\ \bibinfo {author} {\bibfnamefont {E.~E.}\
  \bibnamefont {Kolomeitsev}},\ }\href {\doibase 10.1016/S0375-9474(01)01312-4}
  {\bibfield  {journal} {\bibinfo  {journal} {Nucl. Phys. A}\ }\textbf
  {\bibinfo {volume} {700}},\ \bibinfo {pages} {193}},\ \Eprint
  {http://arxiv.org/abs/nucl-th/0105042} {arXiv:nucl-th/0105042} \BibitemShut
  {NoStop}%
\bibitem [{\citenamefont {Stoica}\ \emph {et~al.}(2011)\citenamefont {Stoica},
  \citenamefont {Lutz},\ and\ \citenamefont {Scholten}}]{Stoica:2011cy}%
  \BibitemOpen
  \bibfield  {author} {\bibinfo {author} {\bibfnamefont {S.}~\bibnamefont
  {Stoica}}, \bibinfo {author} {\bibfnamefont {M.~F.~M.}\ \bibnamefont {Lutz}},
  \ and\ \bibinfo {author} {\bibfnamefont {O.}~\bibnamefont {Scholten}},\
  }\href {\doibase 10.1103/PhysRevD.84.125001} {\bibfield  {journal} {\bibinfo
  {journal} {Phys. Rev. D}\ }\textbf {\bibinfo {volume} {84}},\ \bibinfo
  {pages} {125001} (\bibinfo {year} {2011})},\ \Eprint
  {http://arxiv.org/abs/1106.5619} {arXiv:1106.5619 [hep-ph]} \BibitemShut
  {NoStop}%
\bibitem [{\citenamefont {Heo}\ and\ \citenamefont {Lutz}(2014)}]{Heo:2014cja}%
  \BibitemOpen
  \bibfield  {author} {\bibinfo {author} {\bibfnamefont {Y.}~\bibnamefont
  {Heo}}\ and\ \bibinfo {author} {\bibfnamefont {M.~F.~M.}\ \bibnamefont
  {Lutz}},\ }\href {\doibase 10.1140/epja/i2014-14130-x} {\bibfield  {journal}
  {\bibinfo  {journal} {Eur. Phys. J. A}\ }\textbf {\bibinfo {volume} {50}},\
  \bibinfo {pages} {130} (\bibinfo {year} {2014})},\ \Eprint
  {http://arxiv.org/abs/1405.1597} {arXiv:1405.1597 [hep-ph]} \BibitemShut
  {NoStop}%
\bibitem [{\citenamefont {Lutz}\ \emph {et~al.}(2015)\citenamefont {Lutz},
  \citenamefont {Kolomeitsev},\ and\ \citenamefont {Korpa}}]{Lutz:2015lca}%
  \BibitemOpen
  \bibfield  {author} {\bibinfo {author} {\bibfnamefont {M.~F.~M.}\
  \bibnamefont {Lutz}}, \bibinfo {author} {\bibfnamefont {E.~E.}\ \bibnamefont
  {Kolomeitsev}}, \ and\ \bibinfo {author} {\bibfnamefont {C.~L.}\ \bibnamefont
  {Korpa}},\ }\href {\doibase 10.1103/PhysRevD.92.016003} {\bibfield  {journal}
  {\bibinfo  {journal} {Phys. Rev. D}\ }\textbf {\bibinfo {volume} {92}},\
  \bibinfo {pages} {016003} (\bibinfo {year} {2015})},\ \Eprint
  {http://arxiv.org/abs/1506.02375} {arXiv:1506.02375 [hep-ph]} \BibitemShut
  {NoStop}%
\bibitem [{\citenamefont {Lutz}\ and\ \citenamefont
  {Korpa}(2018)}]{Lutz:2018kaz}%
  \BibitemOpen
  \bibfield  {author} {\bibinfo {author} {\bibfnamefont {M.~F.~M.}\
  \bibnamefont {Lutz}}\ and\ \bibinfo {author} {\bibfnamefont {C.~L.}\
  \bibnamefont {Korpa}},\ }\href {\doibase 10.1103/PhysRevD.98.076003}
  {\bibfield  {journal} {\bibinfo  {journal} {Phys. Rev. D}\ }\textbf {\bibinfo
  {volume} {98}},\ \bibinfo {pages} {076003} (\bibinfo {year} {2018})},\
  \Eprint {http://arxiv.org/abs/1808.08695} {arXiv:1808.08695 [hep-ph]}
  \BibitemShut {NoStop}%
\bibitem [{\citenamefont {Chew}\ and\ \citenamefont
  {Mandelstam}(1960)}]{Chew:1960iv}%
  \BibitemOpen
  \bibfield  {author} {\bibinfo {author} {\bibfnamefont {G.~F.}\ \bibnamefont
  {Chew}}\ and\ \bibinfo {author} {\bibfnamefont {S.}~\bibnamefont
  {Mandelstam}},\ }\href {\doibase 10.1103/PhysRev.119.467} {\bibfield
  {journal} {\bibinfo  {journal} {Phys. Rev.}\ }\textbf {\bibinfo {volume}
  {119}},\ \bibinfo {pages} {467} (\bibinfo {year} {1960})}\BibitemShut
  {NoStop}%
\bibitem [{\citenamefont {Castillejo}\ \emph {et~al.}(1956)\citenamefont
  {Castillejo}, \citenamefont {Dalitz},\ and\ \citenamefont
  {Dyson}}]{Castillejo:1955ed}%
  \BibitemOpen
  \bibfield  {author} {\bibinfo {author} {\bibfnamefont {L.}~\bibnamefont
  {Castillejo}}, \bibinfo {author} {\bibfnamefont {R.~H.}\ \bibnamefont
  {Dalitz}}, \ and\ \bibinfo {author} {\bibfnamefont {F.~J.}\ \bibnamefont
  {Dyson}},\ }\href {\doibase 10.1103/PhysRev.101.453} {\bibfield  {journal}
  {\bibinfo  {journal} {Phys. Rev.}\ }\textbf {\bibinfo {volume} {101}},\
  \bibinfo {pages} {453} (\bibinfo {year} {1956})}\BibitemShut {NoStop}%
\bibitem [{\citenamefont {Aubin}\ \emph {et~al.}(2005)\citenamefont {Aubin},
  \citenamefont {Bernard}, \citenamefont {DeTar}, \citenamefont {Di~Pierro},
  \citenamefont {Freeland} \emph {et~al.}}]{Aubin:2005ar}%
  \BibitemOpen
  \bibfield  {author} {\bibinfo {author} {\bibfnamefont {C.}~\bibnamefont
  {Aubin}}, \bibinfo {author} {\bibfnamefont {C.}~\bibnamefont {Bernard}},
  \bibinfo {author} {\bibfnamefont {C.~E.}\ \bibnamefont {DeTar}}, \bibinfo
  {author} {\bibfnamefont {M.}~\bibnamefont {Di~Pierro}}, \bibinfo {author}
  {\bibfnamefont {E.~D.}\ \bibnamefont {Freeland}},  \emph {et~al.},\ }\href
  {\doibase 10.1103/PhysRevLett.95.122002} {\bibfield  {journal} {\bibinfo
  {journal} {Phys.Rev.Lett.}\ }\textbf {\bibinfo {volume} {95}},\ \bibinfo
  {pages} {122002} (\bibinfo {year} {2005})},\ \Eprint
  {http://arxiv.org/abs/hep-lat/0506030} {arXiv:hep-lat/0506030 [hep-lat]}
  \BibitemShut {NoStop}%
\bibitem [{\citenamefont {Follana}\ \emph {et~al.}(2008)\citenamefont
  {Follana}, \citenamefont {Davies}, \citenamefont {Lepage},\ and\
  \citenamefont {Shigemitsu}}]{Follana:2007uv}%
  \BibitemOpen
  \bibfield  {author} {\bibinfo {author} {\bibfnamefont {E.}~\bibnamefont
  {Follana}}, \bibinfo {author} {\bibfnamefont {C.}~\bibnamefont {Davies}},
  \bibinfo {author} {\bibfnamefont {G.}~\bibnamefont {Lepage}}, \ and\ \bibinfo
  {author} {\bibfnamefont {J.}~\bibnamefont {Shigemitsu}} (\bibinfo
  {collaboration} {HPQCD, UKQCD}),\ }\href {\doibase
  10.1103/PhysRevLett.100.062002} {\bibfield  {journal} {\bibinfo  {journal}
  {Phys.Rev.Lett.}\ }\textbf {\bibinfo {volume} {100}},\ \bibinfo {pages}
  {062002} (\bibinfo {year} {2008})},\ \Eprint {http://arxiv.org/abs/0706.1726}
  {arXiv:0706.1726 [hep-lat]} \BibitemShut {NoStop}%
\bibitem [{\citenamefont {Bazavov}\ \emph {et~al.}(2012)\citenamefont {Bazavov}
  \emph {et~al.}}]{Bazavov:2011aa}%
  \BibitemOpen
  \bibfield  {author} {\bibinfo {author} {\bibfnamefont {A.}~\bibnamefont
  {Bazavov}} \emph {et~al.} (\bibinfo {collaboration} {Fermilab Lattice,
  MILC}),\ }\href {\doibase 10.1103/PhysRevD.85.114506} {\bibfield  {journal}
  {\bibinfo  {journal} {Phys.Rev.}\ }\textbf {\bibinfo {volume} {D85}},\
  \bibinfo {pages} {114506} (\bibinfo {year} {2012})},\ \Eprint
  {http://arxiv.org/abs/1112.3051} {arXiv:1112.3051 [hep-lat]} \BibitemShut
  {NoStop}%
\bibitem [{\citenamefont {Mohler}\ \emph {et~al.}(2013)\citenamefont {Mohler},
  \citenamefont {Lang}, \citenamefont {Leskovec}, \citenamefont {Prelovsek},\
  and\ \citenamefont {Woloshyn}}]{Mohler:2013rwa}%
  \BibitemOpen
  \bibfield  {author} {\bibinfo {author} {\bibfnamefont {D.}~\bibnamefont
  {Mohler}}, \bibinfo {author} {\bibfnamefont {C.}~\bibnamefont {Lang}},
  \bibinfo {author} {\bibfnamefont {L.}~\bibnamefont {Leskovec}}, \bibinfo
  {author} {\bibfnamefont {S.}~\bibnamefont {Prelovsek}}, \ and\ \bibinfo
  {author} {\bibfnamefont {R.}~\bibnamefont {Woloshyn}},\ }\href {\doibase
  10.1103/PhysRevLett.111.222001} {\bibfield  {journal} {\bibinfo  {journal}
  {Phys.Rev.Lett.}\ }\textbf {\bibinfo {volume} {111}},\ \bibinfo {pages}
  {222001} (\bibinfo {year} {2013})},\ \Eprint {http://arxiv.org/abs/1308.3175}
  {arXiv:1308.3175 [hep-lat]} \BibitemShut {NoStop}%
\bibitem [{\citenamefont {Moir}\ \emph {et~al.}(2013)\citenamefont {Moir},
  \citenamefont {Peardon}, \citenamefont {Ryan}, \citenamefont {Thomas},\ and\
  \citenamefont {Liu}}]{Moir:2013ub}%
  \BibitemOpen
  \bibfield  {author} {\bibinfo {author} {\bibfnamefont {G.}~\bibnamefont
  {Moir}}, \bibinfo {author} {\bibfnamefont {M.}~\bibnamefont {Peardon}},
  \bibinfo {author} {\bibfnamefont {S.~M.}\ \bibnamefont {Ryan}}, \bibinfo
  {author} {\bibfnamefont {C.~E.}\ \bibnamefont {Thomas}}, \ and\ \bibinfo
  {author} {\bibfnamefont {L.}~\bibnamefont {Liu}},\ }\href {\doibase
  10.1007/JHEP05(2013)021} {\bibfield  {journal} {\bibinfo  {journal} {JHEP}\
  }\textbf {\bibinfo {volume} {1305}},\ \bibinfo {pages} {021} (\bibinfo {year}
  {2013})},\ \Eprint {http://arxiv.org/abs/1301.7670} {arXiv:1301.7670
  [hep-ph]} \BibitemShut {NoStop}%
\bibitem [{\citenamefont {Lang}\ \emph {et~al.}(2014)\citenamefont {Lang},
  \citenamefont {Leskovec}, \citenamefont {Mohler}, \citenamefont {Prelovsek},\
  and\ \citenamefont {Woloshyn}}]{Lang:2014yfa}%
  \BibitemOpen
  \bibfield  {author} {\bibinfo {author} {\bibfnamefont {C.}~\bibnamefont
  {Lang}}, \bibinfo {author} {\bibfnamefont {L.}~\bibnamefont {Leskovec}},
  \bibinfo {author} {\bibfnamefont {D.}~\bibnamefont {Mohler}}, \bibinfo
  {author} {\bibfnamefont {S.}~\bibnamefont {Prelovsek}}, \ and\ \bibinfo
  {author} {\bibfnamefont {R.}~\bibnamefont {Woloshyn}},\ }\href {\doibase
  10.1103/PhysRevD.90.034510} {\bibfield  {journal} {\bibinfo  {journal}
  {Phys.Rev.}\ }\textbf {\bibinfo {volume} {D90}},\ \bibinfo {pages} {034510}
  (\bibinfo {year} {2014})},\ \Eprint {http://arxiv.org/abs/1403.8103}
  {arXiv:1403.8103 [hep-lat]} \BibitemShut {NoStop}%
\bibitem [{\citenamefont {Bazavov}\ \emph {et~al.}(2014)\citenamefont {Bazavov}
  \emph {et~al.}}]{Bazavov:2014wgs}%
  \BibitemOpen
  \bibfield  {author} {\bibinfo {author} {\bibfnamefont {A.}~\bibnamefont
  {Bazavov}} \emph {et~al.} (\bibinfo {collaboration} {Fermilab Lattice,
  MILC}),\ }\href {\doibase 10.1103/PhysRevD.90.074509} {\bibfield  {journal}
  {\bibinfo  {journal} {Phys.Rev.}\ }\textbf {\bibinfo {volume} {D90}},\
  \bibinfo {pages} {074509} (\bibinfo {year} {2014})},\ \Eprint
  {http://arxiv.org/abs/1407.3772} {arXiv:1407.3772 [hep-lat]} \BibitemShut
  {NoStop}%
\bibitem [{\citenamefont {Aoki}\ \emph {et~al.}(2017)\citenamefont {Aoki} \emph
  {et~al.}}]{Aoki:2016frl}%
  \BibitemOpen
  \bibfield  {author} {\bibinfo {author} {\bibfnamefont {S.}~\bibnamefont
  {Aoki}} \emph {et~al.},\ }\href {\doibase 10.1140/epjc/s10052-016-4509-7}
  {\bibfield  {journal} {\bibinfo  {journal} {Eur. Phys. J.}\ }\textbf
  {\bibinfo {volume} {C77}},\ \bibinfo {pages} {112} (\bibinfo {year}
  {2017})},\ \Eprint {http://arxiv.org/abs/1607.00299} {arXiv:1607.00299
  [hep-lat]} \BibitemShut {NoStop}%
\bibitem [{\citenamefont {Walker-Loud}\ \emph {et~al.}(2009)\citenamefont
  {Walker-Loud}, \citenamefont {Lin}, \citenamefont {Richards}, \citenamefont
  {Edwards}, \citenamefont {Engelhardt} \emph {et~al.}}]{WalkerLoud:2008bp}%
  \BibitemOpen
  \bibfield  {author} {\bibinfo {author} {\bibfnamefont {A.}~\bibnamefont
  {Walker-Loud}}, \bibinfo {author} {\bibfnamefont {H.-W.}\ \bibnamefont
  {Lin}}, \bibinfo {author} {\bibfnamefont {D.}~\bibnamefont {Richards}},
  \bibinfo {author} {\bibfnamefont {R.}~\bibnamefont {Edwards}}, \bibinfo
  {author} {\bibfnamefont {M.}~\bibnamefont {Engelhardt}},  \emph {et~al.},\
  }\href {\doibase 10.1103/PhysRevD.79.054502} {\bibfield  {journal} {\bibinfo
  {journal} {Phys.Rev.}\ }\textbf {\bibinfo {volume} {D79}},\ \bibinfo {pages}
  {054502} (\bibinfo {year} {2009})},\ \Eprint {http://arxiv.org/abs/0806.4549}
  {arXiv:0806.4549 [hep-lat]} \BibitemShut {NoStop}%
\bibitem [{\citenamefont {Orginos}\ \emph {et~al.}(1999)\citenamefont
  {Orginos}, \citenamefont {Toussaint},\ and\ \citenamefont
  {Sugar}}]{Orginos:1999cr}%
  \BibitemOpen
  \bibfield  {author} {\bibinfo {author} {\bibfnamefont {K.}~\bibnamefont
  {Orginos}}, \bibinfo {author} {\bibfnamefont {D.}~\bibnamefont {Toussaint}},
  \ and\ \bibinfo {author} {\bibfnamefont {R.~L.}\ \bibnamefont {Sugar}}
  (\bibinfo {collaboration} {MILC}),\ }\href {\doibase
  10.1103/PhysRevD.60.054503} {\bibfield  {journal} {\bibinfo  {journal} {Phys.
  Rev.}\ }\textbf {\bibinfo {volume} {D60}},\ \bibinfo {pages} {054503}
  (\bibinfo {year} {1999})},\ \Eprint {http://arxiv.org/abs/hep-lat/9903032}
  {arXiv:hep-lat/9903032 [hep-lat]} \BibitemShut {NoStop}%
\bibitem [{\citenamefont {Orginos}\ and\ \citenamefont
  {Toussaint}(1999)}]{Orginos:1998ue}%
  \BibitemOpen
  \bibfield  {author} {\bibinfo {author} {\bibfnamefont {K.}~\bibnamefont
  {Orginos}}\ and\ \bibinfo {author} {\bibfnamefont {D.}~\bibnamefont
  {Toussaint}} (\bibinfo {collaboration} {MILC}),\ }\href {\doibase
  10.1103/PhysRevD.59.014501} {\bibfield  {journal} {\bibinfo  {journal} {Phys.
  Rev.}\ }\textbf {\bibinfo {volume} {D59}},\ \bibinfo {pages} {014501}
  (\bibinfo {year} {1999})},\ \Eprint {http://arxiv.org/abs/hep-lat/9805009}
  {arXiv:hep-lat/9805009 [hep-lat]} \BibitemShut {NoStop}%
\bibitem [{\citenamefont {Bernard}\ \emph {et~al.}(2001)\citenamefont {Bernard}
  \emph {et~al.}}]{MILC2001}%
  \BibitemOpen
  \bibfield  {author} {\bibinfo {author} {\bibfnamefont {C.~W.}\ \bibnamefont
  {Bernard}} \emph {et~al.},\ }\href {\doibase 10.1103/PhysRevD.64.054506}
  {\bibfield  {journal} {\bibinfo  {journal} {Phys. Rev.}\ }\textbf {\bibinfo
  {volume} {D64}},\ \bibinfo {pages} {054506} (\bibinfo {year} {2001})},\
  \Eprint {http://arxiv.org/abs/hep-lat/0104002} {arXiv:hep-lat/0104002}
  \BibitemShut {NoStop}%
\bibitem [{\citenamefont {Aubin}\ \emph {et~al.}(2004)\citenamefont {Aubin}
  \emph {et~al.}}]{MILC2004}%
  \BibitemOpen
  \bibfield  {author} {\bibinfo {author} {\bibfnamefont {C.}~\bibnamefont
  {Aubin}} \emph {et~al.},\ }\href {\doibase 10.1103/PhysRevD.70.094505}
  {\bibfield  {journal} {\bibinfo  {journal} {Phys. Rev.}\ }\textbf {\bibinfo
  {volume} {D70}},\ \bibinfo {pages} {094505} (\bibinfo {year} {2004})},\
  \Eprint {http://arxiv.org/abs/hep-lat/0402030} {arXiv:hep-lat/0402030}
  \BibitemShut {NoStop}%
\bibitem [{\citenamefont {Semke}\ and\ \citenamefont
  {Lutz}(2012)}]{Semke:2011ez}%
  \BibitemOpen
  \bibfield  {author} {\bibinfo {author} {\bibfnamefont {A.}~\bibnamefont
  {Semke}}\ and\ \bibinfo {author} {\bibfnamefont {M.~F.~M.}\ \bibnamefont
  {Lutz}},\ }\href {\doibase 10.1103/PhysRevD.85.034001} {\bibfield  {journal}
  {\bibinfo  {journal} {Phys.Rev.}\ }\textbf {\bibinfo {volume} {D85}},\
  \bibinfo {pages} {034001} (\bibinfo {year} {2012})},\ \Eprint
  {http://arxiv.org/abs/1111.0238} {arXiv:1111.0238 [hep-ph]} \BibitemShut
  {NoStop}%
\bibitem [{\citenamefont {Berlich}\ \emph {et~al.}(2010)\citenamefont
  {Berlich}, \citenamefont {Gabriel}, \citenamefont {Garcia},\ and\
  \citenamefont {Kunze}}]{Geneva}%
  \BibitemOpen
  \bibfield  {author} {\bibinfo {author} {\bibfnamefont {R.}~\bibnamefont
  {Berlich}}, \bibinfo {author} {\bibfnamefont {S.}~\bibnamefont {Gabriel}},
  \bibinfo {author} {\bibfnamefont {A.}~\bibnamefont {Garcia}}, \ and\ \bibinfo
  {author} {\bibfnamefont {M.}~\bibnamefont {Kunze}},\ }\href {www.gemfony.eu}
  {\bibfield  {journal} {\bibinfo  {journal} {Data Driven e-Science, Conference
  proceedings of ISGC 2010, Springer New York}\ ,\ \bibinfo {pages} {303}}
  (\bibinfo {year} {2010})}\BibitemShut {NoStop}%
\bibitem [{\citenamefont {Aoki}\ \emph {et~al.}(2021)\citenamefont {Aoki} \emph
  {et~al.}}]{Aoki:2021kgd}%
  \BibitemOpen
  \bibfield  {author} {\bibinfo {author} {\bibfnamefont {Y.}~\bibnamefont
  {Aoki}} \emph {et~al.},\ }\href@noop {} {\  (\bibinfo {year} {2021})},\
  \Eprint {http://arxiv.org/abs/2111.09849} {arXiv:2111.09849 [hep-lat]}
  \BibitemShut {NoStop}%
\bibitem [{\citenamefont {Nakamura}\ \emph {et~al.}(2010)\citenamefont
  {Nakamura} \emph {et~al.}}]{PDG}%
  \BibitemOpen
  \bibfield  {author} {\bibinfo {author} {\bibfnamefont {K.}~\bibnamefont
  {Nakamura}} \emph {et~al.} (\bibinfo {collaboration} {Particle Data Group}),\
  }\href {\doibase 10.1088/0954-3899/37/7A/075021} {\bibfield  {journal}
  {\bibinfo  {journal} {J. Phys.}\ }\textbf {\bibinfo {volume} {G37}},\
  \bibinfo {pages} {075021} (\bibinfo {year} {2010})}\BibitemShut {NoStop}%
\bibitem [{\citenamefont {Liu}\ \emph {et~al.}(2009)\citenamefont {Liu},
  \citenamefont {Liu},\ and\ \citenamefont {Zhu}}]{Liu:2009uz}%
  \BibitemOpen
  \bibfield  {author} {\bibinfo {author} {\bibfnamefont {Y.-R.}\ \bibnamefont
  {Liu}}, \bibinfo {author} {\bibfnamefont {X.}~\bibnamefont {Liu}}, \ and\
  \bibinfo {author} {\bibfnamefont {S.-L.}\ \bibnamefont {Zhu}},\ }\href
  {\doibase 10.1103/PhysRevD.79.094026} {\bibfield  {journal} {\bibinfo
  {journal} {Phys.Rev.}\ }\textbf {\bibinfo {volume} {D79}},\ \bibinfo {pages}
  {094026} (\bibinfo {year} {2009})},\ \Eprint {http://arxiv.org/abs/0904.1770}
  {arXiv:0904.1770 [hep-ph]} \BibitemShut {NoStop}%
\bibitem [{\citenamefont {Guo}\ and\ \citenamefont
  {Mei{\ss}ner}(2011)}]{Guo:2011dd}%
  \BibitemOpen
  \bibfield  {author} {\bibinfo {author} {\bibfnamefont {F.-K.}\ \bibnamefont
  {Guo}}\ and\ \bibinfo {author} {\bibfnamefont {U.-G.}\ \bibnamefont
  {Mei{\ss}ner}},\ }\href {\doibase 10.1103/PhysRevD.84.014013} {\bibfield
  {journal} {\bibinfo  {journal} {Phys. Rev.}\ }\textbf {\bibinfo {volume}
  {D84}},\ \bibinfo {pages} {014013} (\bibinfo {year} {2011})},\ \Eprint
  {http://arxiv.org/abs/1102.3536} {arXiv:1102.3536 [hep-ph]} \BibitemShut
  {NoStop}%
\bibitem [{\citenamefont {Guo}\ \emph {et~al.}(2019)\citenamefont {Guo},
  \citenamefont {Heo},\ and\ \citenamefont {Lutz}}]{Guo:2018gyd}%
  \BibitemOpen
  \bibfield  {author} {\bibinfo {author} {\bibfnamefont {X.-Y.}\ \bibnamefont
  {Guo}}, \bibinfo {author} {\bibfnamefont {Y.}~\bibnamefont {Heo}}, \ and\
  \bibinfo {author} {\bibfnamefont {M.~F.~M.}\ \bibnamefont {Lutz}},\ }\href
  {\doibase 10.1016/j.physletb.2019.02.022} {\bibfield  {journal} {\bibinfo
  {journal} {Phys. Lett. B}\ }\textbf {\bibinfo {volume} {791}},\ \bibinfo
  {pages} {86} (\bibinfo {year} {2019})},\ \Eprint
  {http://arxiv.org/abs/1809.01311} {arXiv:1809.01311 [hep-ph]} \BibitemShut
  {NoStop}%
\bibitem [{\citenamefont {Workman}\ and\ \citenamefont
  {Others}(2022)}]{Workman:2022ynf}%
  \BibitemOpen
  \bibfield  {author} {\bibinfo {author} {\bibfnamefont {R.~L.}\ \bibnamefont
  {Workman}}\ and\ \bibinfo {author} {\bibnamefont {Others}} (\bibinfo
  {collaboration} {Particle Data Group}),\ }\href {\doibase
  10.1093/ptep/ptac097} {\bibfield  {journal} {\bibinfo  {journal} {PTEP}\
  }\textbf {\bibinfo {volume} {2022}},\ \bibinfo {pages} {083C01} (\bibinfo
  {year} {2022})}\BibitemShut {NoStop}%
\end{thebibliography}%
\bibliographystyle{apsrev4-1}
\end{document}